\documentclass[backref, breaklinks, twocolumn, colorlinks]{aastex63}   
\hypersetup{linkcolor=blue,citecolor=blue,filecolor=cyan,urlcolor=black}

\usepackage{times,color,natbib,url,amsmath}
\usepackage{graphicx,float,psfrag}
\usepackage{grffile}
\usepackage{tabularx}
\usepackage{latexsym,amsmath,amssymb}
\usepackage{natbib}
\usepackage{verbatim}
\usepackage{multirow}
\usepackage{color}
\usepackage{longtable}
\usepackage[hang]{footmisc}
\usepackage{soul,xcolor}

\bibpunct{(}{)}{;}{a}{}{,}

\newcommand {\xmm} {\textsl{XMM-Newton}}
\newcommand {\chandra} {\textsl{Chandra}}
\newcommand {\swift} {\textsl{Swift}}

\newcommand {\fermi} {\textsl{Fermi}}

\newcommand {\nicer} {\textsl{NICER}}

\def \rsun {\ifmmode$R$_{\odot}\else R$_{\odot}$}

\def \hcm {\hbox {\ifmmode $ atoms cm$^{-2}\else atoms cm$^{-2}$\fi}}

\def\approxgt{\mathrel{\hbox{\rlap{\lower.55ex \hbox {$\sim$}}
        \kern-.3em \raise.4ex \hbox{$>$}}}}
\def\approxlt{\mathrel{\hbox{\rlap{\lower.55ex \hbox {$\sim$}}
        \kern-.3em \raise.4ex \hbox{$<$}}}}

\def \arcmin {\hbox{$^\prime$}}

\newcommand\T{\rule{0pt}{2.6ex}}
\newcommand\B{\rule[-1.2ex]{0pt}{0pt}}

\def \src {SGR~1935$+$2154}

\begin{document}
\tighten
\setstcolor{red}

\title{\rm \uppercase{The {\it NICER} View of the 2020 Burst Storm and Persistent Emission of \src\ }}

\author[0000-0002-7991-028X]{George~Younes}
\affiliation{Department of Physics, The George Washington University, Washington, DC 20052, USA, gyounes@gwu.edu}
\affiliation{Astronomy, Physics and Statistics Institute of Sciences (APSIS), The George Washington University, Washington, DC 20052, USA}

\author[0000-0002-3531-9842]{Tolga G\"uver}
\affiliation{Istanbul University, Science Faculty, Department of Astronomy and Space Sciences, Beyaz\i t, 34119, Istanbul, Turkey}
\affiliation{Istanbul University Observatory Research and Application Center, Istanbul University 34119, Istanbul Turkey}

\author[0000-0003-1443-593X]{Chryssa~Kouveliotou}
\affiliation{Department of Physics, The George Washington University, Washington, DC 20052, USA, gyounes@gwu.edu}
\affiliation{Astronomy, Physics and Statistics Institute of Sciences (APSIS), The George Washington University, Washington, DC 20052, USA}

\author[0000-0003-4433-1365]{Matthew~G.~Baring}
\affiliation{Science and Technology Institute, Universities Space Research Association, Huntsville, AL 35805, USA}

\author[0000-0001-8551-2002]{Chin-Ping Hu}
\affiliation{Department of Physics, National Changhua University of Education, Changhua 50007, Taiwan}

\author[0000-0002-9249-0515]{Zorawar~Wadiasingh}
\affiliation{Astrophysics Science Division, NASA Goddard Space Flight Center, Greenbelt, MD 20771}

\author[0000-0001-5072-8444]{Beste Begi\c{c}arslan}
\affiliation{Istanbul University, Science Faculty, Department of Astronomy and Space Sciences, Beyaz\i t, 34119, Istanbul, Turkey}

\author[0000-0003-1244-3100]{Teruaki Enoto}
\affiliation{Extreme Natural Phenomena RIKEN Hakubi Research Team, Cluster for Pioneering Research, RIKEN, 2-1 Hirosawa, Wako, Saitama 351-0198, Japan}

\author[0000-0002-5274-6790]{Ersin G\"o\u{g}\"u\c{s}}
\affiliation{Sabanc\i~University, Faculty of Engineering and Natural
  Sciences, \.Istanbul 34956 Turkey}

\author[0000-0002-0633-5325]{Lin Lin}
\affiliation{Department of Astronomy, Beijing Normal University, Beijing 100875, China}

\author{Alice~K.~Harding}
\affiliation{Astrophysics Science Division, NASA Goddard Space Flight Center, Greenbelt, MD 20771}

\author[0000-0001-9149-6707]{Alexander J. van der Horst}
\affiliation{Department of Physics, The George Washington University, Washington, DC 20052, USA, gyounes@gwu.edu}
\affiliation{Astronomy, Physics and Statistics Institute of Sciences (APSIS), The George Washington University, Washington, DC 20052, USA}

\author[0000-0002-4694-4221]{Walid A. Majid}
\affiliation{Jet Propulsion Laboratory, California Institute of Technology, Pasadena, CA 91109, USA}
\affiliation{California Institute of Technology, Pasadena, CA 91125, USA}

\author{Sebastien Guillot}
\affiliation{IRAP, CNRS, 9 avenue du Colonel Roche, BP 44346, F-31028
  Toulouse Cedex 4, France}

\author[0000-0002-0380-0041]{Christian~Malacaria}
\affiliation{NASA Marshall Space Flight Center, NSSTC, 320 Sparkman Drive, Huntsville, AL 35805, USA}
\affiliation{Universities Space Research Association, Science and Technology Institute, 320 Sparkman Drive, Huntsville, AL 35805, USA}

\begin{abstract}


  We report on \nicer\ observations of the magnetar \src, covering its
  2020 burst storm and long-term persistent emission evolution up to
  $\sim90$ days post outburst. During the first 1120~seconds taken on
  April 28 00:40:58 UTC we detect over 217 bursts, corresponding to a
  burst rate of $>0.2$~bursts~s$^{-1}$. Three hours later the rate is
  at 0.008~bursts~s$^{-1}$, remaining at a comparatively low level
  thereafter. The $T_{90}$ burst duration distribution peaks at
  840~ms; the distribution of waiting times to the next burst is fit
  with a log-normal with an average of 2.1~s. The 1-10~keV burst
  spectra are well fit by a blackbody, with an average temperature and
  area of $kT=1.7$~keV and $R^2=53$~km$^2$. The differential burst
  fluence distribution over $\sim3$ orders of magnitude is well
  modeled with a power-law form $dN/dF\propto F^{-1.5\pm0.1}$. The
  source persistent emission pulse profile is double-peaked hours
  after the burst storm. We find that the bursts peak arrival
    times follow a uniform distribution in pulse phase, though the
    fast radio burst associated with the source aligns in phase with
    the brighter peak. We measure the source spin-down from
  heavy-cadence observations covering days 21 to 39 post-outburst,
  $\dot\nu=-3.72(3)\times10^{-12}$~Hz~s$^{-1}$; a factor 2.7 larger
  than the value measured after the 2014 outburst. Finally, the
  persistent emission flux and blackbody temperature decrease rapidly
  in the early stages of the outburst, reaching quiescence 40 days
  later, while the size of the emitting area remains unchanged.

\end{abstract}

\section{Introduction}
\label{Intro}

Large variability patterns over a broad range of time-scales
(milliseconds to hours) is a defining property of magnetars, rarely
shared with other classes of the isolated neutron star family. Most
common are the short (average duration $\sim200$~ms) very bright
($L_{\rm X}\lesssim10^{42}$~erg~s$^{-1}$) hard X-ray bursts,
ubiquitously detected from the majority of the magnetar
population. These bursts can occur in isolation, with a single to a
few bursts observed,  \citep[e.g.,][]{an15ApJ:1841,younes20ApJ:1708},
or during a burst storm, when hundreds are detected within hours to
days from the start of the source activity
\citep[e.g.,][]{israel08ApJ:1900,vanderhorst12ApJ:1550}. The least
common form of magnetar bursting activity is the emission of a Giant
Flare (GF). GFs consist of an initial sub-second hard ``spike''
reaching luminosities of $\sim10^{47}$~erg~s$^{-1}$, followed by a
softer tail pulsating at the spin-period of the source and lasting for
several minutes. These events have so far been detected on three
occasions from three known magnetars
\citep[e.g.,][]{Mazets1979Nat,hurley99Natur,palmer05Natur}. On longer
time-scales, magnetars randomly enter active episodes, usually
associated contemporaneously with bursting activity, where their
persistent flux level increases by factors of few to a thousand,
accompanied by spectral and temporal variability. These properties
often recover to their pre-outburst levels months to years after
activity cessation \citep[e.g.,][]{woods04ApJ:1E2259,rea13ApJ:0418,
  scholz14ApJ:1822,younes17ApJ:1806,cotizelati18MNRAS}. Magnetars are
widely believed to be powered by the decay of their super-critical
external magnetic fields, often in excess of $10^{14}$~G
\citep{kouveliotou98Nat:1806}, and perhaps larger internal ones
\citep[][see also \citealt{turolla15:mag,kaspi17:magnetars} for
reviews]{thompson96ApJ:magnetar}.

\src\ was discovered in 2014 when \swift-BAT triggered on
magnetar-like bursts from the Galactic plane direction \citep{
  stamatikos14:1935}, close to the geometrical center of the SNR
G57.2$+$0.8 \citep{kothes18ApJ:1935}. Subsequent \chandra\ and \xmm\
monitoring revealed a source spin-period $P=3.24\,$s and a spin-down
rate $\dot{P}=1.43\times10^{-11}$~s~s$^{-1}$, implying a magnetar-like
dipolar B-field strength, $B\approx 2.2\times10^{14}$~G, at the
equator, which, together with the bursts, cemented its identification
as a magnetar source \citep{israel16mnras:1935}. Since discovery,
\src\ has been very active, showing outbursts in 2015 and 2016, each
more intense than the preceding one, in terms of total number of
bursts per active episode and total energy emitted in bursts and in
persistent emission from the source \citep{younes17:1935,lin20ApJ:1935}. 

On 2020 April 27, a multitude of wide field-of-view instruments
detected intense bursting activity from \src, comprising its most
prolific episode since discovery \citep{fletcher20GCN:1935,
  palmer20ATel13675,younes20ATel13678}. Hours after the initial
trigger, an intense radio burst from the direction of \src\ was
independently detected with the Canadian Hydrogen Intensity Mapping
Experiment \citep[CHIME,][]{chime20arXiv200510324T}  and the Survey
for Transient Astronomical Radio Emission 2 \citep[STARE2,][]{
  Bochenek20arXiv200510828B} radio telescopes at 400–800 MHz and 1.4
GHz, respectively. This radio burst had fluence of the order of
1~MJy~ms, bright enough to be potentially detectable at distances of
several tens of Mpc by existing large radio facilities
\citep{Bochenek20arXiv200510828B}; this places it close to the faint end
of the extragalactic fast radio burst (FRB) population. Simultaneous
to the radio burst, multiple hard X-ray telescopes detected a
magnetar-like burst from \src, with a spectrum somewhat harder
than previously observed from the source \citep{mereghetti20ApJ:1935,
  li20arXiv200511071L,Ridnaia20arXiv200511178R,tavani20arXiv200512164T}.
This exceptional FRB$-$X-ray burst association placed magnetars at
center stage as the potential origin of at least some extragalactic
fast radio bursts. Interestingly, \src\ has since shown several 
millisecond radio bursts with fluences between 3 and 7 orders of
magnitude smaller than the FRB-like burst \citep{zhang20ATel13699,
  2020arXiv200705101K,good20ATel14074,pleunis20ATel14080,
  zhu20ATel14084}.

Here, we report on the burst storm of \src\ as observed with \nicer\
hours after the activity onset, as well as the ensuing outburst
evolution of the source persistent emission. Section 2 summarizes the
observations and data reduction. Section 3 presents the temporal and
spectral analyses of the burst storm, while Section 4 discusses the
analysis of the persistent emission up to 90 days following the
outburst onset. We discuss our findings in Section 5.

Throughout the paper, we adopt a fiducial distance towards \src\ of
9~kpc due to the large uncertainties in its distance estimate
\citep{kothes18ApJ:1935,zhou20arXiv200503517Z,mereghetti20ApJ:1935,
  zhong20ApJ}.

\vspace{1cm}

\section{Observations and data reduction}
\label{obs}

\begin{figure*}[!th]
\begin{center}
\includegraphics[angle=0,width=0.90\textwidth]{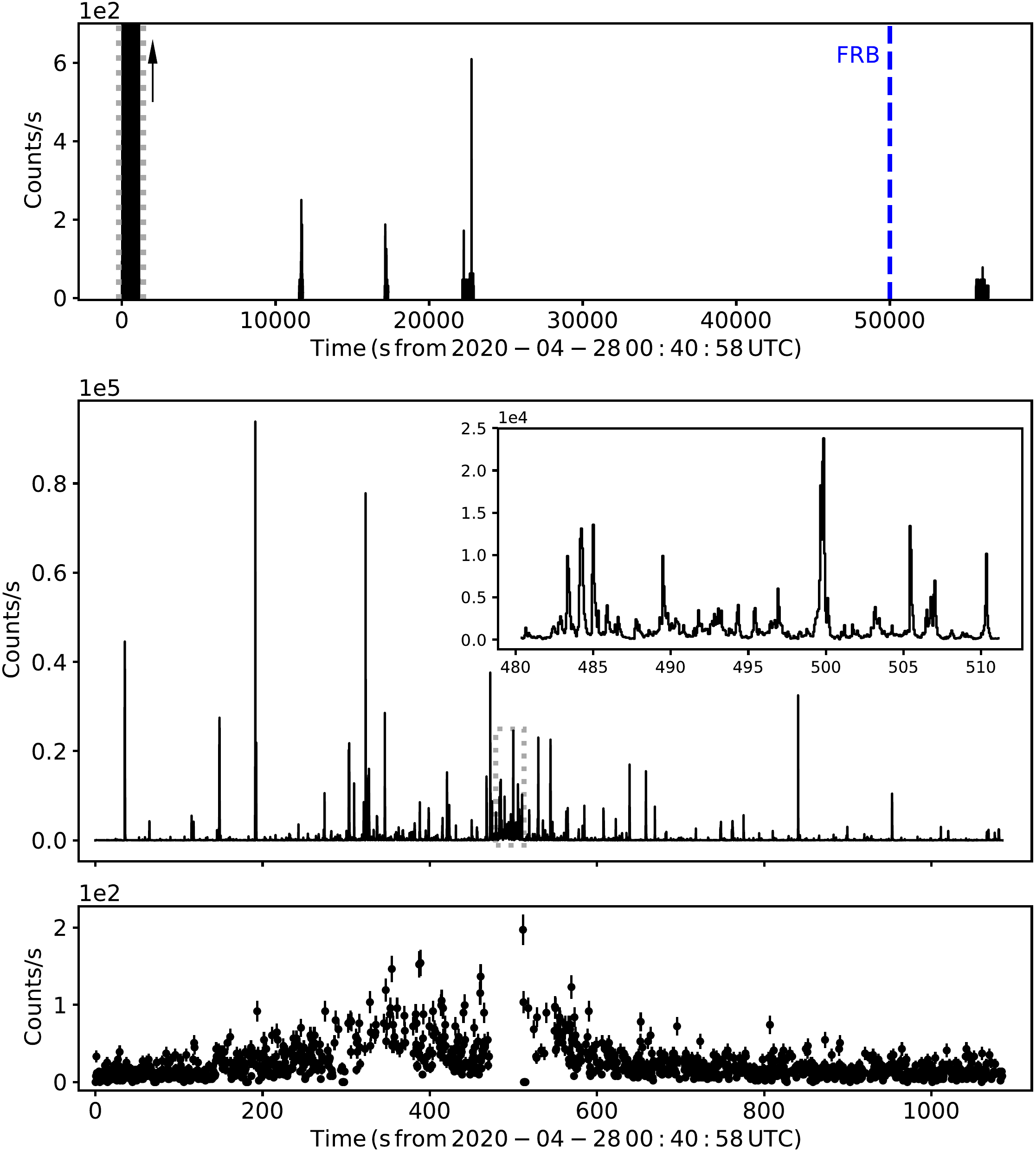}
\caption{{\sl Upper panel.} \nicer\ light curve of observation ID
  3020560101 shown at 64~ms resolution in the 1-10~keV energy range. The
  dashed blue vertical line is the time of FRB~200428. The gray dashed
  vertical lines delimit the first GTI when the burst forest
  occurred. The arrow indicates that
  the count rate is outside the y-axis limit. {\sl Middle panel.} A ``zoom-in'' view of the burst forest. We
  detect more than 217  bursts during $\sim$1120 seconds. The inset is
  a zoom-in at the area delimited with a dotted gray box , representing
  the most intense bursting period. {\sl Lower panel.} The light curve (with 0.5\,s resolution)
  of the burst forest after eliminating all identified bursts.}
\label{stormLC}
\end{center}
\end{figure*}

\nicer\ is a non-imaging X-ray timing and spectral instrument
providing a collecting area of 1900~cm$^2$ at 1.5 keV. It consists of
56 co-aligned X-ray concentrating optics covering a (30\arcmin)$^2$
field of view \citep{gendreau16SPIE}, 52 of which are currently
operating. We utilize all 52 detectors for our burst analyses. \nicer\
started observing \src\ on 2020 April 28 at 00:40:58 UTC, six hours
after the initial \swift-BAT and \fermi-GBM triggers that signaled the
start of another burst active period from the source, and just under
14 hours prior to the FRB \citep{Barthelmy20GCN,fletcher20GCN:1935}.
The first observation, with ID 3020560101, had an exposure of 3.1 ks
spread over a large portion of April 28. We show the 1-10~keV
light curve at the 64~ms resolution in the upper panel of
Figure~\ref{stormLC}. The first uninterrupted good time interval
(GTI), shown in the middle panel of Figure~\ref{stormLC} and totalling
1120 seconds, caught the tail end of the burst storm from the
source. We focus on this observation, and more specifically on the
first GTI, for the analysis of the burst forest. For the outburst
evolution, however, we analyze all publicly available \nicer\
observations, as summarized in Table~\ref{tab:pers_sp_res}.

We processed the \nicer\ data using NICERDAS version 7a, as part of
HEASOFT version 6.27.2. We start our data reduction from level 1 event
files. We create good time intervals using standard filtering criteria
as described in the \nicer\ Data Analysis Guide
\footnote{\href{https://heasarc.gsfc.nasa.gov/docs/nicer/data_analysis/nicer_analysis_guide.html}{https://heasarc.gsfc.nasa.gov/docs/nicer/data\_analysis/nicer\_analysis\_guide.html}}.
Due to the relatively large absorbing hydrogen column density in
  the direction of \src\ and the reduced \nicer\ sensitivity at high
  energies, we only consider photons in the energy range $1-10$~keV
  for our temporal and spectral analyses. For the analysis of the
bursts, we correct for the loss of exposure fraction due to deadtime
following the steps described in \citet{younes20arXiv200611358Y}. We
use the response matrices and ancillary files given in the latest
\nicer\ calibration files, version 20200722. The background is
estimated from a 40-second long interval centered on each burst peak
time, after excluding all identified bursts (Figure~\ref{stormLC},
bottom panel). As for the persistent emission, we estimate the
background using the \nicer\ tool
\texttt{nibackgen3C50}\footnote{\href{https://heasarc.gsfc.nasa.gov/docs/nicer/tools/nicer\_bkg\_est\_tools.html}{https://heasarc.gsfc.nasa.gov/docs/nicer/tools/nicer\_bkg\_est\_tools.html}}. A
bright dust scattering halo was detected in \swift-XRT on April 27,
which decayed very rapidly to almost background level on April 28
\citep[][their Figure 5]{mereghetti20ApJ:1935}. Hence, this halo
emission may marginally contribute to the sky background, especially
during the first GTI of our first \nicer\ observation
(Figure~\ref{stormLC}), which we only use to analyze the bursts. This
excess emission is automatically accounted for since it is part of the
background of each burst as defined above.

We use XSPEC version 12.11.0k \citep{arnaud96conf} to perform all
spectral analyses. For the burst spectral analysis, we group the
spectra to have five counts per energy bin, and use the W-statistic
(command \texttt{statistic Cstat} in XSPEC) for model parameter
estimation and error calculation. For the persistent emission
spectra, we group each spectrum to have at least 50 counts per energy
channel for the use of the $\chi^2$ statistics. We use the
T\"{u}bingen-Boulder model (\texttt{tbabs}) to account for
interstellar absorption in the direction of \src, along with the
abundances of \citet{wilms00ApJ} and the photoelectric cross-sections
of \citet{verner96ApJ:crossSect}. We report all parameter
uncertainties at the $1\sigma$ level, unless quoted otherwise.

\section{Results}
\label{res}

\subsection{Burst identification}

\begin{figure*}[th!]
\begin{center}
\includegraphics[angle=0,width=0.45\textwidth]{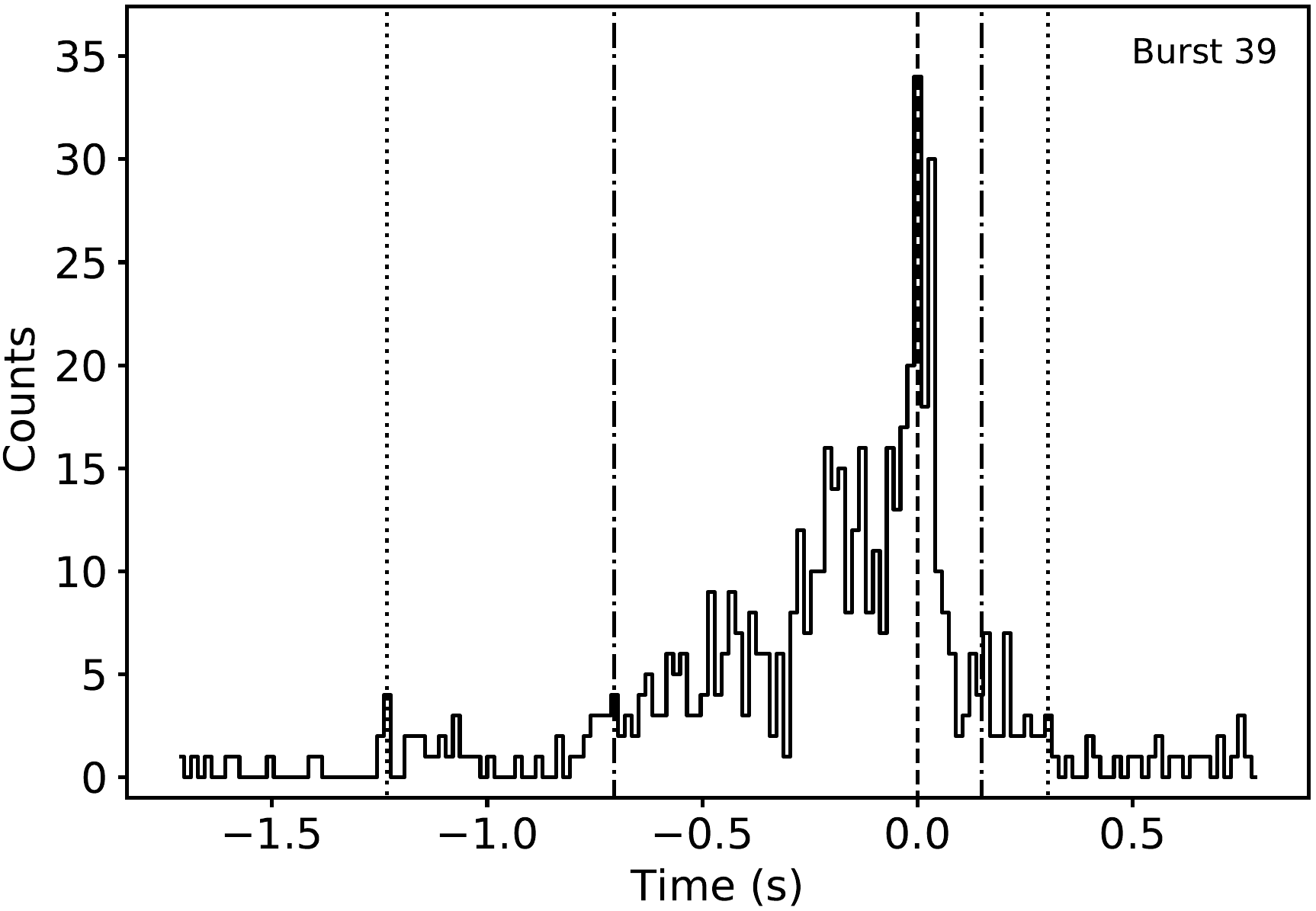}
\includegraphics[angle=0,width=0.48\textwidth]{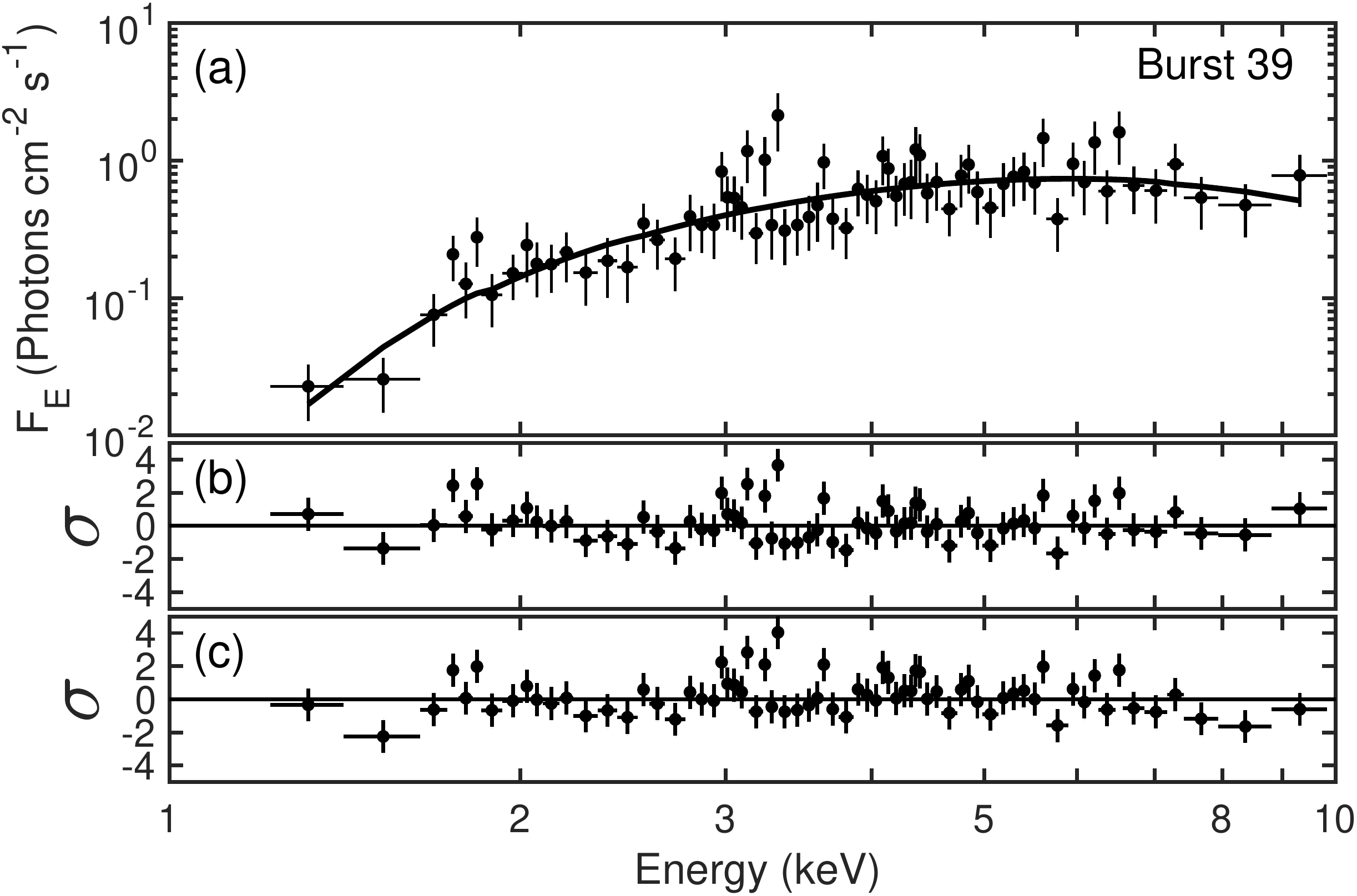}\\
\includegraphics[angle=0,width=0.45\textwidth]{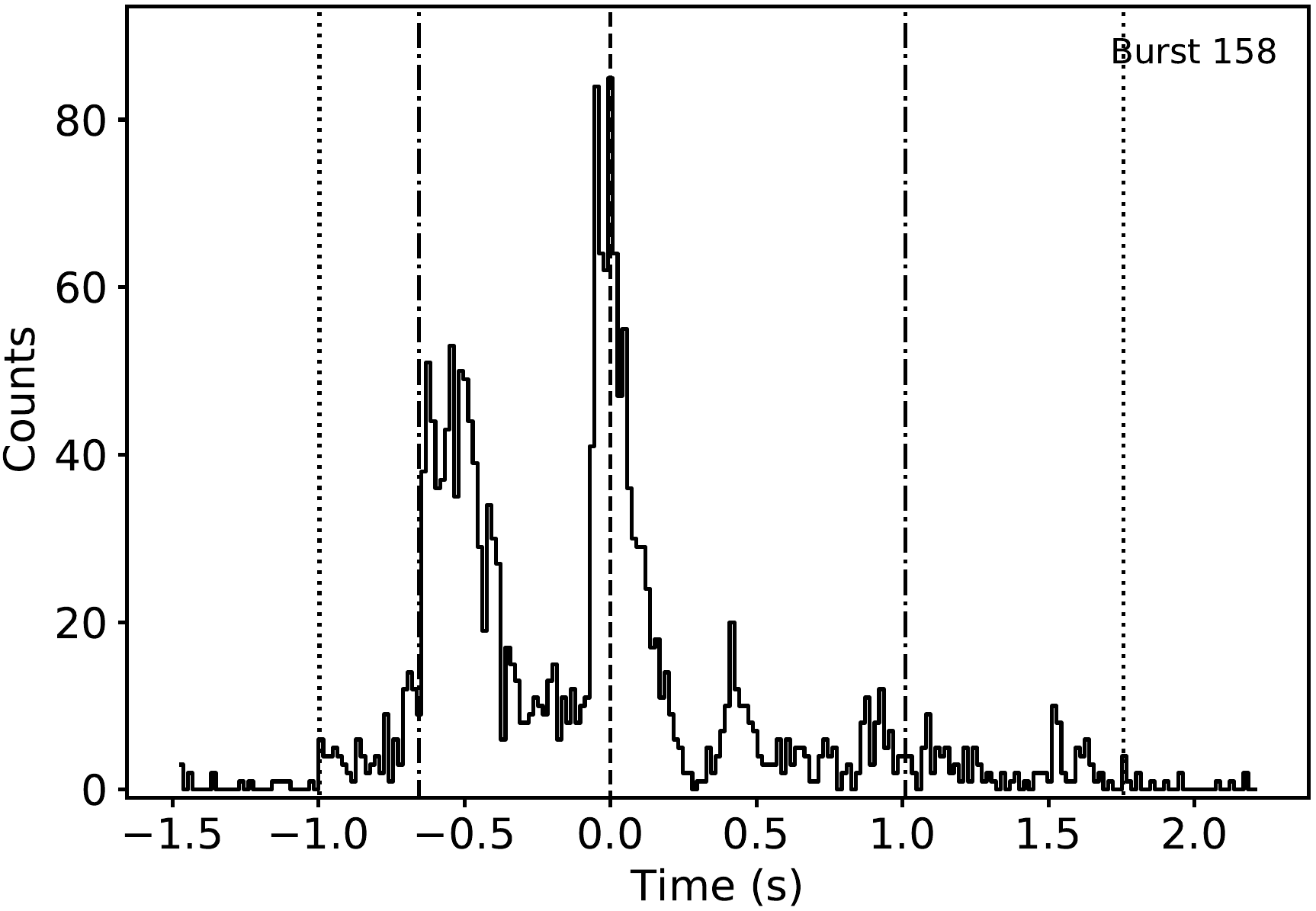}
\includegraphics[angle=0,width=0.48\textwidth]{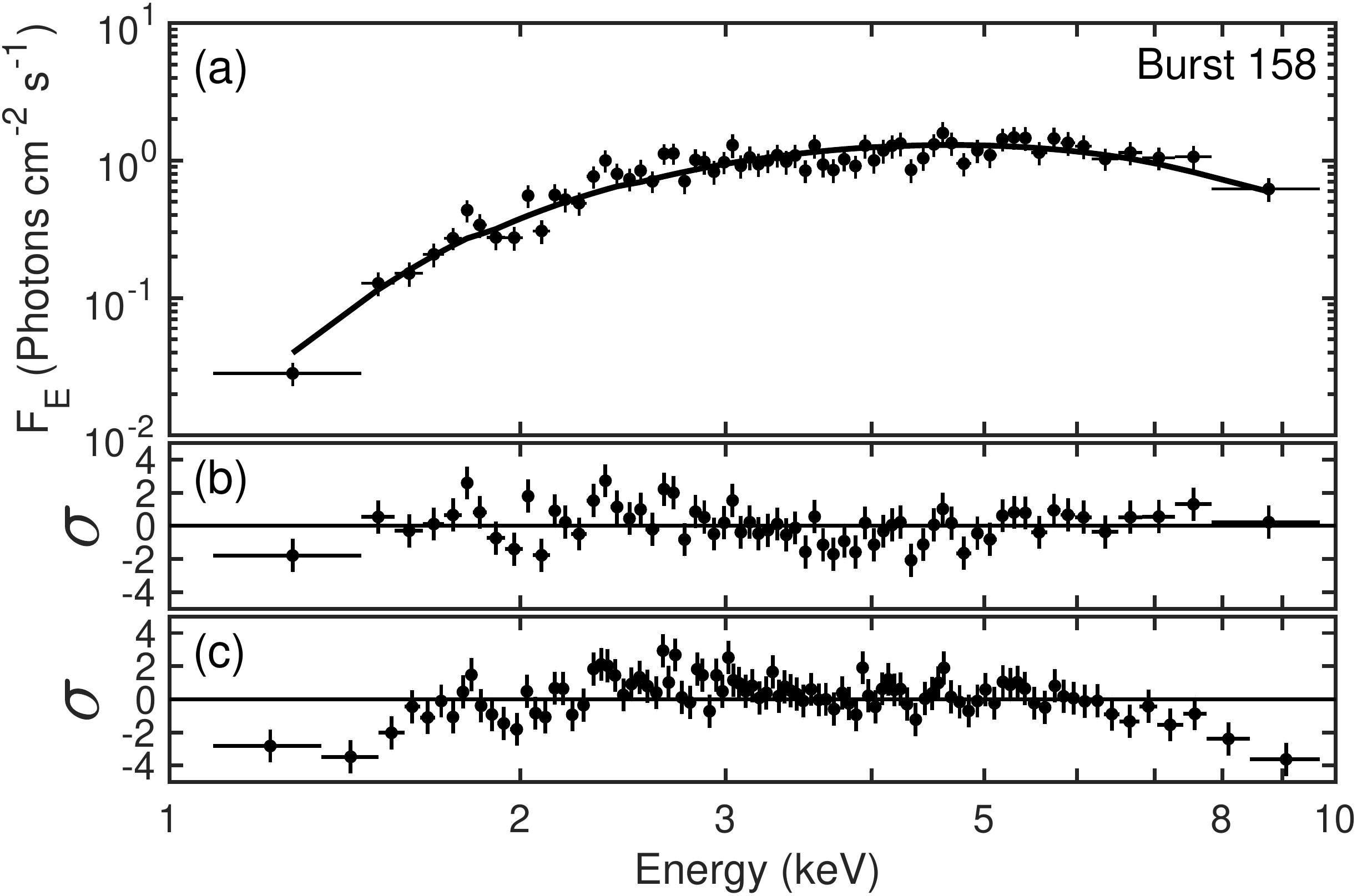}\\
\includegraphics[angle=0,width=0.45\textwidth]{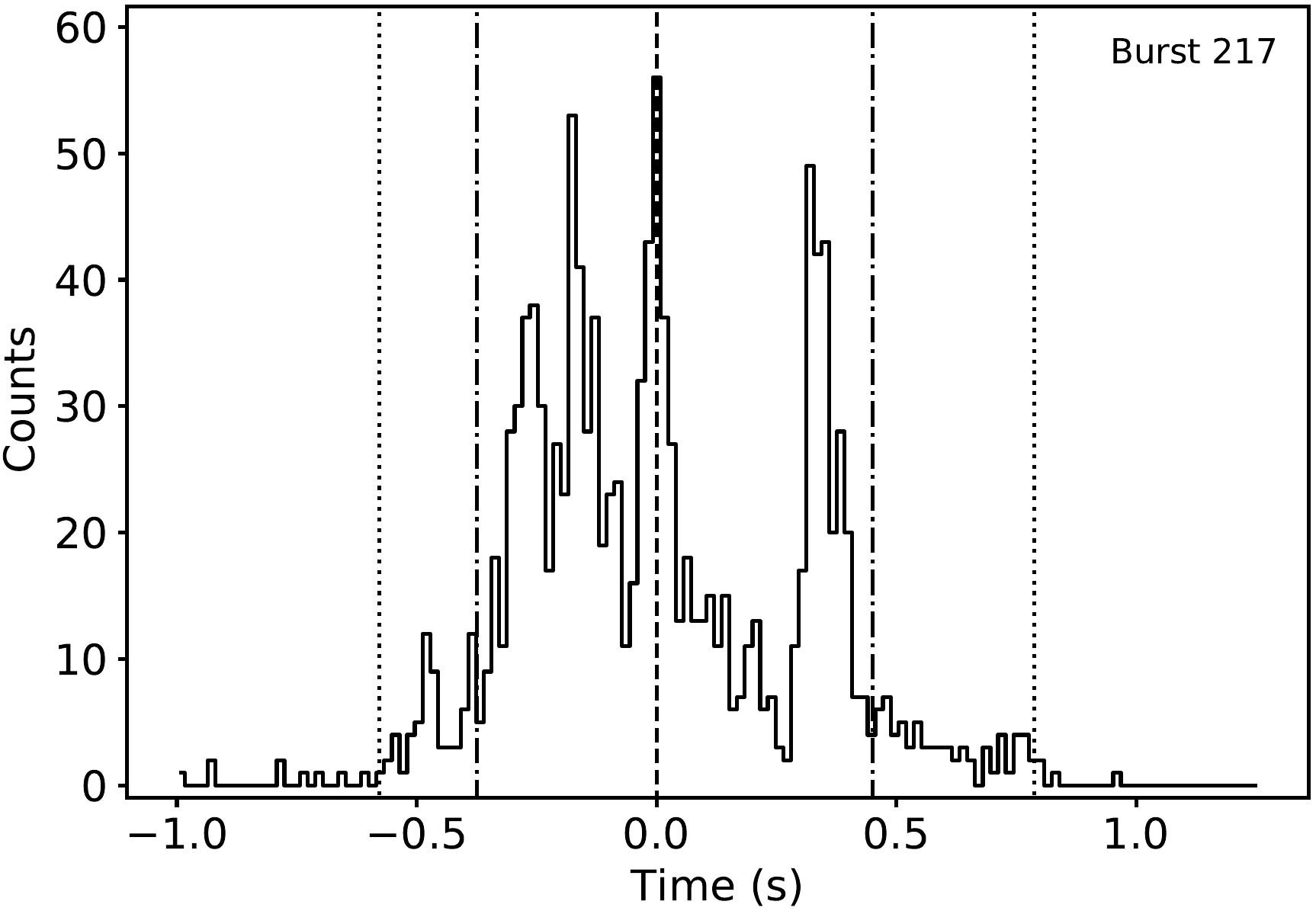}
\includegraphics[angle=0,width=0.48\textwidth]{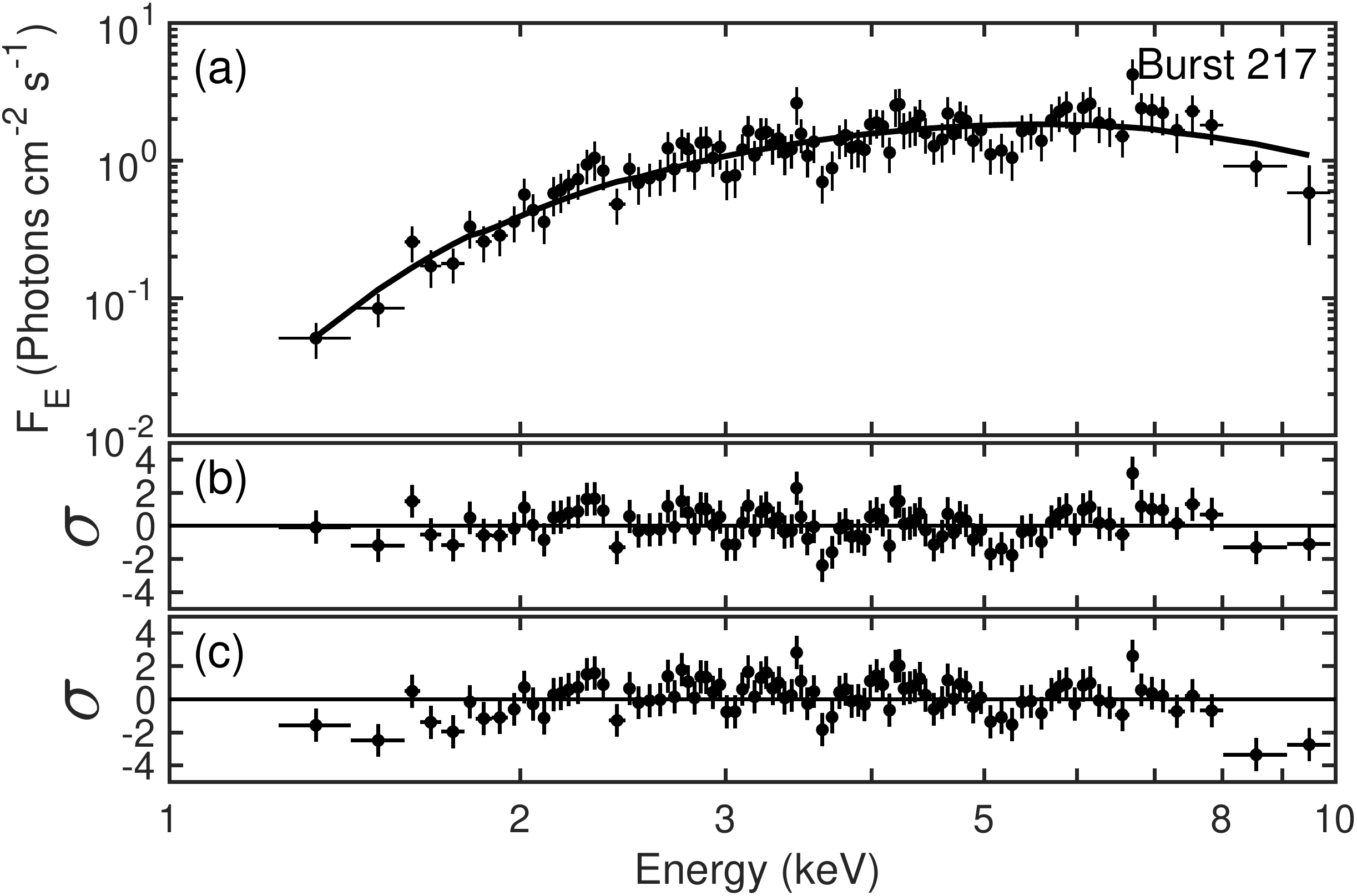}
\caption{{\sl Left panels.} Three examples of bursts emitted during
  the burst forest plotted with 16~ms resolution in the 1-10~keV energy range. The vertical
  dotted lines delimit the $T_{\rm 100}$ start and end times of the bursts. The
  dot-dashed lines delimit the start and end times of the $T_{90}$
  interval. The peak time of the bursts is shown as a dashed
  line and corresponds to t=0. {\sl Right panels.} Spectra of each of
  the bursts. Panels (a) show the data in $F_{\rm E}$ space and
  the best fit absorbed BB model. Panels (b) are the
  corresponding residuals in terms of $\sigma$. Panels (c) show
  the residuals of an absorbed PL model for comparison. See text for details.}
\label{exLCs}
\end{center}
\end{figure*}

\begin{figure*}[!th]
\begin{center}
  \includegraphics[angle=0,width=0.325\textwidth]{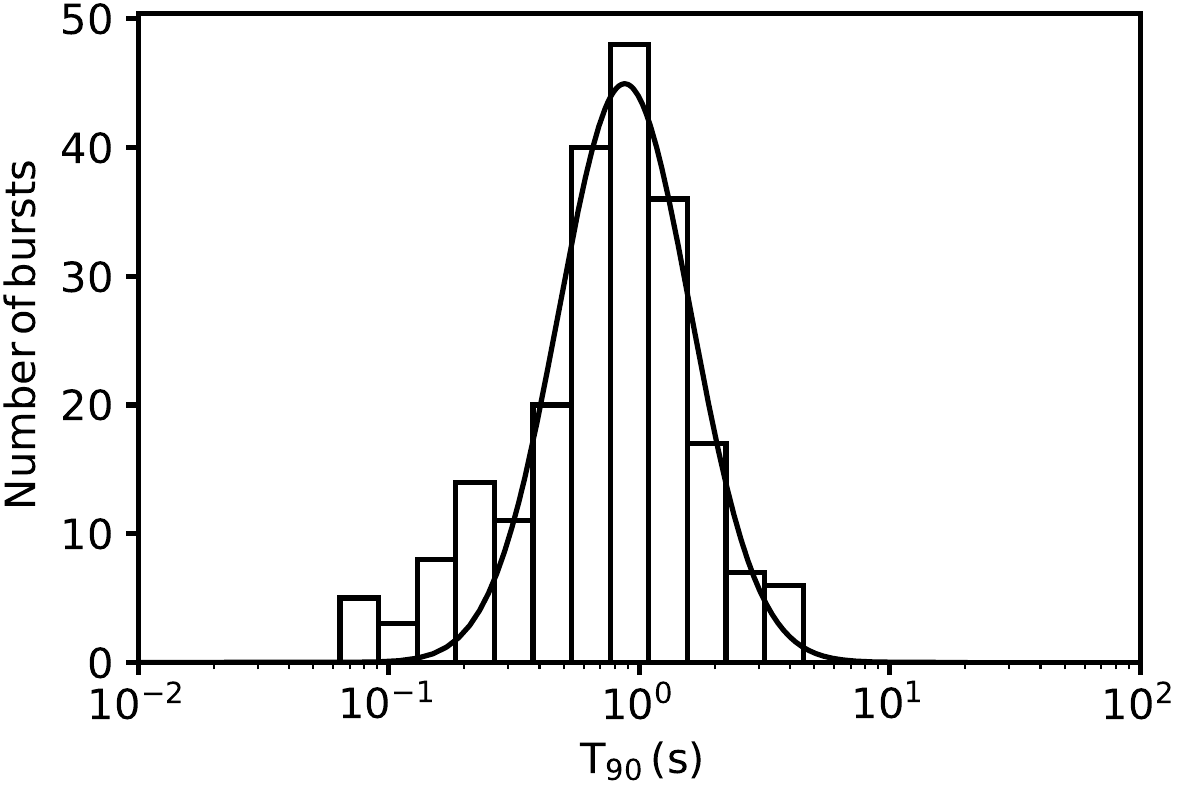}
  \includegraphics[angle=0,width=0.33\textwidth]{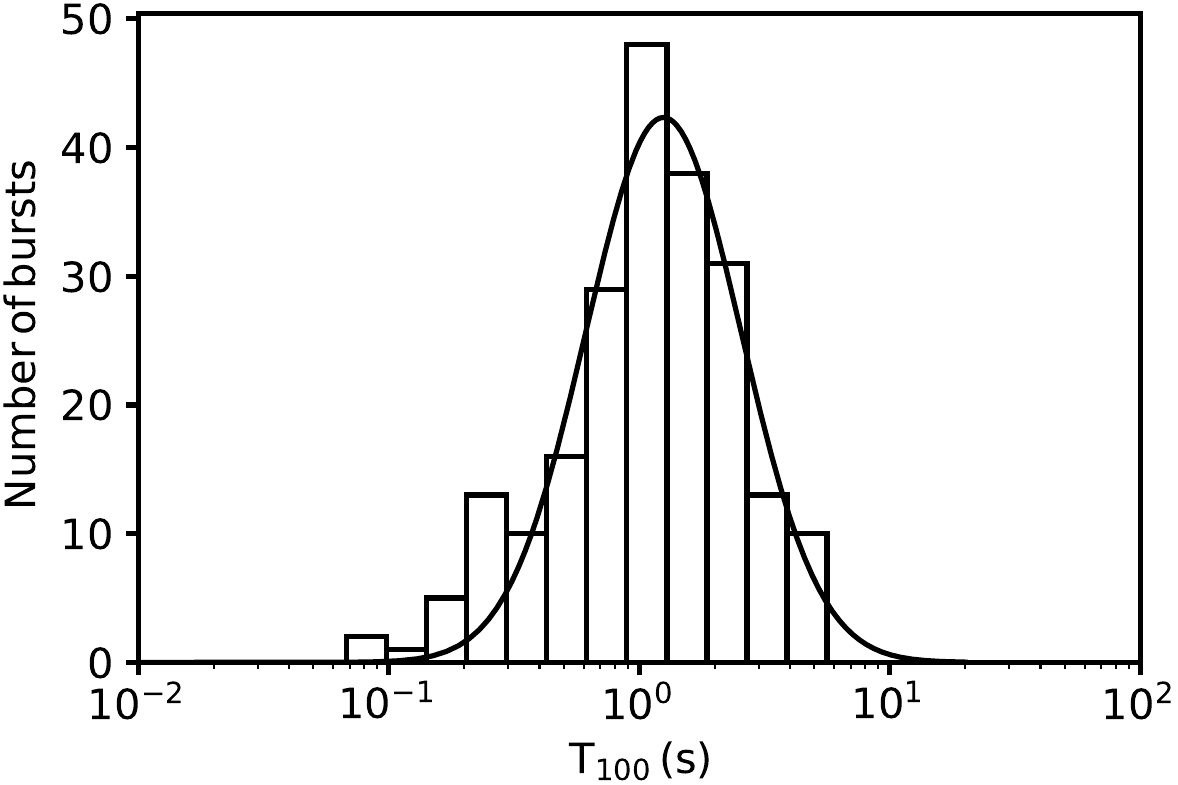}
  \includegraphics[angle=0,width=0.325\textwidth]{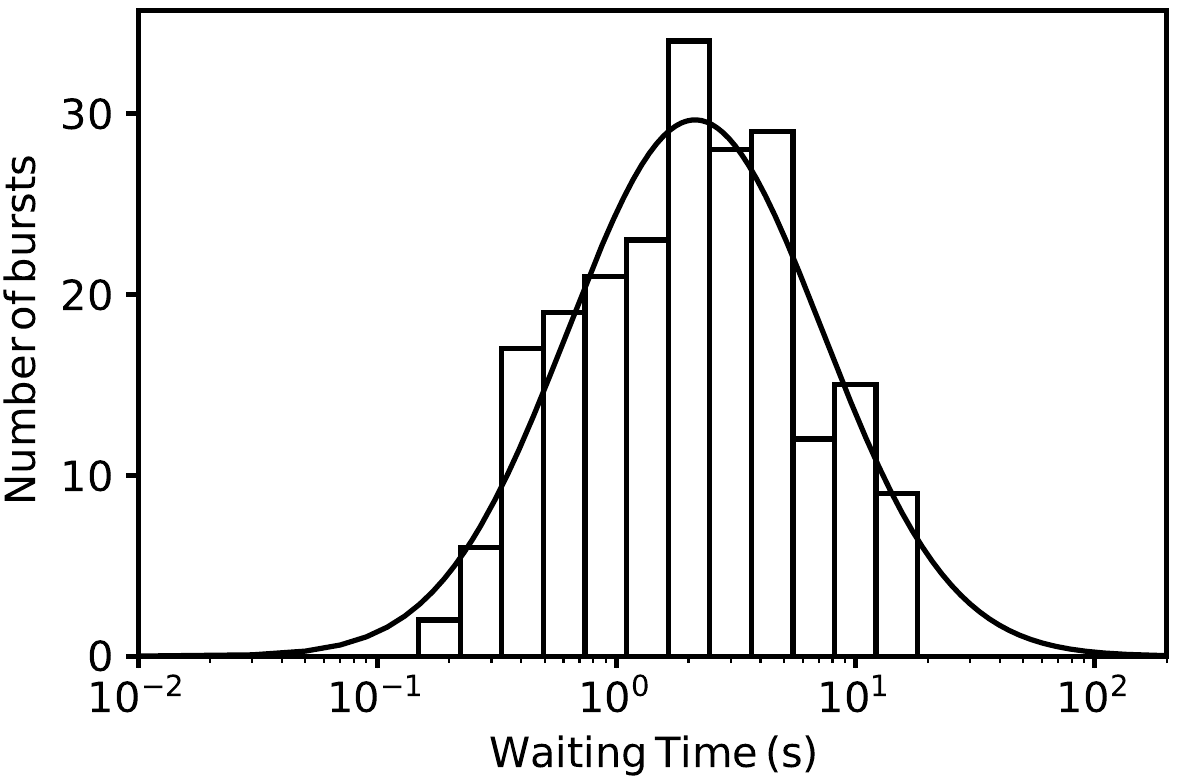}
\caption{{\sl Left panel.} $T_{90}$ distribution of the \src\ bursts
  detected during the burst storm. The solid black line is the best
  fit log-normal function to the distribution with a mean of
  840~ms. {\sl Right panel.} $T_{100}$ distribution along with the best
  fit log-normal function (solid black line) with a mean of
  1.27~s. {\sl Right panel.} Waiting-time distribution and the
  best fit log-normal function shown as black solid line. The average
  waiting time between bursts is 2.1~s. See text for more details.}
\label{distT90WT}
\end{center}
\end{figure*}

\begin{figure*}[]
\begin{center}
\includegraphics[angle=0,width=0.33\textwidth]{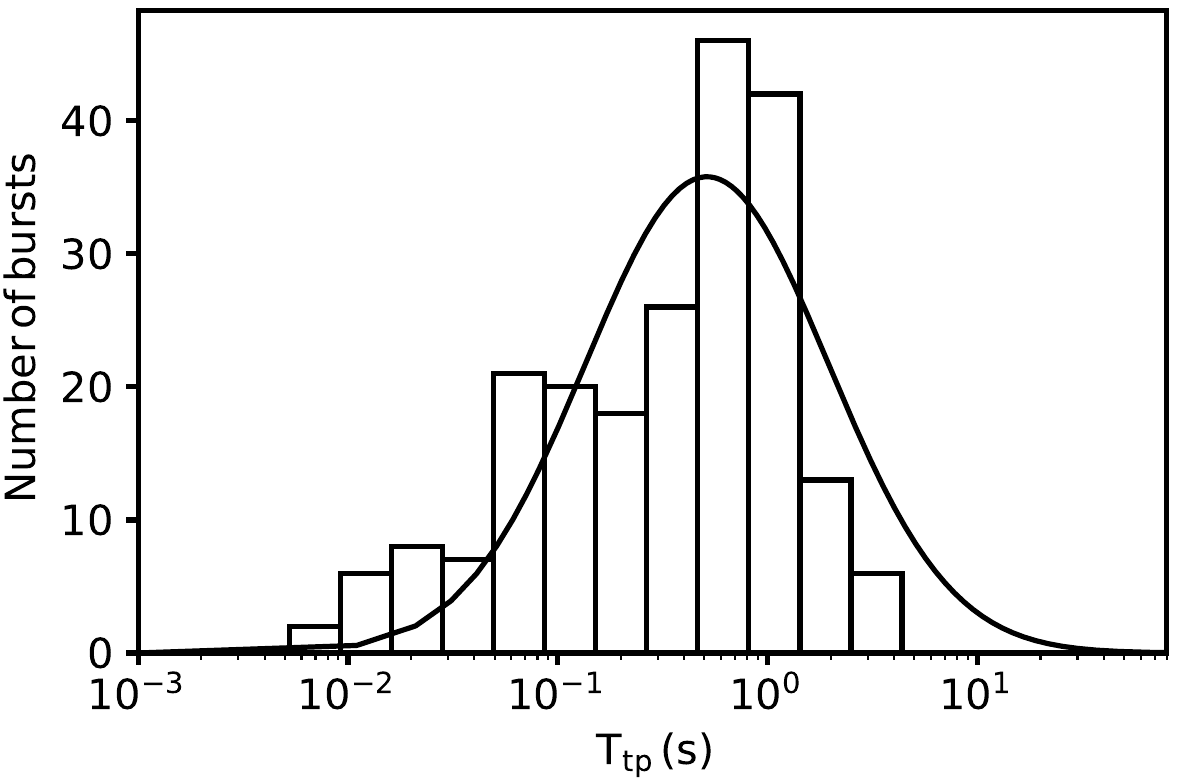}
\includegraphics[angle=0,width=0.33\textwidth]{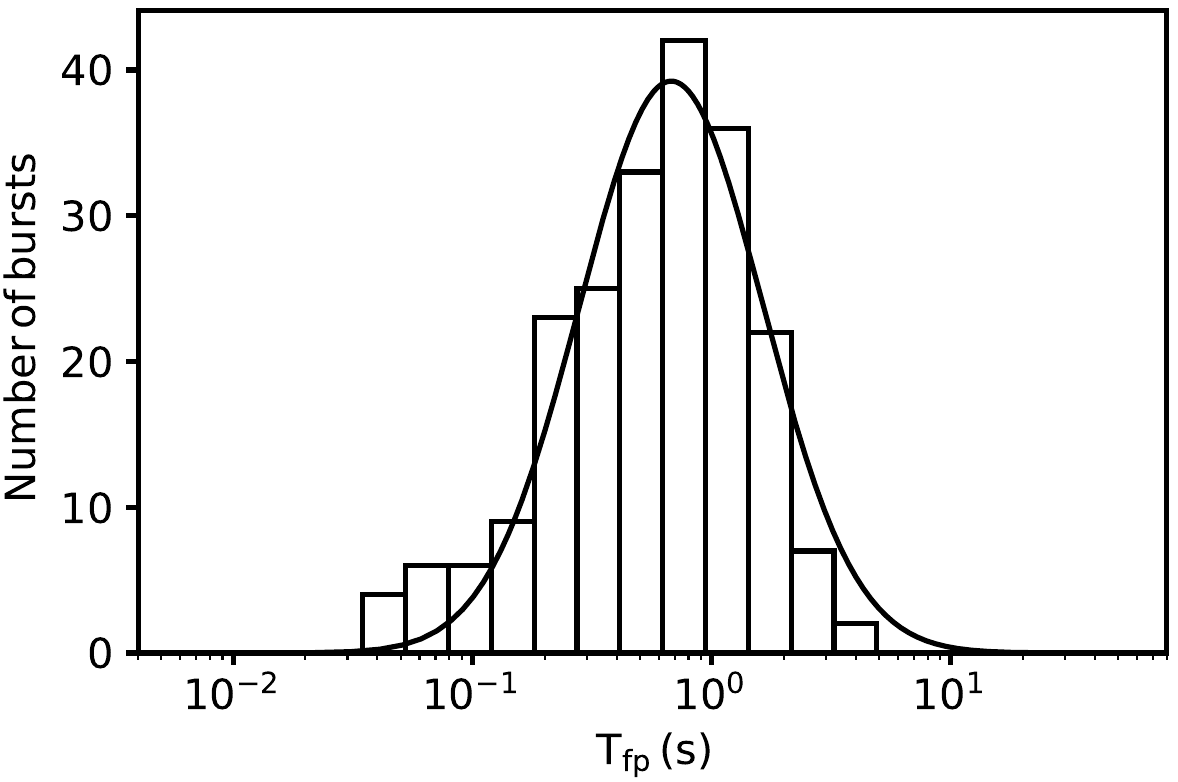}
\includegraphics[angle=0,width=0.33\textwidth]{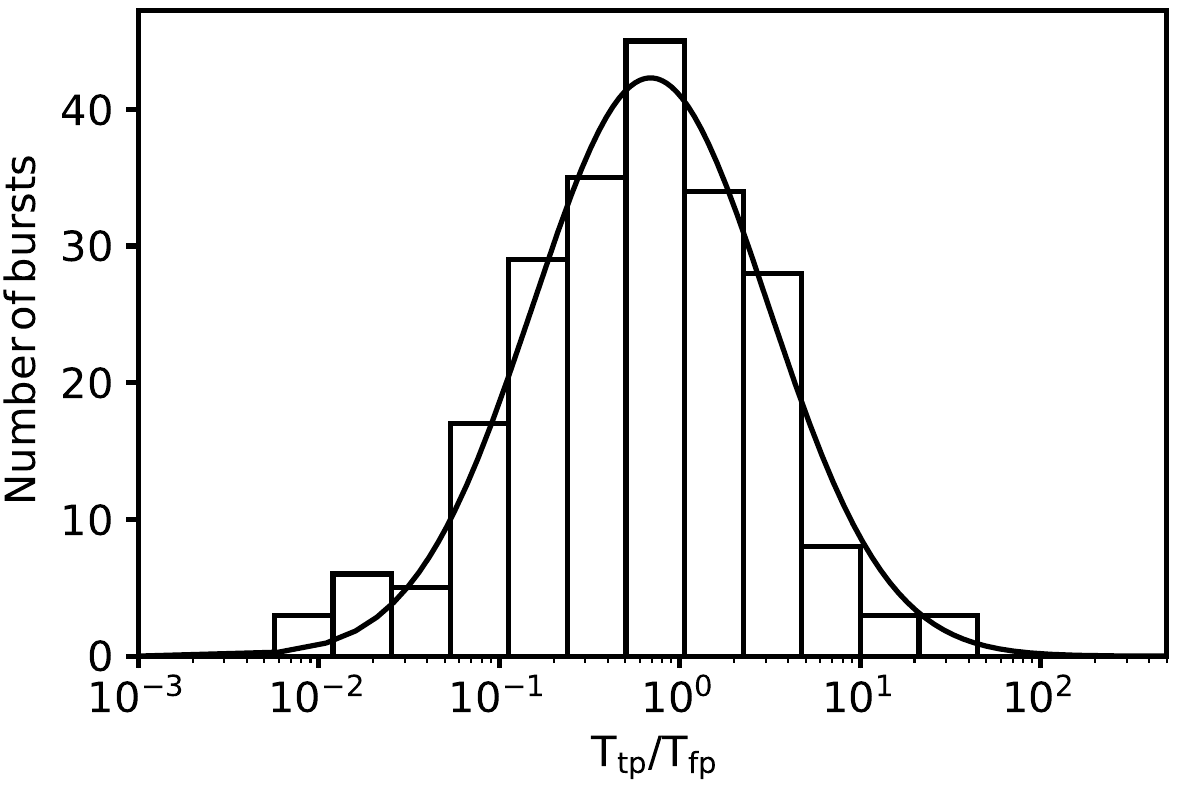}
\caption{Time-to-peak ($T_{\rm tp}$, {\sl left panel}), Time-from-peak
  ($T_{\rm fp}$, {\sl middle panel}), and the ratio $T_{\rm tp}/T_{\rm
    fp}$ ({\sl right panel}) distributions. The best fit log-normal
  function to each distribution is shown as black-solid line with means of 510~ms, 680~ms, and 0.69, respectively.}
\label{risefallT}
\end{center}
\end{figure*}

We apply a Poissonian procedure to identify bursts within each GTI
starting with dividing it into multiple $100$~s duration time
intervals, $\Delta T$ \citep[e.g.,][]{gavriil04ApJ:1E2259}. We then
create a light curve with 4~ms resolution within each $\Delta T$,
resulting in a total number of $N=25000$ bins per interval. Next, we
calculate the probability $P_{\rm i}$ of the total counts in each 4~ms
time bin, $n_{\rm i}$, to be a random fluctuation around the average
$\lambda$ (the ratio of the total counts within $\Delta T$ over
$\Delta T$) as $P_{\rm i}=(\lambda^{n_{\rm i}}\exp(-\lambda))/n_{\rm
  i}!$. Any time bin satisfying the criterion $P_{\rm i}<0.01/N$, is
flagged as part of a burst. The procedure is reiterated until no more
bins are identified in a $\Delta T$. To capture the weaker tails of
bursts as well as fainter bursts that are not resolved at high
resolution, we repeat the above procedure using time resolutions of
32~ms, 128~ms, and 512~ms, after excluding all flagged bins of 4~ms,
as discussed above. We note that at lower resolutions, the most
intense parts of the burst forest, i.e., between 300 and 600 seconds
after the start of the observation (Figure~\ref{stormLC}, bottom
panel), appear to be sitting on a bed of elevated emission. We correct
for this variable background by fitting a non-parametric function to
local minima within each $\Delta T$ \citep{eilers2005baseline}. We
repeated our burst search algorithm for different $\Delta T$s, ranging
from 20 to 200~s in steps of 20~s. We found that our search algorithm
is only weakly dependent on the interval duration.

We define the start of a new burst to be when the emission of the
previous burst drops and remains at the background level, after
subtracting the elevated emission level, for 0.5~s (i.e., 15\% of the
source spin period). This establishes a $T_{\rm start}$ and $T_{\rm
  end}$ for each burst and provides enough background before and
after to derive the burst temporal properties. We identify a total of
217 bursts within the first GTI. At the peak of the burst forest, we
identify a 31~s interval (about 10 rotational periods) where the
emission never reaches the background level (designated as burst 100
in Table~\ref{bstTempProp}). This is shown in the inset of the middle
panel of Figure~\ref{stormLC}. The length of this bursting interval, which
contains many individual bursts, is $>6$~times the duration of the
second longest burst. We exclude this interval from all burst
analyses. In the remaining four GTIs we identify a total of 6 bursts.

\subsection{Burst temporal results}
\label{btr}

To derive burst temporal properties, we consider the 1-10~keV unbinned
events within the interval $T_{\rm start}$ and $T_{\rm end}$ for each
burst as estimated above, as well as a background interval just before
and after these times, respectively. These background intervals range
between 0.3 and 2 seconds. We fit the cumulative count distribution of
the background interval with a linear function, and then correct the
burst cumulative count distribution using this background estimate.

We consider the start (end) time of the burst $T_{\rm s,100} (T_{\rm
  e,100})$ the time at which the cumulative sum rises (drops) to
$3\sigma$ above (below) the average level of the pre- (post-) burst
background-corrected interval; we calculate the burst fluence as the
total number of counts between $T_{\rm s,100}$ and $T_{\rm
  e,100}$, and we define $T_{100}=T_{\rm e,100}-T_{\rm s,100}$. The
$T_{\rm 90}$ burst duration is estimated as the time interval during
which 5\% to 95\% of the burst fluence is accumulated
\citep{kouveliotou93ApJ:GRBs}. Figure~\ref{exLCs} shows three examples 
of burst light curves along with their $T_{\rm s,100}$ and $T_{\rm
  e,100}$ (dotted vertical lines), and the start and end times of the
$T_{\rm 90}$ interval, $T_{\rm s,90}$ and $T_{\rm e,90}$, respectively
(dot-dashed vertical lines). The bottom panel of Figure~\ref{stormLC}
shows the burst storm at 0.5~s resolution, after excluding the
$T_{100}$ of all bursts.

To establish the peak time of each burst, $T_{\rm peak}$, we start
from the unbinned list of events between $T_{\rm s,100}$ and $T_{\rm
  e,100}$. We create light curves with a resolution of $2^n$~ms where
$n$ is an integer iteratively increased from 1 to 6 (i.e., 2 to
64~ms). Starting with $n=1$, we define $T_{\rm peak}$ as the first
time bin that reaches a total count with significance $>7\sigma$
above the background. If this condition is not met, $n$ is increased
and the procedure is repeated until $T_{\rm peak}$ is established. The
peak times of the three burst-examples are shown as dashed lines in
Figure~\ref{exLCs}. We define the time-to-peak $T_{\rm tp}$ and
time-from-peak $T_{\rm fp}$ as $T_{\rm peak}-T_{\rm s,100}$ and
$T_{\rm e,100}-T_{\rm peak}$, respectively. Finally, we define the
waiting time until the next burst as $T_{\rm i+1~s,100}-T_{\rm
  i~e,100}$, where $i=1,2...$, is the burst number.

In the following, we give the statistical properties of the bursts
detected during the first GTI, i.e., the burst storm. We later compare
their results to the bursts detected in the subsequent four GTIs
(Figure~\ref{stormLC}). 

The $T_{\rm 90}$ and $T_{\rm 100}$ distributions of the bursts are
shown in the left and middle panel of Figure~\ref{distT90WT},
respectively. The best fit log-normal distributions are shown as the
solid black line, and results in a $T_{\rm 90}$ and $T_{\rm 100}$
means of 840~ms and 1270~ms. The $1\sigma$ interval ranges from 430
to 1630~ms for $T_{\rm 90}$ and 620 to 2580~ms for $T_{\rm 100}$. The
waiting time distribution for the bursts identified within the first
GTI is shown in Figure~\ref{distT90WT}, right panel. A fit to the
distribution with a log-normal function is shown as a black solid
line. We measure a mean of 2.1~s with an $1\sigma$ interval range of
0.6 to 7.2~s. Figure~\ref{risefallT} shows the $T_{\rm tp}$, $T_{\rm
  fp}$, and $T_{\rm tp}/T_{\rm fp}$ distributions. The best fit
log-normal distributions to each are shown as black solid lines. We
measure a mean time-to-peak $<T_{\rm tp}>=0.51_{-0.36}^{+1.64}$~s and
a mean time-from-peak $<T_{\rm fp}>=0.68_{-0.40}^{+0.96}$~s. The
average of the distribution of the ratio of these two parameters is
$0.69_{-0.54}^{+2.40}$, indicating a steeper rise than decay in the
burst profiles. The uncertainties on each parameter represent the
$1\sigma$ standard deviation of the best-fit log-normal function to
the respective distribution.

\begin{figure}[]
\begin{center}
  \includegraphics[angle=0,width=0.49\textwidth]{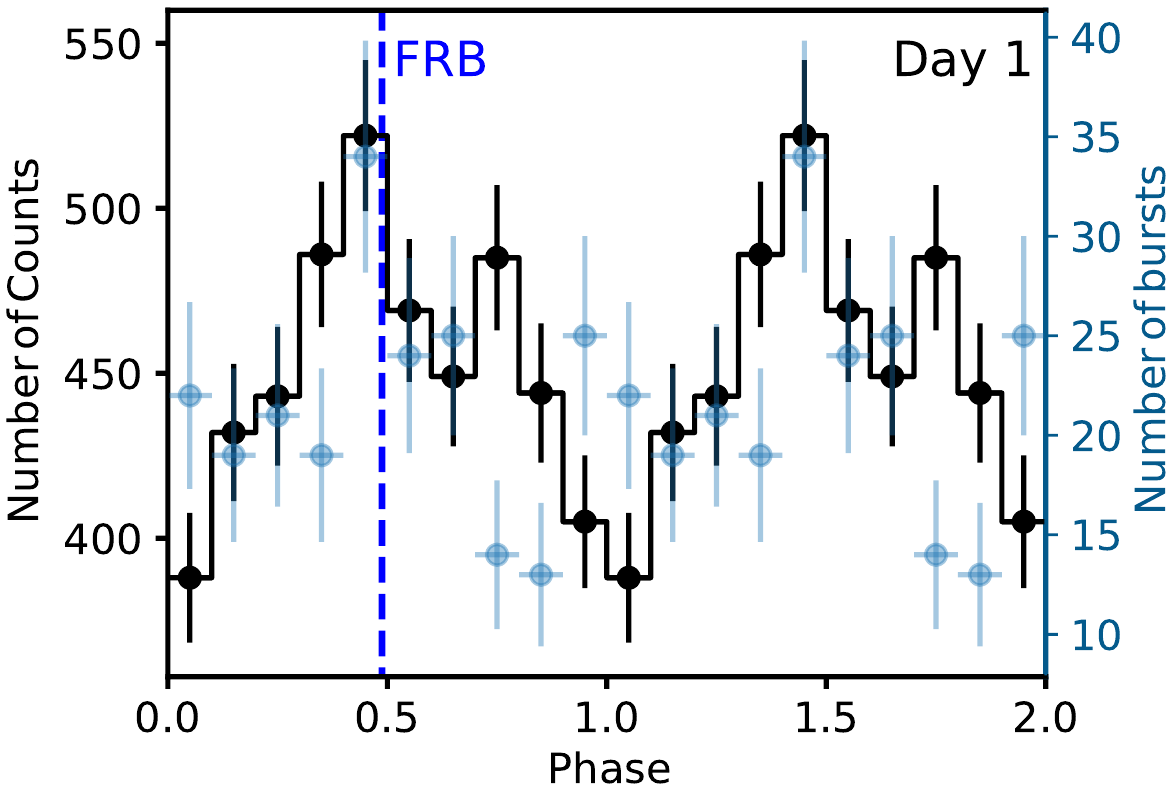}\\
  \hspace{-0.23in}
  \includegraphics[angle=0,width=0.46\textwidth]{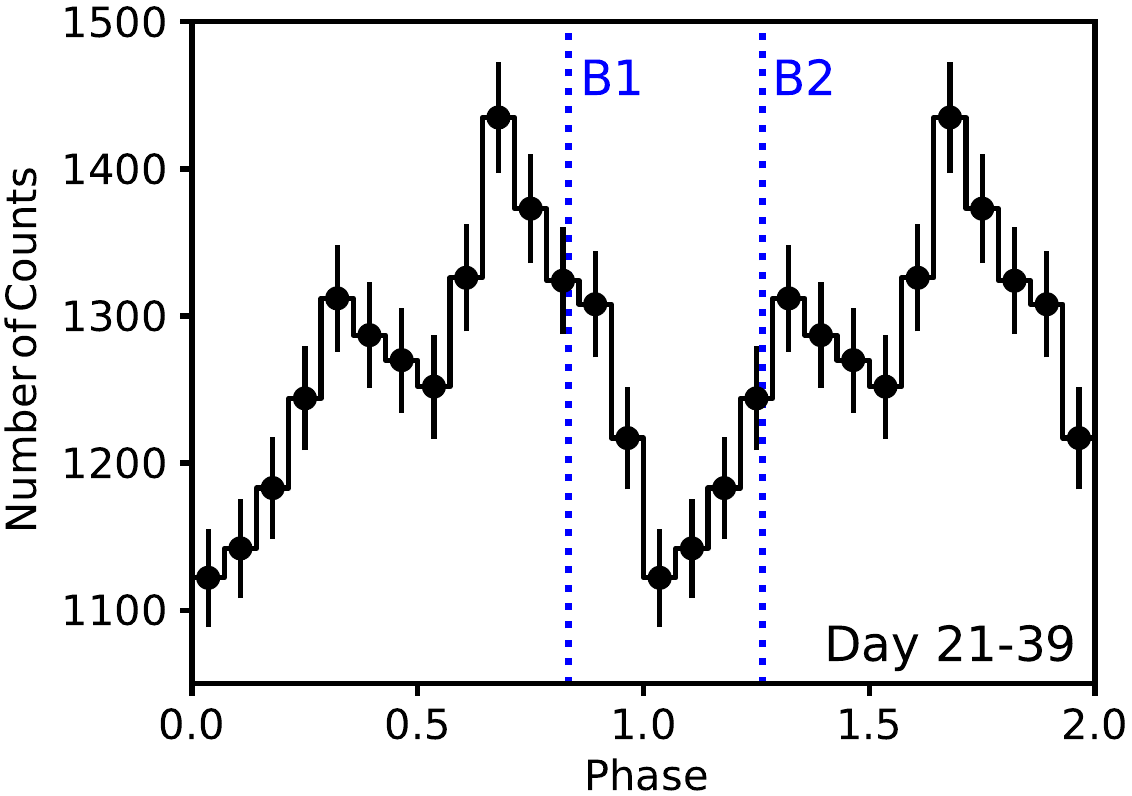}
\caption{{\sl Upper panel.} Persistent emission pulse profile of
    the \nicer\ data taken on 2020 April 28 (observation ID
    3020560101), day 1 after the outburst onset (black dots and solid
    line). We exclude the first GTI during which the burst forest
    occurred. The rms pulsed fraction is $8\pm2\%$. The light blue
    points represent the peak times of the \nicer\ bursts folded at
    the spin period of the source. We find no preference for burst
    peak arrival time with phase. The vertical dashed-blue line is the
    phase of the FRB arrival time. {\sl Bottom panel.} Pulse profile
    of the persistent emission as observed during days 21 to 39 post
    outburst (observation IDs 3020560105 to 3020560119). The dotted
    lines are the phases of the two radio bursts observed by
    \citet{2020arXiv200705101K}. The rms pulsed fraction is
    $6.7\pm0.8\%$. The two profiles, shown in the energy range
    1.5-5~keV, are not phase-connected; their respective minima are
    shifted to phase 0. See text for more details.}
\label{pulseprof}
\end{center}
\end{figure}

\begin{figure}[!t]
\begin{center}
\includegraphics[angle=0,width=0.48\textwidth]{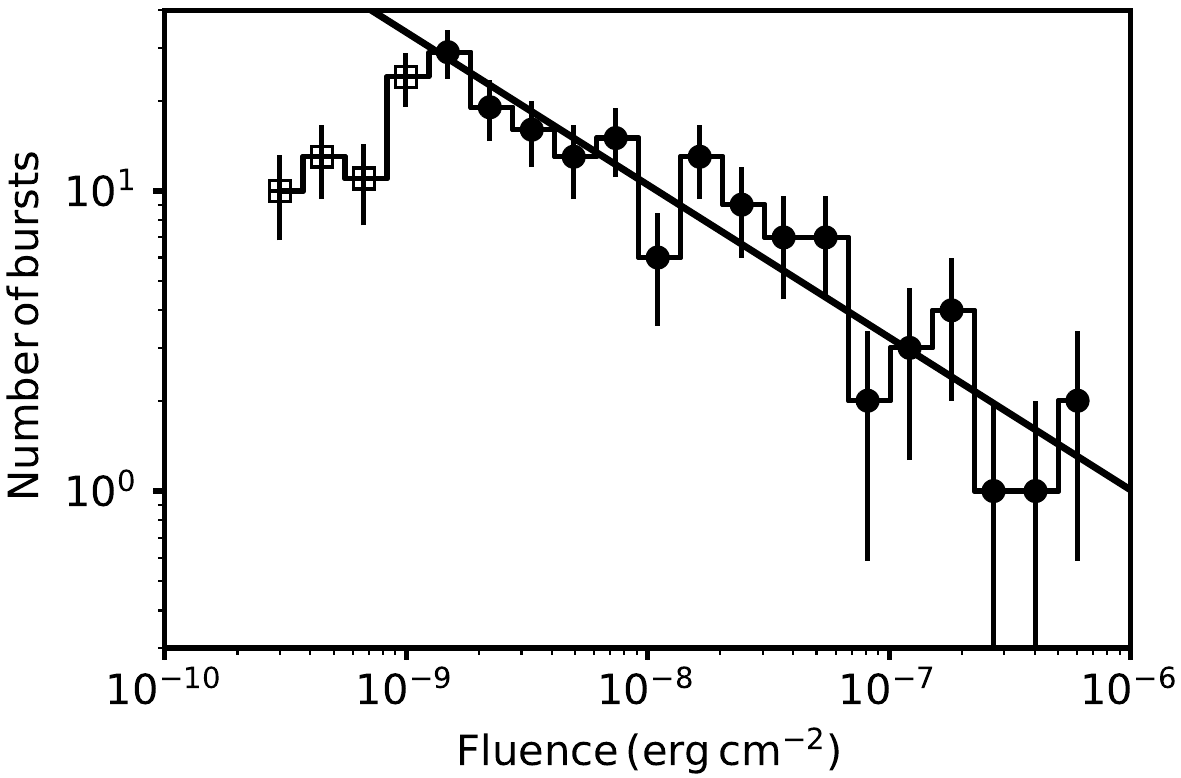}
\caption{Burst fluence distribution of the bursts detected during the \src\ burst
  storm. The filled dots represent the bursts for which our detection
  efficiency is $\geq99\%$. Open squares represent fluences
  with decreased detection efficiency. The solid line is a PL fit to
  the black dots only, $N\propto F^{-0.5\pm0.1}$.}
\label{diffFl}
\end{center}
\end{figure}

\begin{figure*}[!t]
\begin{center}
\includegraphics[angle=0,width=0.32\textwidth]{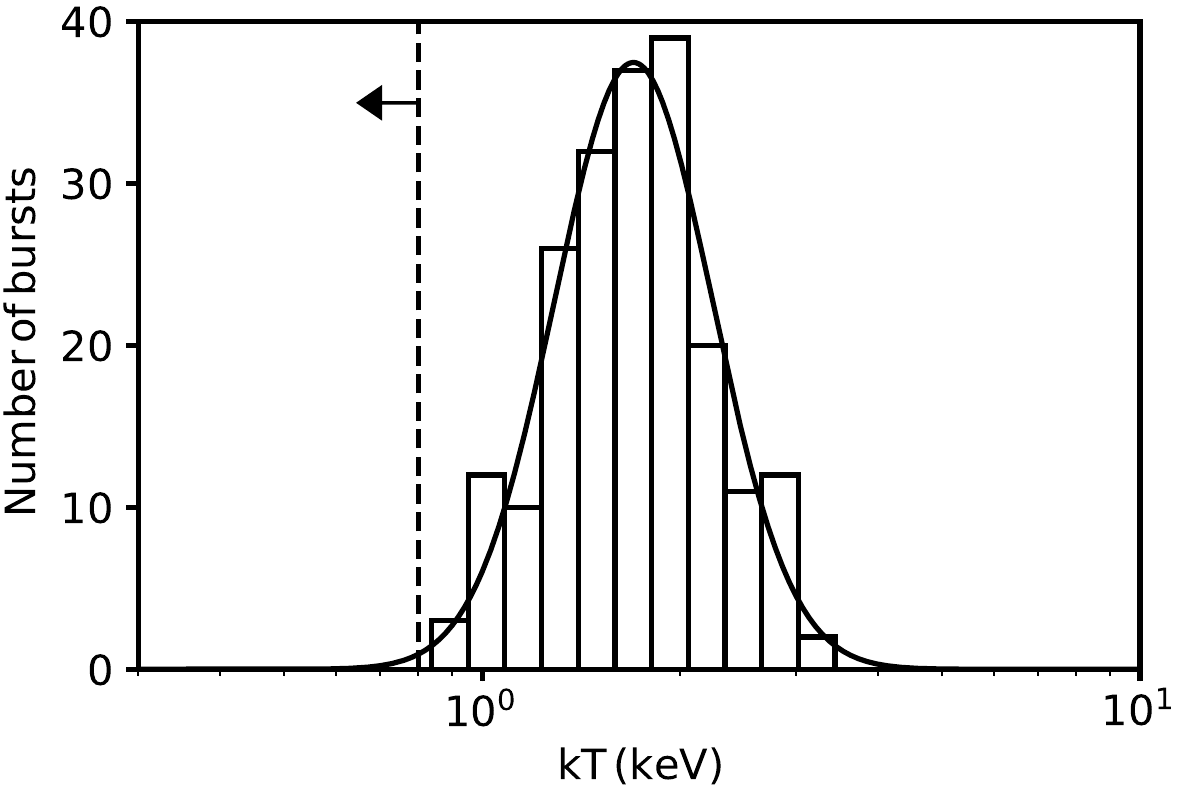}
\includegraphics[angle=0,width=0.33\textwidth]{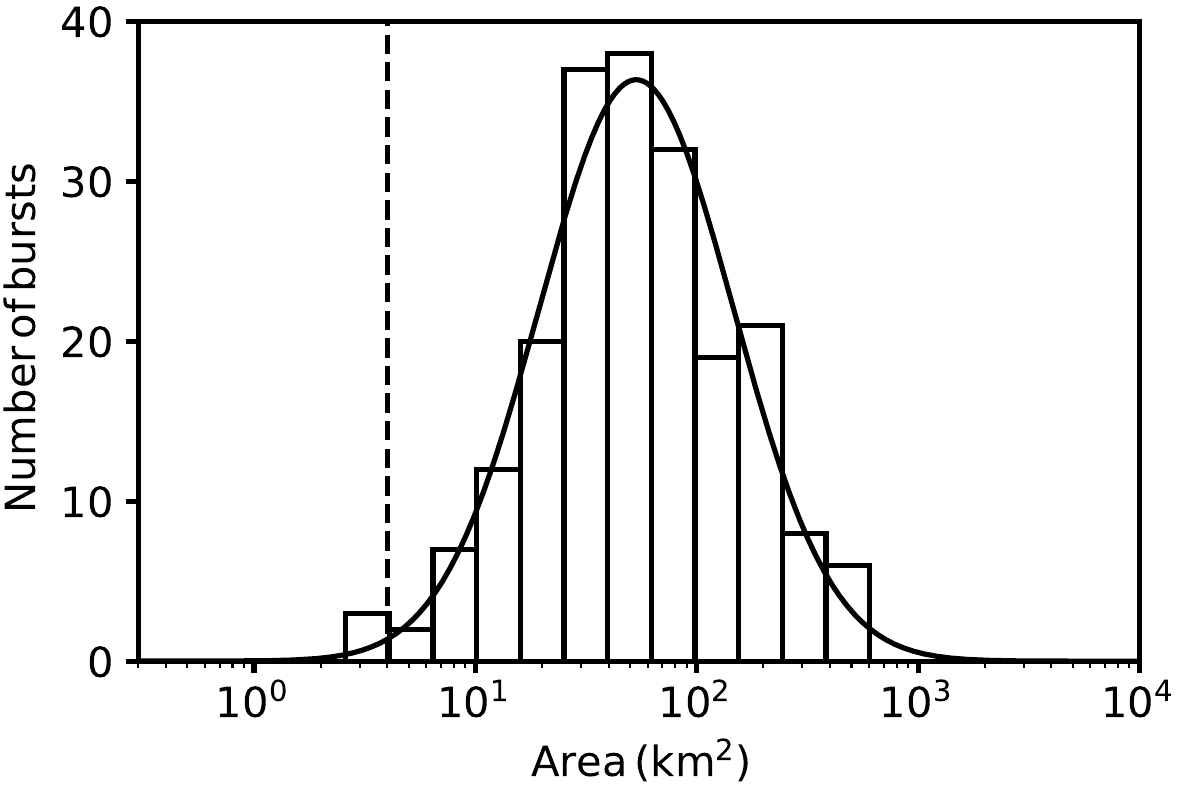}
\includegraphics[angle=0,width=0.325\textwidth]{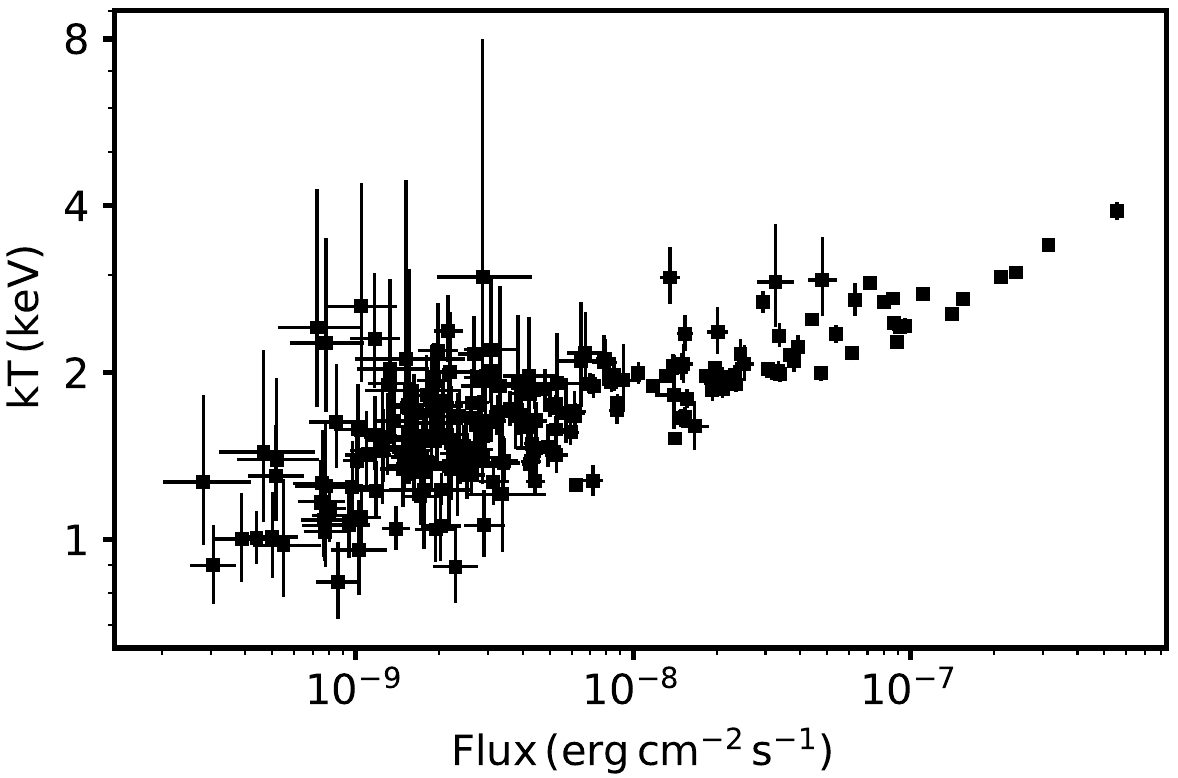}
\caption{{\sl Left panel.} Distribution of the burst BB temperatures
  along with the best fit lognormal function (solid line). The
  vertical dashed line is the highest BB temperature measured for the
  persistent emission. The arrow indicates the surface temperature
  evolution trend during the outburst decay. {\sl Middle panel.}
  Distribution of the burst BB area along with the best fit lognormal
  function (solid line). The vertical dashed line is the BB area of
  the persistent emission. {\sl Right panel.} Burst BB temperature vs
  flux, measured in the 0.5-10~keV energy band.}
\label{specDist}
\end{center}
\end{figure*}

We searched for any correlation of the bursts peak arrival time (after
applying a barycentric-correction) with rotational phase
(Figure~\ref{pulseprof}, light blue dots), given that we detect the
source spin period in the four GTIs after the burst storm (see
Section~\ref{timAna}). No clustering is obvious at any particular
phase. We find a $\chi^2$ of 15 for 9 dof when fitting the peak
arrival data to a horizontal line. We also apply the Anderson-Darling
(AD) test to compare the cumulative distribution function (CDF) of the
burst phases to a uniform distribution. We find an AD test statistic
of 1.0 and p-value of 0.35, implying that the phases of the bursts are
consistent with a uniform distribution. We also tested for the
dependence of $R^2$, kT, $T_{90}$, waiting time, and flux on phase,
and found no significant correlations. Considering very short bursts
with $T_{90}<0.5$~s also results in no significant dependence of any
of the burst temporal or spectral parameters with phase. Since the FRB
time lands within our observation at a barycenteric-corrected time of
58967.60857593\footnote{We converted the FRB geocentric arrival
    time to the barycentric reference frame using the tool
    \texttt{pintbary} from the precision timing software \texttt{PINT}
    \citep[][https://github.com/nanograv/PINT]{lu19ascl}. We considered
  the same source position and JPL ephemerides (DE405) as the ones
  used to barycenter the X-ray data.} MJD (Modified Julian Date), we
also show its arrival time in phase space as a dashed blue line in
Figure~\ref{pulseprof}. We discuss the implications of its phase
association in Section~5.3.

\subsection{Burst spectral results}
\label{specBurSec}

We fit the 1-10~keV spectra of each burst with a simple model
consisting of an absorbed blackbody (BB) or PL function. We do not
attempt more complex models such as a cutoff PL or a 2BB model since
\nicer\ only covers a small range of the magnetar bursts broad-band
energy spectrum ($\sim1-200$~keV). We could not constrain the
hydrogen column density in the direction of \src\ due to the low total
number of counts in the majority of the bursts. Hence, we fix $N_{\rm
  H}$ to $2.4\times10^{22}$~cm$^{-2}$, which is the best fit value as
derived with high signal-to-noise persistent emission spectra of the
source \citep[e.g.,][]{younes17:1935}. This value is also consistent
with the one derived for bursts with enough counts to enable a
measurement of $N_{\rm H}$. Finally, 12 bursts had very few counts to
allow for any spectral analysis (Table~\ref{bstTempProp}).

We find that most bursts were better fit with the BB model as opposed
to the PL one, which consistently resulted in residuals at the lower
and upper-end of the \nicer\ energy range. We show three examples of
burst spectra in the right panel of Figure~\ref{exLCs}. The spectra
and the best fit absorbed BB model are shown in panels (a). The
residuals of the absorbed BB and of the absorbed PL are displayed in
panels (b) and (c), respectively. As can be seen, especially in the
second and third bursts, significant residuals remain at the lower and
upper end of the energy coverage when the bursts are fit with the PL
model.

The distribution of the BB fit temperatures is exhibited in
Figure~\ref{specDist} (left). We find an average BB temperature of
$1.7$ and $1\sigma$ range of 1.3-2.2~keV. Assuming a spherically
emitting region obeying the Stefan-Boltzmann law $R^2T^4= $constant,
we also display the distribution of emission areas $R^2$ in
Figure~\ref{specDist}, and estimate an average of 54~km$^2$. The
1$\sigma$ interval is 30 to 154~km$^2$. Both are substantially larger
than the corresponding values for the persistent emission as presented
in Section~\ref{sec:persistent}.  Finally, the positive correlation
between the BB temperatures, $kT$, and burst fluxes
(Figure~\ref{specDist}, right panel), indicates a hardening of burst
spectra with increasing flux. However, we should stress the caveat
here, that we are using a very narrow energy range to derive these
results, which might under-represent the source spectral properties.

We measure burst fluences, $F$,  by multiplying the time-averaged flux
of each burst with the corresponding $T_{90}$ (measured in
Section~\ref{btr}). We show in Figure~\ref{diffFl} the  differential
fluence distribution of the bursts in the first GTI, uniformly binned
on a logarithmic scale. The turnover at fluences
$\lesssim1.5\times10^{-9}$ (open squares) reflects our inability to
recover bursts with lower fluences. To verify this assumption, we
perform simple simulations as follows. We assume that a burst is
approximated with a top-hat profile of a certain width $w\sim T_{\rm
  90}$. We assume that the simulated bursts have a total number of
counts ranging from 50 to 500, which we iteratively increase in steps
of 5 counts. We distribute these counts along $w$ following an
exponential distribution, i.e., the expected waiting time between
events for a Poisson process. For each total number of counts, we
simulate $10^4$ bursts with $w$ drawn from the log-normal distribution
that best fit the $T_{90}$ distribution. We add a background
contribution to the burst in a $\pm5$~s interval around its
centroid. The  background count rate is drawn from a normal
distribution with a mean and standard deviation of $75\pm
5$~counts~s$^{-1}$, which is the maximum average background count rate
of the first GTI (Figure~\ref{stormLC}, lower-panel). Running our
burst search algorithm on the simulated light curves, we find that for
a 110 total burst counts, we detect the bursts at the 99\% rate. For a
typical burst spectrum (blackbody with $kT=1.7$~keV, see below), this
corresponds to a burst fluence
$F\sim1.6\times10^{-9}$~erg~cm$^{-2}$. Ignoring the bins in
Figure~\ref{diffFl} with fluences lower than this value, we find that
the differential fluence distribution for \src\ bursts can be well
modeled with a power-law (PL) $N\propto F^{-0.5\pm0.1}$ or
$dN/dF\propto F^{-1.5\pm0.1}$, where $N$ represents the total number
of bursts within a fluence bin, over approximately three orders of
magnitude of fluence from $10^{-6}-10^{-9}$~erg~cm$^{-2}$.

Finally, we find that the total bursts fluence emitted during the
burst storm $F_{\rm tot}=(5.0\pm0.1)\times10^{-6}$~erg~cm$^{-2}$ in
the 0.5-10~keV energy range, which translates to a total energy emitted
in the bursts of $4.8\times10^{40}$~erg. This measurement should be
considered a lower limit since, for at least the brighter bursts, the
spectral peak is at energy $>10$~keV, beyond the coverage of \nicer\
\citep{younes20arXiv200611358Y}.

\subsection{Properties of bursts beyond the burst storm}

The April 28 \nicer\ observation exhibits five GTIs of which the last
four occur 2.9, 4.4 , 5.8, and 15 hours after the first GTI (i.e., the
burst storm, Figure~\ref{stormLC}). We find two bursts in each of GTIs
2, 3, and 4 and none in the very last. Given their similar exposure,
we derive for each a burst rate of 0.008~bursts~s$^{-1}$. This is in
contrast to the burst rate $>0.2$~bursts~s$^{-1}$ we detect during the
burst storm, just 3 hours earlier. The average $T_{90}$ of these
bursts is 0.6~s, roughly consistent with the average burst storm
$T_{90}$. Their spectra are well fit with a BB model with temperatures
also consistent with the bulk of the bursts detected during the burst
storm. In contrast, the average fluence of these bursts is
$4.5\times10^{-10}$~erg~cm$^{-2}$, which is only consistent with the
very faint end of the fluence distribution shown in
Figure~\ref{diffFl}.

\section{Persistent Emission}
 \label{sec:persistent}

\subsection{Timing}
\label{timAna}

\begin{figure*}[!th]
\begin{center}
\includegraphics[angle=0,width=0.82\textwidth]{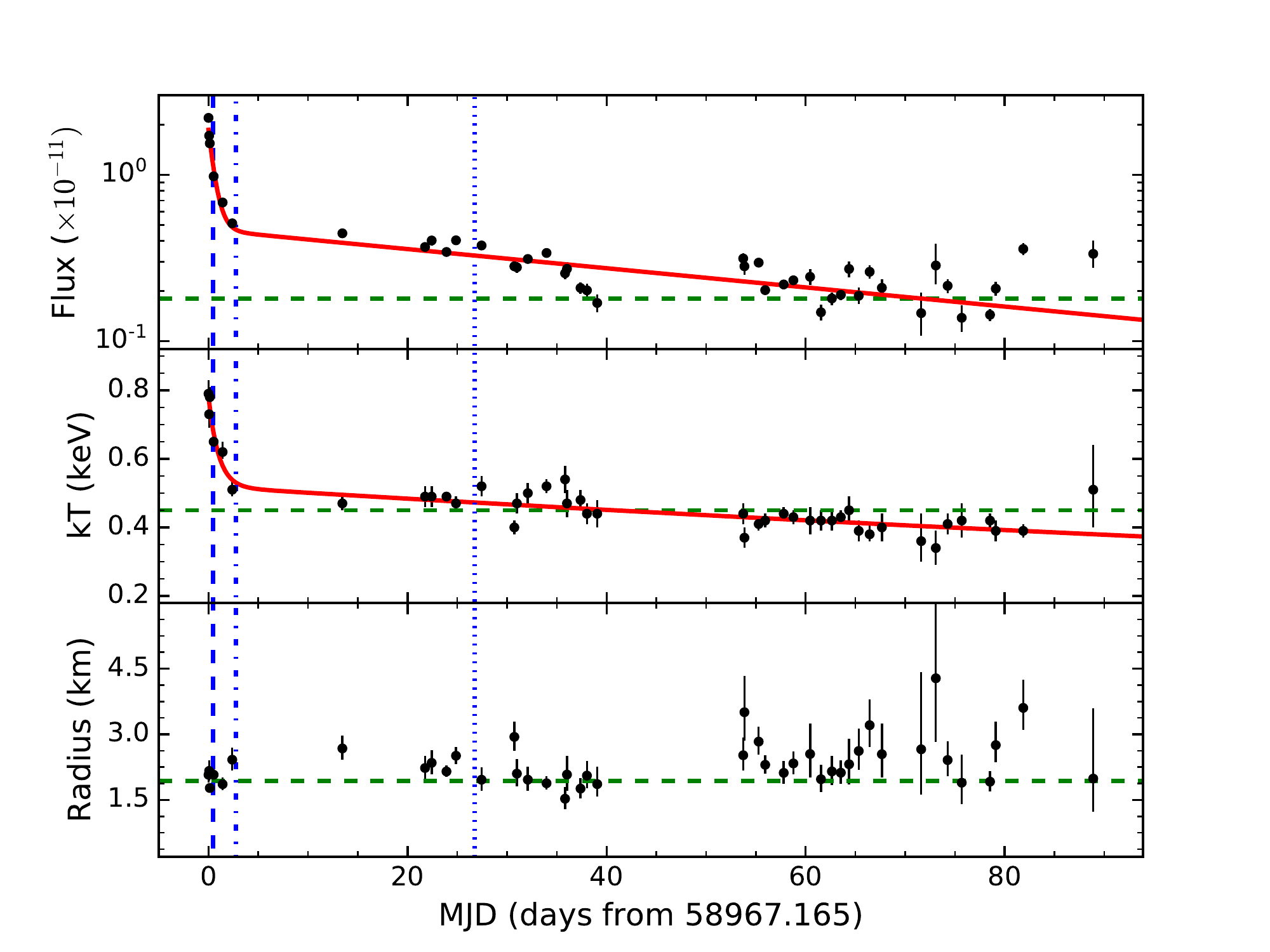}
\caption{Spectral evolution of the persistent emission of \src\
  observed with {\it NICER}, after the initial observation obtained on
  2020, April 28. From upper to lower, panels show the evolution in
  the $0.3-10$~keV range of the unabsorbed flux (in units of
  erg~$^{-1}$~cm$^{-2}$), the BB temperature, and the emitting radius
  assuming a distance of 9~kpc. In all panels green dashed lines show
  the average of the values inferred from historical \nicer\
  observations obtained in $2017-2019$, before the source burst active
  period. The blue vertical dashed, dotted and dash-dotted lines,
  respectively, mark the times of the FRB like event, two additional
  weak radio bursts \citep{2020arXiv200705101K}, and another weak
  radio burst reported by FAST \citep{zhang20ATel13699} from the
  source. The red solid curves in the upper two panels constitute the
  best fit double exponential decay models to the flux and BB
  temperature, displaying an initial very rapid rise and then a much
  slower decline.}
\label{pers_sp_res_figure}
\end{center}
\end{figure*}

We searched for the spin period of \src\ using the
barycenter-corrected events detected during the first \nicer\
observation, ID 3020560101, after excluding all identified bursts. To
increase our sensitivity for period detection, we excluded the first
GTI of the observation since it is strongly contaminated by bursts
that are not resolved from the underlying persistent emission
(Section~\ref{specBurSec}). We also only consider events in the
energy range 1.5 to 5~keV, and we restricted our search
interval to the period range $3.236~{\rm s}<P<3.249$~s, which
encapsulates the source spin period, $P=3.24731(1)$~s, as derived from
\nicer\ observations performed on April 29 and 30 \citep{
  borghese20arXiv200600215B}. We find the largest $Z_1^2$
\citep{buccheri83AApulse} power of $24.5$ at a frequency
$\nu=0.307946(2)$ (the number in parenthesis represents the $1\sigma$
uncertainty, which corresponds to a decrease in $Z_1^2$ by
  2.3). This corresponds to a period $P=3.24732(2)$ at a reference
time 58967.423047 MJD. This period is consistent at the $1\sigma$
level with the one in \citet{borghese20arXiv200600215B}, which implies
a $3\sigma$ upper limit on any spin-down or spin-up event
$|\dot{P}|\gtrsim6.0\times10^{-10}$~s~s$^{-1}$. Figure~\ref{pulseprof},
upper-panel, shows the $1.5-5$\,keV pulse profile in black. We derive
an rms pulse fraction of $8\pm2\%$, however, we note that this value
is not background-corrected and should be considered a lower-limit.

The fast decay trend of the persistent emission flux beyond the
  first exposure resulted in very few counts detected in each separate
  observation for pulsation searches. However, \nicer\ performed
  heavy cadence observations on two occasions during the outbursts, observing \src\
  almost daily. The first (interval 1) covered days 21 to 39 from the
  outburst onset (observation IDs 3020560105 to 3020560119), while the
  second (interval 2) extended from days 54 to 68 (observation IDs
  3020560120 to 3020560133). Hence, we searched simultaneously for the
  source spin-frequency and its derivative, $0.3079<\nu<0.3080$ and
  $-12<\log{\dot{\nu}}<-11$ in steps of $\Delta\nu/1000$ and
  $\Delta\log\dot\nu/500$, for each of these time spans using two
  different methods, the $Z^2$-test \citep{buccheri83AApulse} and an
  epoch-folding or $\chi^2$-test \citep[e.g.,][]{Staelin69EF}.

During interval 1, the total \nicer\ exposure is 24.3~ks,
  detecting about 17800 counts in the energy range 1.5-5~keV. Both
  methods resulted in the detection of a strong signal, $Z^2=77$
  and $\chi^2=91$, with spin parameters consistent at the $1\sigma$
  level. We then refined this timing model through a phase-coherence
  analysis by dividing the interval into four portions each with about
  4500 counts. From the latter, we estimate the spin frequency
  $\nu=0.30794014(1)$ and frequency derivative $\dot{\nu} =
  -3.72(3)\times10^{-12}$~Hz~s$^{-1}$ (reference time 58997.571 MJD).
  The photon arrival times of interval 1 folded with the above timing
  model is shown in the bottom panel of Figure~\ref{pulseprof}. We
  measure an rms pulsed fraction of $6.7\pm0.8\%$. The dotted-lines
  are the corresponding phases of the two radio bursts detected on
  2020 May 24 \citep{2020arXiv200705101K}.

\nicer\ exposure of \src\ during interval 2 is 21.5~ks, detecting
  a total of 12100 counts in the energy range 1.5-5~keV. Applying the
  same methodology as above, we cannot detect significant pulsations
  within the searched ($\nu$, $\dot{\nu}$) range. We also tried to
  phase-connect interval 1 and 2, however we do not detect the source
  pulse in interval 2 when folding with the timing model derived for
  interval 1. Assuming a similar pulse shape compared to interval 1,
  we estimate a 3$\sigma$ upper limit on the rms pulsed fraction of
  about $5\%$.

\subsection{Spectroscopy}
\label{specAna}

We perform spectral analysis on the persistent emission of \src\
starting with the second GTI of the first \nicer\ observation
(Figure~\ref{stormLC}). We exclude the first GTI due to strong
contamination from unresolved burst emission. We fit the X-ray spectra
in the energy range of $1.0-5.0$~keV. The background starts to
dominate beyond 5~keV due to the softness and relatively low flux of
the source. In this energy range, the X-ray spectra are well described
by an absorbed BB model. We fix the hydrogen column density  to
2.4$\times$10$^{22}$~cm$^{-2}$, which is inferred from earlier high
S/N {\it Chandra} and {\it XMM-Newton} data \citep{younes17:1935}. Our
results are given in Table \ref{tab:pers_sp_res}, and the spectral
parameter evolution is shown in Figure~\ref{pers_sp_res_figure}. In
the latter, we also show the average of the spectral parameters as
obtained from earlier \nicer\ observations of \src\ during 2017-2019.

\section{Discussion}
\label{discuss}

We have analyzed the \nicer\ monitoring of \src\ following its most
intense burst-active period. We report on the statistical
characteristics of 217 bursts detected in the first observation taken
on 2020 April 28, six hours after the start of the latest activity
episode of the source. We also report the timing analysis results of
the persistent emission of the source on that day, as well as its
spectral evolution up to 90~days after the outburst onset. In the
following we discuss our results in comparison to other magnetar burst
storms and active episodes.

\subsection{\src\ burst storm comparison to other magnetars}
\label{disburStorm}

\src\ entered its sixth and most intense burst active episode on 2020
April 27 emitting tens of bright bursts detected in the span of
minutes.  \nicer\ started observing \src\ just six hours after the
initial trigger and caught the tail end of the burst storm during a
span of 1120~s, detecting bursts at a rate $>0.2$~bursts~s$^{-1}$.
Large burst rates have previously been observed from several magnetars
such as SGR 1900+14 \citep{gogus99apjl:1900,israel08ApJ:1900}, SGR
1806$-$20 \citep{woods07ApJ:1806}, SGR 1627$-$41
\citep{woods99ApJ:1627,esposito08MNRAS:1627}, 1E~2259+586
\citep{gavriil04ApJ:1E2259}, and SGR~J1550$-$5418
\citep{mereghetti09ApJ,israel10MNRAS,scholz11ApJ:1547,vanderhorst12ApJ:1550}.
We discuss below a qualitative comparison across magnetars, as a
quantitative comparison is not feasible, given the different
characteristics (such as, e.g., energy ranges and sensitivity) of the
instruments with which they were observed.

The average $T_{90}$ duration of 840~ms for \src\ bursts is among the
highest within the magnetar burst family
\citep[e.g.,][]{collazzi15ApJS}. Yet, most of the $T_{90}$ values for
other magnetars have come from large field of view, high-background
instruments operating above 5\,keV (such as e.g.,  {\it CGRO}/BATSE
and {\it Fermi}/GBM), that may have skewed durations towards lower
values \citep[e.g.,][]{israel08ApJ:1900,younes14ApJ:1550}. For
instance, some of the bursts we report here were also detected with
\fermi/GBM above 8\,keV. It is quite evident by comparing the \nicer\
and GBM light curves of these bursts that the latter misses the weak
tails of the bursts, and hence results in an under-estimate of their
$T_{90}$s (see, e.g., Figure~1 in \citealt{younes20arXiv200611358Y}).
Hence, our larger than usual $T_{90}$ measurement is likely a
reflection of the large sensitivity and low background of \nicer\
rather than an indication of an intrinsic source property.

Regarding burst morphology, we find very few single-peaked bursts,
with the majority showing multi-peaked profiles. Roughly 65\% of our
bursts have shorter rise than fall times, commensurate with the bulk
of magnetar bursts, especially the ones observed during burst storms
\citep[e.g.,][]{vanderhorst12ApJ:1550}. Notably, the average
rise-to-fall time ratio of 0.69 we find for \src\ is quite similar to
the 0.59 value measured for another prolific burster, 1E~1547.0$-$5408
\citep{scholz11ApJ:1547}. We note that we cannot exclude the
  possibility that some bursts in our sample are the superposition of
  two or more independent events, which may impact the results of our
  distributions somewhat.

The waiting time distributions of magnetar bursts during burst storms
have been documented for several magnetars. For instance,
\citet{gogus99apjl:1900} studied SGR~$1900+14$ bursts observed with
{\it RXTE} during its 1998 burst storm, while
\citet{gavriil04ApJ:1E2259} studied the ones from 1E~$2259+586$,
during its 2002 outburst. Both studies found that waiting times follow
a log-normal distribution with a mean value of $\sim50$~s. We also
find that the waiting times for the \src\ bursts follow a log-normal
distribution, but with a mean of about $2.1_{-1.4}^{+5.2}$~s, albeit
capturing the latter part of its burst storm.  This marked difference
could reflect a peculiar character of the \src\ storm, yet it could
also be partly due to how the waiting times are
determined. \citet{gavriil04ApJ:1E2259} notes a positive correlation
between the waiting time and the time of the next burst (from a
fiducial start time) over their 10~ks exposure of 1E~$2259+586$. In
fact, the average waiting time within the first 1~ks of their
observation (comparable to the length of our burst storm observation)
is about $10$~s. The larger burst rate observed from \src\ during this
burst storm relative to that for 1E~$2259+586$ may yield shorter
waiting times on average, and this may be in part due to the intrinsic
nature of the \src\ storm, yet it may also reflect the excellent
sensitivity below 5\,keV of the \nicer\ detector. Interestingly, a
log-normal waiting time distribution has also been derived for the
repeating FRB 121102 \citep{Wadiasingh2019ApJ,katz19MNRAS}.

Several previous works have shown that magnetar burst fluences (and
hence energies) follow a power-law distribution $dN/dF\propto
F^{-\alpha}$. This relation holds over several orders of magnitude in
fluence ranging from $\sim10^{-10}$~erg~cm$^{-2}$ to
$10^{-5}$~erg~cm$^{-2}$ \citep[e.g.,][]{gogus00ApJ:1806,
  gavriil04ApJ:1E2259,lin13ApJ:magnetarsBB}. For most magnetars the
index of this relation has been shown to cluster around -1.6
\citep[e.g.,][]{cheng96Natur, gogus99apjl:1900,gogus00ApJ:1806,
  aptekar01ApJS:SGR,gavriil04ApJ:1E2259,scholz11ApJ:1547,
  vanderhorst12ApJ:1550,lin13ApJ:magnetarsBB}. This is similar to the
energy distribution shape of terrestrial earthquakes
\citep[e.g.,][]{Gutenberg56} and solar flares
\citep[e.g.,][]{crosby93SoPh}, both of which result in an index of
$\sim-1.6$. The \src\ burst fluence distribution is also well modeled
with a PL function with an index $\alpha=-1.5\pm0.1$, clearly
commensurate with the values inferred for other magnetars. This
suggests a universal property of the burst energetics for the whole
magnetar class, perhaps underpinned by the phenomenon of
self-organized criticality \citep{gogus99apjl:1900}.

Magnetar burst spectra are best described either with a cutoff PL or a
2BB model, both of which result in a turnover in the $20-50$~keV
energy range \citep[e.g.,][]{lin12ApJ:1550,enoto12MNRAS:1550}. \src\
spectra in the 8-200~keV range from the previous activations are well 
fit with both models, with spectral parameters broadly consistent with
the rest of the magnetar family, albeit softer on average than some
\citep{lin20ApJ:1935}. During the 2020 burst storm,
\citet{younes20arXiv200611358Y} presented the \nicer+GBM spectral
analysis of 24 bursts, also showing that they are commensurate with
the previous activations. These authors, as well as others
\citep[e.g.,][]{Ridnaia20arXiv200511178R}, found a positive
correlation between the high energy cutoff and the burst flux. This
implies that, for most of the bursts we analyze here, the cutoff
energy is around the few keV range, well within the \nicer\ 1-10~keV
energy band. This may explain why a simple PL cannot explain the
spectra well, whereas a BB model, which mimics in shape a cutoff PL
when restrained to small energy ranges, gives a good fit to the bulk
of the bursts. However, the spectral analysis presented here, is
restricted to the \nicer\ energy range, which is less sensitive above
5\,keV, thus possibly imposing a bias in the burst spectral analysis.

Finally, the lack of dependency of the burst peak arrival times with
phase is also consistent with the majority of magnetar sources
\citep[e.g.,][]{scholz11ApJ:1547}; see also the literature survey in
\cite{2018MNRAS.476.1271E}. This implies that magnetar bursts occur
approximately randomly in magnetic colatitudes in the magnetosphere,
perhaps close to the stellar surface.  Yet the bursts are not
spatially proximate to the surface locale of the persistent pulsed
emission, though there may be a physical association between the
transient and quiescent signals mediated by field line flux tubes in
either dipolar or twisted field geometries.

\subsection{Persistent emission}
\label{disPerEmiss}

Following strong bursting activity, the increase in the persistent
X-ray flux level is ubiquitous in magnetars. This increase is often
accompanied by hardening of the X-ray spectra, usually in the form of
higher surface thermal temperature and/or a decrease in the PL index
\citep[e.g.,][]{cotizelati18MNRAS}. These characteristics are evident
in the case of the \src\ previous activations
\citep{israel16mnras:1935,younes17:1935}, as well as the current one
\citep{borghese20arXiv200600215B}.  \citet{younes17:1935} noted that 
the \src\ persistent emission flux increased in proportion to the
total energy emitted in the bursts, with the largest increase (by a
factor 7) detected following the 2016 bursting episode, the most
intense up to that time. The initial flux increase of $\gtrsim10$
during the 2020 bursting activity is consistent with the above
picture.

The flux evolution during the previous episodes showed an initial
rapid exponential decay with a characteristic time-scale of few days,
possibly followed by a shallower return to quiescence. Our \nicer\
monitoring of the 2020 activation also reveals two decay trends, which
can be well characterized with a double exponential function, with
very different e-folding times. As can be seen from
Figure~\ref{pers_sp_res_figure}, \src\ shows an initial rapid decay
with a best fitting e-folding time of 0.65$\pm$0.08 days, followed by
a long-term flux decay and cooling whose e-folding time is 75$\pm$5
days. The decay in flux is accompanied by cooling (middle panel in
Figure~\ref{pers_sp_res_figure}). In fact a similar double exponential
decay model can be used to fit the decay in the inferred BB
temperature. The best fit parameters of such a model are two e-folding
times of 0.99$\pm$0.3 and 285$\pm$45 days, respectively. On the other
hand the apparent emitting radius only exhibits a marginal increase
with time with a slope of 0.46$\pm$0.15 km per hundred days.

Such an initially rapid flux decrease followed by a more shallow
decline has previously been observed from a few magnetars
\citep[e.g.,][]{kargaltsev12apj:1834,an12ApJ:1627}, and is indicative
of cooling hot spots. These spots can develop at the onset of the
outburst either through surface bombardment by relativistic particles
that are energized in the magnetosphere by toroidal twists to the
global field structure
\citep{thompson02ApJ:magnetars,beloborodov13ApJ,Gonzalez-2019-MNRAS},
or via an internal seismic process such as crustal shearing induced by
stresses imposed by the enormous fields
\citep{thompson96ApJ:magnetar,pons09AA}. This cooling picture is
clearly demonstrated in the middle panel of
Figure~\ref{pers_sp_res_figure}, and is accompanied by a stability of
the inferred radius of the emitting area
(Figure~\ref{pers_sp_res_figure}, bottom).   It is notable that after
about 50 days, the spectral fitting yields systematically lower
temperatures than the historical ones obtained in the 2017--2019
period that are indicated via the green dashed lines in
Figure~\ref{pers_sp_res_figure}.  Yet, due to the lower flux of the
source and the known degeneracy between the emitting area and the
temperature of the BB model, the uncertainties in these determinations
are relatively high.

The source broad double-peaked pulse shape in the soft X-ray band
(Figure~\ref{pulseprof}, \citealt{borghese20arXiv200600215B}) differs
markedly from the single-peaked, quasi-sinusoidal pulse profile
following its 2014 outburst \citep{israel16mnras:1935}, implying that
distinct regions are heated on the surface of the star during each
outburst. We also detect clear evolution in the pulse profile during
this outburst; the prominent peak led the secondary one immediately
after outburst onset \citep[see also][]{borghese20arXiv200600215B},
but flipped 20 days later. Such pulse shape evolution during outburst
epochs is rather common to magnetars \citep[e.g.,][]{woods99ApJ:1900,
  gogus02ApJ:1806,esposito10MNRAS,ng11ApJ:1547,castillo14MNRAS,
  younes15ApJ:1806,castillo16MNRAS}, possibly pointing to a complex
magnetic field topology. For \src, the separation by $\sim0.3$ in
phase of the two peaks provides evidence in favor of this assertion.

Finally, the spin-down rate that we measure between days 21 and 39
post-outburst is a factor 2.7 larger than the rate measured in 2014
\citep[$\dot\nu=-1.36\times10^{-12}$~Hz~s$^{-1}$,][]{israel16mnras:1935},
indicating a larger torque on the magnetar. This is yet another
distinct characteristic of magnetar in outbursts \citep[e.g.,][]{
  woods07ApJ:1806}, which is believed to emanate from
increased particle wind in the magnetosphere due to the strong
bursting activity \citep{harding99ApJ,tong2013ApJ}.

\subsection{The FRB Connection}

The pulse period detection of \src\ during the first \nicer\
observation, which brackets the time of the FRB (see
Figure~\ref{stormLC}), enabled us to place it, and by extension the
peak of the X-ray associated burst, on the pulse profile; this is a
crucial piece of information that had not been achieved
before. Figure~6 demonstrates that the FRB time aligns with the
principal peak of the pulse profile. It is common in the magnetar
literature to attribute the peak to an observer viewing a hot region
on the neutron star surface
\citep[e.g.,][]{Perna-2008-ApJ,Albano-2010-ApJ,younes20ApJ:1708}. If
the heating originates internally, the hot spot would naturally be
associated with the magnetic poles since heat conduction upwards from
the crust is efficient when the field is oriented vertically. As an
alternative possibility, twists in magnetic field loops can lead to
the development of surface hot spots via particle bombardment,
discussed in Section~\ref{disPerEmiss}.  Such twists can also favor
quasi-polar hot spots or annuli since stresses in the crust which
drive twists are generally larger in polar regions
\citep[e.g.,][]{perna11ApJ}. In either scenario, the pulse peak for
\src\ can then be realized if the magnetar is instantaneously viewed
almost down the polar axis, thereby concluding that the FRB is somehow
intimately connected to the polar field lines. Yet, we remark that
non-polar surface hot spot locales can also be entertained.
  
This special observational perspective is in fact the picture that was
drawn by \cite{younes20arXiv200611358Y} via a comparison of the
broadband spectroscopic signal of \nicer+{\it Fermi}/GBM bursts and
the FRB-associated burst as observed with {\sl HXMT}
\citep{li20arXiv200511071L}. This scenario of an ephemeral (due to
stellar rotation) polar viewing of the FRB and its contemporaneous,
spectroscopically-unique X-ray burst could help explain the rarity of
both, and restricts the range of possible viewing and rotational
geometries for \src.

A physical connection between the surface pole and a radio emission
zone has a precedent in canonical, young/middle-aged radio pulsars,
which also exhibit phase-aligned persistent radio and surface thermal
X-ray signals. In those systems, global magnetospheric solutions 
\citep[e.g.,][]{1999ApJ...511..351C,2010ApJ...715.1282B,2014ApJ...793...97K}
require pair cascades that both generate return currents that bombard
and heat polar cap zones \citep{Harding2001,2013MNRAS.429...20T}, and
seed coherent radio emission \citep{PhysRevLett.124.245101}. Yet
magnetar magnetospheres differ profoundly from those of pulsars, with
their currents generally being associated with twisted field
geometries in closed field zones
\citep{thompson02ApJ:magnetars,Chen-2017-ApJ}. Moreover, much of the
magnetar activity associated with such twists is ephemeral. The
ability here to determine the X-ray pulse phase associated with the
FRB provides an important advance towards understanding the
FRB-magnetar connection, with potential implications for the
extragalactic FRB paradigm.

We find that the two, order-of-magnitudes, fainter radio
  bursts detected by \citet{2020arXiv200705101K}, which were separated
  by 1.4~s, are offset from the X-ray pulse-peaks. As pointed by those
  authors, it is not yet clear what the origin of such radio bursts
  is, whether they are driven by similar physical and emission
  mechanisms that resulted in the FRB or rather differing ones.
  Interestingly, three more radio bursts with comparable fluence have
  recently been detected from \src\ during a single rotational period
  with peak separation of 0.95~s between the first and the second and
  1.95~s between the second and the third bursts \citep{
    pleunis20ATel14080}. This indicates that these radio bursts 
  are likely occurring sporadically at largely distinct phases. This
  is unlike the more-persistent radio magnetar emission that mostly
  clusters in a small rotational phase-space \citep[e.g.,
  XTE~J1810$-$197;][and now perhaps \src; \citealt{zhu20ATel14084}]{
    maan19ApJ}. Assuming that repeating FRBs are produced in the
  close environs of magnetars, the above result fits with the fact
  that no magnetar-like periodic behavior have so far been detected
  from these sources \citep[e.g.,][]{zhangYG18ApJ}.

  
\section*{Acknowledgments}

A portion of this work was supported by NASA through the \nicer\
mission and the Astrophysics Explorers Program. This research has made
use of data and software provided by the High Energy Astrophysics
Science Archive Research Center (HEASARC), which is a service of the
Astrophysics Science Division at NASA/GSFC and the High Energy
Astrophysics Division of the Smithsonian Astrophysical
Observatory. The authors are grateful to the referee for constructive
comments that improved the quality of the manuscript. G.Y. sincerely
thanks Jason Hessels for his request to perform more detailed
late-time temporal analysis which proved highly fruitful.
G.Y. acknowledges support from NASA under NICER Guest Observer cycle-1
program 2098, grant number 80NSSC19K1452. M.G.B. acknowledges the
generous support of the National Science Foundation through grant
AST-1813649. Z.W. is supported by the NASA postdoctoral
program. C.K. acknowledges support from NASA under grant
80NSSC17K0761. This work has made use of the NASA Astrophysics Data
System.


\begin{thebibliography}{79}
\expandafter\ifx\csname natexlab\endcsname\relax\def\natexlab#1{#1}\fi

\bibitem[{{Albano} {et~al.}(2010){Albano}, {Turolla}, {Israel}, {Zane},
  {Nobili}, \& {Stella}}]{Albano-2010-ApJ}
{Albano}, A., {Turolla}, R., {Israel}, G.~L., {et~al.} 2010, \apj, 722, 788

\bibitem[{{An} {et~al.}(2015){An}, {Archibald}, {Hasco{\"e}t}, {Kaspi},
  {Beloborodov}, {Archibald}, {Beardmore}, {Boggs}, {Christensen}, {Craig},
  {Gehrels}, {Hailey}, {Harrison}, {Kennea}, {Kouveliotou}, {Stern}, {Younes},
  \& {Zhang}}]{an15ApJ:1841}
{An}, H., {Archibald}, R.~F., {Hasco{\"e}t}, R., {et~al.} 2015, \apj, 807, 93

\bibitem[{{An} {et~al.}(2012){An}, {Kaspi}, {Tomsick}, {Cumming}, {Bodaghee},
  {Gotthelf}, \& {Rahoui}}]{an12ApJ:1627}
{An}, H., {Kaspi}, V.~M., {Tomsick}, J.~A., {et~al.} 2012, \apj, 757, 68

\bibitem[{{Aptekar} {et~al.}(2001){Aptekar}, {Frederiks}, {Golenetskii},
  {Il'inskii}, {Mazets}, {Pal'shin}, {Butterworth}, \&
  {Cline}}]{aptekar01ApJS:SGR}
{Aptekar}, R.~L., {Frederiks}, D.~D., {Golenetskii}, S.~V., {et~al.} 2001,
  \apjs, 137, 227

\bibitem[{{Arnaud}(1996)}]{arnaud96conf}
{Arnaud}, K.~A. 1996, in ASP Conf. Ser. 101: Astronomical Data Analysis
  Software and Systems V, 17

\bibitem[{{Bai} \& {Spitkovsky}(2010)}]{2010ApJ...715.1282B}
{Bai}, X.-N. \& {Spitkovsky}, A. 2010, \apj, 715, 1282

\bibitem[{{Barthelmy} {et~al.}(2020){Barthelmy}, {Bernardini}, {D'Avanzo},
  {Gropp}, {Kennea}, {Lien}, {Meland ri}, {Palmer}, {Sbarrato}, {Siegel}, \&
  {Neil Gehrels Swift Observatory Team}}]{Barthelmy20GCN}
{Barthelmy}, S.~D., {Bernardini}, M.~G., {D'Avanzo}, P., {et~al.} 2020, GRB
  Coordinates Network, 27657, 1

\bibitem[{{Beloborodov}(2013)}]{beloborodov13ApJ}
{Beloborodov}, A.~M. 2013, \apj, 762, 13

\bibitem[{{Bochenek} {et~al.}(2020){Bochenek}, {Ravi}, {Belov}, {Hallinan},
  {Kocz}, {Kulkarni}, \& {McKenna}}]{Bochenek20arXiv200510828B}
{Bochenek}, C.~D., {Ravi}, V., {Belov}, K.~V., {et~al.} 2020, arXiv e-prints,
  arXiv:2005.10828

\bibitem[{{Borghese} {et~al.}(2020){Borghese}, {Coti Zelati}, {Rea},
  {Esposito}, {Israel}, {Mereghetti}, \& {Tiengo}}]{borghese20arXiv200600215B}
{Borghese}, A., {Coti Zelati}, F., {Rea}, N., {et~al.} 2020, arXiv e-prints,
  arXiv:2006.00215

\bibitem[{{Buccheri} {et~al.}(1983){Buccheri}, {Bennett}, {Bignami}, {Bloemen},
  {Boriakoff}, {Caraveo}, {Hermsen}, {Kanbach}, {Manchester}, {Masnou},
  {Mayer-Hasselwander}, {Ozel}, {Paul}, {Sacco}, {Scarsi}, \&
  {Strong}}]{buccheri83AApulse}
{Buccheri}, R., {Bennett}, K., {Bignami}, G.~F., {et~al.} 1983, \aap, 128, 245

\bibitem[{{Chen} \& {Beloborodov}(2017)}]{Chen-2017-ApJ}
{Chen}, A.~Y. \& {Beloborodov}, A.~M. 2017, \apj, 844, 133

\bibitem[{{Cheng} {et~al.}(1996){Cheng}, {Epstein}, {Guyer}, \&
  {Young}}]{cheng96Natur}
{Cheng}, B., {Epstein}, R.~I., {Guyer}, R.~A., \& {Young}, A.~C. 1996, \nat,
  382, 518

\bibitem[{{Collazzi} {et~al.}(2015){Collazzi}, {Kouveliotou}, {van der Horst},
  {Younes}, {Kaneko}, {G{\"o}{\u{g}}{\"u}{\textcommabelow s}}, {Lin}, {Granot},
  {Finger}, {Chaplin}, {Huppenkothen}, {Watts}, {von Kienlin}, {Baring},
  {Gruber}, {Bhat}, {Gibby}, {Gehrels}, {McEnery}, {van der Klis}, \&
  {Wijers}}]{collazzi15ApJS}
{Collazzi}, A.~C., {Kouveliotou}, C., {van der Horst}, A.~J., {et~al.} 2015,
  The Astrophysical Journal Supplement Series, 218, 11

\bibitem[{{Contopoulos} {et~al.}(1999){Contopoulos}, {Kazanas}, \&
  {Fendt}}]{1999ApJ...511..351C}
{Contopoulos}, I., {Kazanas}, D., \& {Fendt}, C. 1999, \apj, 511, 351

\bibitem[{{Coti Zelati} {et~al.}(2018){Coti Zelati}, {Rea}, {Pons}, {Campana},
  \& {Esposito}}]{cotizelati18MNRAS}
{Coti Zelati}, F., {Rea}, N., {Pons}, J.~A., {Campana}, S., \& {Esposito}, P.
  2018, \mnras, 474, 961

\bibitem[{{Crosby} {et~al.}(1993){Crosby}, {Aschwanden}, \&
  {Dennis}}]{crosby93SoPh}
{Crosby}, N.~B., {Aschwanden}, M.~J., \& {Dennis}, B.~R. 1993, \solphys, 143,
  275
  
\bibitem[{Eilers \& Boelens(2005)}]{eilers2005baseline}
Eilers, P.~H. \& Boelens, H.~F. 2005, Leiden University Medical Centre Report,
  1, 5

\bibitem[{{Elenbaas} {et~al.}(2018){Elenbaas}, {Watts}, \&
  {Huppenkothen}}]{2018MNRAS.476.1271E}
{Elenbaas}, C., {Watts}, A.~L., \& {Huppenkothen}, D. 2018, \mnras, 476, 1271

\bibitem[{{Enoto} {et~al.}(2012){Enoto}, {Nakagawa}, {Sakamoto}, \&
  {Makishima}}]{enoto12MNRAS:1550}
{Enoto}, T., {Nakagawa}, Y.~E., {Sakamoto}, T., \& {Makishima}, K. 2012,
  \mnras, 427, 2824

\bibitem[{{Esposito} {et~al.}(2008){Esposito}, {Israel}, {Zane}, {Senziani},
  {Starling}, {Rea}, {Palmer}, {Gehrels}, {Tiengo}, {de Luca}, {G{\"o}tz},
  {Mereghetti}, {Romano}, {Sakamoto}, {Barthelmy}, {Stella}, {Turolla},
  {Feroci}, \& {Mangano}}]{esposito08MNRAS:1627}
{Esposito}, P., {Israel}, G.~L., {Zane}, S., {et~al.} 2008, \mnras, 390, L34

\bibitem[Esposito et al.(2010)]{esposito10MNRAS} Esposito, P.,
  Israel, G.~L., Turolla, R., et al.\ 2010, \mnras, 405,
  1787. doi:10.1111/j.1365-2966.2010.16551.x

\bibitem[{{Fletcher} \& {Fermi GBM Team}(2020)}]{fletcher20GCN:1935}
{Fletcher}, C. \& {Fermi GBM Team}. 2020, GRB Coordinates Network, 27659, 1

\bibitem[{{Gavriil} {et~al.}(2004){Gavriil}, {Kaspi}, \&
  {Woods}}]{gavriil04ApJ:1E2259}
{Gavriil}, F.~P., {Kaspi}, V.~M., \& {Woods}, P.~M. 2004, \apj, 607, 959

\bibitem[{{Gendreau} {et~al.}(2016){Gendreau}, {Arzoumanian}, {Adkins},
  {Albert}, {Anders}, {Aylward}, {Baker}, {Balsamo}, {Bamford}, {Benegalrao},
  {Berry}, {Bhalwani}, {Black}, {Blaurock}, {Bronke}, {Brown}, {Budinoff},
  {Cantwell}, {Cazeau}, {Chen}, {Clement}, {Colangelo}, {Coleman},
  {Coopersmith}, {Dehaven}, {Doty}, {Egan}, {Enoto}, {Fan}, {Ferro}, {Foster},
  {Galassi}, {Gallo}, {Green}, {Grosh}, {Ha}, {Hasouneh}, {Heefner}, {Hestnes},
  {Hoge}, {Jacobs}, {J{\o}rgensen}, {Kaiser}, {Kellogg}, {Kenyon}, {Koenecke},
  {Kozon}, {LaMarr}, {Lambertson}, {Larson}, {Lentine}, {Lewis}, {Lilly},
  {Liu}, {Malonis}, {Manthripragada}, {Markwardt}, {Matonak}, {Mcginnis},
  {Miller}, {Mitchell}, {Mitchell}, {Mohammed}, {Monroe}, {Montt de Garcia},
  {Mul{\'e}}, {Nagao}, {Ngo}, {Norris}, {Norwood}, {Novotka}, {Okajima},
  {Olsen}, {Onyeachu}, {Orosco}, {Peterson}, {Pevear}, {Pham}, {Pollard},
  {Pope}, {Powers}, {Powers}, {Price}, {Prigozhin}, {Ramirez}, {Reid},
  {Remillard}, {Rogstad}, {Rosecrans}, {Rowe}, {Sager}, {Sanders}, {Savadkin},
  {Saylor}, {Schaeffer}, {Schweiss}, {Semper}, {Serlemitsos}, {Shackelford},
  {Soong}, {Struebel}, {Vezie}, {Villasenor}, {Winternitz}, {Wofford},
  {Wright}, {Yang}, \& {Yu}}]{gendreau16SPIE}
{Gendreau}, K.~C., {Arzoumanian}, Z., {Adkins}, P.~W., {et~al.} 2016, Society
  of Photo-Optical Instrumentation Engineers (SPIE) Conference Series, Vol.
  9905, {The Neutron star Interior Composition Explorer (NICER): design and
  development}, 99051H

\bibitem[{{Gonz{\'a}lez-Caniulef} {et~al.}(2019){Gonz{\'a}lez-Caniulef},
  {Zane}, {Turolla}, \& {Wu}}]{Gonzalez-2019-MNRAS}
{Gonz{\'a}lez-Caniulef}, D., {Zane}, S., {Turolla}, R., \& {Wu}, K. 2019,
  \mnras, 483, 599

\bibitem[Good \& Chime/Frb Collaboration(2020)]{good20ATel14074}
  Good, D. \& Chime/Frb Collaboration\ 2020, The Astronomer's
  Telegram, 14074
  
\bibitem[{{G{\"o}{\u g}{\"u}{\c s} } {et~al.}(1999){G{\"o}{\u g}{\"u}{\c s} },
  {Woods}, {Kouveliotou}, {van Paradijs}, {Briggs}, {Duncan}, \&
  {Thompson}}]{gogus99apjl:1900}
{G{\"o}{\u g}{\"u}{\c s} }, E., {Woods}, P.~M., {Kouveliotou}, C., {et~al.}
  1999, \apjl, 526, L93

\bibitem[{{G{\"o}{\u g}{\"u}{\c s}} {et~al.}(2000){G{\"o}{\u g}{\"u}{\c s}},
  {Woods}, {Kouveliotou}, {van Paradijs}, {Briggs}, {Duncan}, \&
  {Thompson}}]{gogus00ApJ:1806}
{G{\"o}{\u g}{\"u}{\c s}}, E., {Woods}, P.~M., {Kouveliotou}, C., {et~al.}
  2000, \apjl, 532, L121

\bibitem[{{G{\"o}{\v g}{\"u}{\c s}} {et~al.}(2002){G{\"o}{\v g}{\"u}{\c s}},
  {Kouveliotou}, {Woods}, {Finger}, \& {van der Klis}}]{gogus02ApJ:1806}
{G{\"o}{\v g}{\"u}{\c s}}, E., {Kouveliotou}, C., {Woods}, P.~M., {Finger},
  M.~H., \& {van der Klis}, M. 2002, \apj, 577, 929

\bibitem[{Gutenberg \& Richter(1956)}]{Gutenberg56}
Gutenberg, B. \& Richter, C.~F. 1956, Bulletin of the Seismological Society of
  America, 46, 105

\bibitem[Harding et al.(1999)]{harding99ApJ} Harding, A.~K.,
   Contopoulos, I., \& Kazanas, D.\ 1999, \apjl, 525,
   L125. doi:10.1086/312339 
  
\bibitem[{{Harding} \& {Muslimov}(2001)}]{Harding2001}
{Harding}, A.~K. \& {Muslimov}, A.~G. 2001, \apj, 556, 987

\bibitem[{{Hurley} {et~al.}(1999){Hurley}, {Cline}, {Mazets}, {Barthelmy},
  {Butterworth}, {Marshall}, {Palmer}, {Aptekar}, {Golenetskii}, {Il'Inskii},
  {Frederiks}, {McTiernan}, {Gold}, \& {Trombka}}]{hurley99Natur}
{Hurley}, K., {Cline}, T., {Mazets}, E., {et~al.} 1999, \nat, 397, 41

\bibitem[{{Israel} {et~al.}(2016){Israel}, {Esposito}, {Rea}, {Coti Zelati},
  {Tiengo}, {Campana}, {Mereghetti}, {Rodriguez Castillo}, {G{\"o}tz},
  {Burgay}, {Possenti}, {Zane}, {Turolla}, {Perna}, {Cannizzaro}, \&
  {Pons}}]{israel16mnras:1935}
{Israel}, G.~L., {Esposito}, P., {Rea}, N., {et~al.} 2016, \mnras, 457, 3448

\bibitem[{{Israel} {et~al.}(2008){Israel}, {Romano}, {Mangano}, {Dall'Osso},
  {Chincarini}, {Stella}, {Campana}, {Belloni}, {Tagliaferri}, {Blustin},
  {Sakamoto}, {Hurley}, {Zane}, {Moretti}, {Palmer}, {Guidorzi}, {Burrows},
  {Gehrels}, \& {Krimm}}]{israel08ApJ:1900}
{Israel}, G.~L., {Romano}, P., {Mangano}, V., {et~al.} 2008, \apj, 685, 1114

\bibitem[Israel et al.(2010)]{israel10MNRAS} Israel, G.~L.,
  Esposito, P., Rea, N., et al.\ 2010, \mnras, 408,
  1387. doi:10.1111/j.1365-2966.2010.17001.x

\bibitem[{{Kalapotharakos} {et~al.}(2014){Kalapotharakos}, {Harding}, \&
  {Kazanas}}]{2014ApJ...793...97K}
{Kalapotharakos}, C., {Harding}, A.~K., \& {Kazanas}, D. 2014, \apj, 793, 97

\bibitem[{{Kargaltsev} {et~al.}(2012){Kargaltsev}, {Kouveliotou}, {Pavlov},
  {G{\"o}{\u g}{\"u}{\c s}}, {Lin}, {Wachter}, {Griffith}, {Kaneko}, \&
  {Younes}}]{kargaltsev12apj:1834}
{Kargaltsev}, O., {Kouveliotou}, C., {Pavlov}, G.~G., {et~al.} 2012, \apj, 748,
  26

\bibitem[{{Kaspi} \& {Beloborodov}(2017)}]{kaspi17:magnetars}
{Kaspi}, V.~M. \& {Beloborodov}, A.~M. 2017, \araa, 55, 261

\bibitem[Katz(2019)]{katz19MNRAS} Katz, J.~I.\ 2019, \mnras,
  487, 491. doi:10.1093/mnras/stz1250

\bibitem[{{Kirsten} {et~al.}(2020){Kirsten}, {Snelders}, {Jenkins}, {Nimmo},
  {van den Eijnden}, {Hessels}, {Gawronski}, \& {Yang}}]{2020arXiv200705101K}
{Kirsten}, F., {Snelders}, M., {Jenkins}, M., {et~al.} 2020, arXiv e-prints,
  arXiv:2007.05101

\bibitem[{{Kothes} {et~al.}(2018){Kothes}, {Sun}, {Gaensler}, \&
  {Reich}}]{kothes18ApJ:1935}
{Kothes}, R., {Sun}, X., {Gaensler}, B., \& {Reich}, W. 2018, \apj, 852, 54

\bibitem[{{Kouveliotou} {et~al.}(1998){Kouveliotou}, {Dieters}, {Strohmayer},
  {van Paradijs}, {Fishman}, {Meegan}, {Hurley}, {Kommers}, {Smith}, {Frail},
  \& {Murakami}}]{kouveliotou98Nat:1806}
{Kouveliotou}, C., {Dieters}, S., {Strohmayer}, T., {et~al.} 1998, \nat, 393,
  235

\bibitem[{{Kouveliotou} {et~al.}(1993){Kouveliotou}, {Meegan}, {Fishman},
  {Bhat}, {Briggs}, {Koshut}, {Paciesas}, \&
  {Pendleton}}]{kouveliotou93ApJ:GRBs}
{Kouveliotou}, C., {Meegan}, C.~A., {Fishman}, G.~J., {et~al.} 1993, \apjl,
  413, L101
  
\bibitem[{{Li} {et~al.}(2020){Li}, {Lin}, {Xiong}, {Ge}, {Li}, {Li}, {Lu},
  {Zhang}, {Tuo}, {Nang}, {Zhang}, {Xiao}, {Chen}, {Song}, {Xu}, {Liu}, {Jia},
  {Cao}, {Zhang}, {Qu}, {Liao}, {Zhao}, {Tan}, {Nie}, {Zhao}, {Zheng}, {Zheng},
  {Luo}, {Cai}, {Li}, {Xue}, {Bu}, {Chang}, {Chen}, {Chen}, {Chen}, {Chen},
  {Chen}, {Cui}, {Cui}, {Deng}, {Dong}, {Du}, {Fu}, {Gao}, {Gao}, {Gao}, {Gu},
  {Guan}, {Guo}, {Han}, {Huang}, {Huo}, {Jiang}, {Jiang}, {Jin}, {Jin}, {Kong},
  {Li}, {Li}, {Li}, {Li}, {Li}, {Li}, {Li}, {Liang}, {Liu}, {Liu}, {Liu},
  {Liu}, {Liu}, {Lu}, {Lu}, {Luo}, {Ma}, {Meng}, {Ou}, {Sai}, {Shang}, {Song},
  {Sun}, {Tao}, {Wang}, {Wang}, {Wang}, {Wang}, {Wang}, {Wen}, {Wu}, {Wu},
  {Wu}, {Xiao}, {Yang}, {Yang}, {Yang}, {Yang}, {Yi}, {Yin}, {You}, {Zhang},
  {Zhang}, {Zhang}, {Zhang}, {Zhang}, {Zhang}, {Zhang}, {Zhang}, {Zhang},
  {Zhang}, {Zhang}, {Zhang}, {Zhang}, {Zhang}, {Zhang}, {Zhang}, {Zhou},
  {Zhou}, {Zhu}, {Zhu}, \& {Zhuang}}]{li20arXiv200511071L}
{Li}, C.~K., {Lin}, L., {Xiong}, S.~L., {et~al.} 2020, arXiv e-prints,
  arXiv:2005.11071

\bibitem[{{Lin} {et~al.}(2012){Lin}, {G{\"o}{\u g}{\"u}{\c s}}, {Baring},
  {Granot}, {Kouveliotou}, {Kaneko}, {van der Horst}, {Gruber}, {von Kienlin},
  {Younes}, {Watts}, \& {Gehrels}}]{lin12ApJ:1550}
{Lin}, L., {G{\"o}{\u g}{\"u}{\c s}}, E., {Baring}, M.~G., {et~al.} 2012, \apj,
  756, 54

\bibitem[{{Lin} {et~al.}(2020){Lin}, {G{\"o}{\u{g}}{\"u}{\textcommabelow s}},
  {Roberts}, {Kouveliotou}, {Kaneko}, {van der Horst}, \&
  {Younes}}]{lin20ApJ:1935}
{Lin}, L., {G{\"o}{\u{g}}{\"u}{\textcommabelow s}}, E., {Roberts}, O.~J.,
  {et~al.} 2020, \apj, 893, 156

\bibitem[{{Lin} {et~al.}(2013){Lin}, {G{\"o}{\v g}{\"u}{\c s}}, {Kaneko}, \&
  {Kouveliotou}}]{lin13ApJ:magnetarsBB}
{Lin}, L., {G{\"o}{\v g}{\"u}{\c s}}, E., {Kaneko}, Y., \& {Kouveliotou}, C.
  2013, \apj, 778, 105

\bibitem[Luo et al.(2019)]{lu19ascl} Luo, J., Ransom, S.,
  Demorest, P., et al.\ 2019, Astrophysics Source Code
  Library. ascl:1902.007

\bibitem[Maan et al.(2019)]{maan19ApJ} Maan, Y., Joshi,
  B.~C., Surnis, M.~P., et al.\ 2019, \apjl, 882,
  L9. doi:10.3847/2041-8213/ab3a47
  
\bibitem[{{Mazets} {et~al.}(1979){Mazets}, {Golentskii}, {Ilinskii}, {Aptekar},
  \& {Guryan}}]{Mazets1979Nat}
{Mazets}, E.~P., {Golentskii}, S.~V., {Ilinskii}, V.~N., {Aptekar}, R.~L., \&
  {Guryan}, I.~A. 1979, \nat, 282, 587

\bibitem[Mereghetti et al.(2009)]{mereghetti09ApJ} Mereghetti,
    S., G{\"o}tz, D., Weidenspointner, G., et al.\ 2009, \apjl, 696,
    L74. doi:10.1088/0004-637X/696/1/L74
    
\bibitem[{{Mereghetti} {et~al.}(2020){Mereghetti}, {Savchenko}, {Ferrigno},
  {G{\"o}tz}, {Rigoselli}, {Tiengo}, {Bazzano}, {Bozzo}, {Coleiro},
  {Courvoisier}, {Doyle}, {Goldwurm}, {Hanlon}, {Jourdain}, {Kienlin},
  {Lutovinov}, {Martin-Carrillo}, {Molkov}, {Natalucci}, {Onori}, {Panessa},
  {Rodi}, {Rodriguez}, {S{\'a}nchez-Fern{\'a}ndez}, {Sunyaev}, \&
  {Ubertini}}]{mereghetti20ApJ:1935}
{Mereghetti}, S., {Savchenko}, V., {Ferrigno}, C., {et~al.} 2020, \apjl, 898,
  L29

\bibitem[{{Ng} {et~al.}(2011){Ng}, {Kaspi}, {Dib}, {Olausen}, {Scholz},
  {G{\"u}ver}, {{\"O}zel}, {Gavriil}, \& {Woods}}]{ng11ApJ:1547}
{Ng}, C.-Y., {Kaspi}, V.~M., {Dib}, R., {et~al.} 2011, \apj, 729, 131

\bibitem[{{Palmer}(2020)}]{palmer20ATel13675}
{Palmer}, D.~M. 2020, The Astronomer's Telegram, 13675, 1

\bibitem[{{Palmer} {et~al.}(2005){Palmer}, {Barthelmy}, {Gehrels}, {Kippen},
  {Cayton}, {Kouveliotou}, {Eichler}, {Wijers}, {Woods}, {Granot}, {Lyubarsky},
  {Ramirez-Ruiz}, {Barbier}, {Chester}, {Cummings}, {Fenimore}, {Finger},
  {Gaensler}, {Hullinger}, {Krimm}, {Markwardt}, {Nousek}, {Parsons}, {Patel},
  {Sakamoto}, {Sato}, {Suzuki}, \& {Tueller}}]{palmer05Natur}
{Palmer}, D.~M., {Barthelmy}, S., {Gehrels}, N., {et~al.} 2005, \nat, 434, 1107

\bibitem[{{Perna} \& {Gotthelf}(2008)}]{Perna-2008-ApJ}
{Perna}, R. \& {Gotthelf}, E.~V. 2008, \apj, 681, 522

\bibitem[{{Perna} \& {Pons}(2011)}]{perna11ApJ}
{Perna}, R. \& {Pons}, J.~A. 2011, \apjl, 727, L51

\bibitem[{Philippov {et~al.}(2020)Philippov, Timokhin, \&
  Spitkovsky}]{PhysRevLett.124.245101}
Philippov, A., Timokhin, A., \& Spitkovsky, A. 2020, Phys. Rev. Lett., 124,
  245101

\bibitem[Pleunis \& CHIME/FRB
  Collaboration(2020)]{pleunis20ATel14080} Pleunis, Z. \& CHIME/FRB
  Collaboration\ 2020, The Astronomer's Telegram, 14080
  
\bibitem[{{Pons} {et~al.}(2009){Pons}, {Miralles}, \& {Geppert}}]{pons09AA}
{Pons}, J.~A., {Miralles}, J.~A., \& {Geppert}, U. 2009, Astronomy and
  Astrophysics, 496, 207

\bibitem[{{Rea} {et~al.}(2013){Rea}, {Israel}, {Pons}, {Turolla}, {Vigan{\`o}},
  {Zane}, {Esposito}, {Perna}, {Papitto}, {Terreran}, {Tiengo}, {Salvetti},
  {Girart}, {Palau}, {Possenti}, {Burgay}, {G{\"o}{\u g}{\"u}{\c s}},
  {Caliandro}, {Kouveliotou}, {G{\"o}tz}, {Mignani}, {Ratti}, \&
  {Stella}}]{rea13ApJ:0418}
{Rea}, N., {Israel}, G.~L., {Pons}, J.~A., {et~al.} 2013, \apj, 770, 65

\bibitem[{{Ridnaia} {et~al.}(2020){Ridnaia}, {Svinkin}, {Frederiks}, {Bykov},
  {Popov}, {Aptekar}, {Golenetskii}, {Lysenko}, {Tsvetkova}, {Ulanov}, \&
  {Cline}}]{Ridnaia20arXiv200511178R}
{Ridnaia}, A., {Svinkin}, D., {Frederiks}, D., {et~al.} 2020, arXiv e-prints,
  arXiv:2005.11178

\bibitem[Rodr{\'\i}guez Castillo et al.(2014)]{castillo14MNRAS}
  Rodr{\'\i}guez Castillo, G.~A., Israel, G.~L., Esposito, P., et al.\
  2014, \mnras, 441, 1305. doi:10.1093/mnras/stu603

\bibitem[Rodr{\'\i}guez Castillo et al.(2016)]{castillo16MNRAS}
    Rodr{\'\i}guez Castillo, G.~A., Israel, G.~L., Tiengo, A., et al.\
    2016, \mnras, 456, 4145. doi:10.1093/mnras/stv2490
    
\bibitem[{{Scholz} \& {Kaspi}(2011)}]{scholz11ApJ:1547}
{Scholz}, P. \& {Kaspi}, V.~M. 2011, \apj, 739, 94

\bibitem[{{Scholz} {et~al.}(2014){Scholz}, {Kaspi}, \&
  {Cumming}}]{scholz14ApJ:1822}
{Scholz}, P., {Kaspi}, V.~M., \& {Cumming}, A. 2014, \apj, 786, 62

\bibitem[Staelin(1969)]{Staelin69EF} Staelin, D.~H.\ 1969,
  IEEE Proceedings, 57, 724. doi:10.1109/PROC.1969.7051

\bibitem[{{Stamatikos} {et~al.}(2014){Stamatikos}, {Malesani}, {Page}, \&
  {Sakamoto}}]{stamatikos14:1935}
{Stamatikos}, M., {Malesani}, D., {Page}, K.~L., \& {Sakamoto}, T. 2014, GRB
  Coordinates Network, 16520

\bibitem[Tavani et al.(2020)]{tavani20arXiv200512164T} Tavani, M.,
  Casentini, C., Ursi, A., et al.\ 2020, arXiv:2005.12164
  
\bibitem[{{The CHIME/FRB Collaboration} {et~al.}(2020){The CHIME/FRB
  Collaboration}, {:}, {Andersen}, {Band ura}, {Bhardwaj}, {Bij}, {Boyce},
  {Boyle}, {Brar}, {Cassanelli}, {Chawla}, {Chen}, {Cliche}, {Cook},
  {Cubranic}, {Curtin}, {Denman}, {Dobbs}, {Dong}, {Fandino}, {Fonseca},
  {Gaensler}, {Giri}, {Good}, {Halpern}, {Hill}, {Hinshaw}, {H{\"o}fer},
  {Josephy}, {Kania}, {Kaspi}, {Landecker}, {Leung}, {Li}, {Lin}, {Masui},
  {Mckinven}, {Mena-Parra}, {Merryfield}, {Meyers}, {Michilli}, {Milutinovic},
  {Mirhosseini}, {M{\"u}nchmeyer}, {Naidu}, {Newburgh}, {Ng}, {Patel}, {Pen},
  {Pinsonneault-Marotte}, {Pleunis}, {Quine}, {Rafiei-Ravandi}, {Rahman},
  {Ransom}, {Renard}, {Sanghavi}, {Scholz}, {Shaw}, {Shin}, {Siegel}, {Singh},
  {Smegal}, {Smith}, {Stairs}, {Tan}, {Tendulkar}, {Tretyakov}, {Vanderlinde},
  {Wang}, {Wulf}, \& {Zwaniga}}]{chime20arXiv200510324T}
{The CHIME/FRB Collaboration}, {:}, {Andersen}, B.~C., {et~al.} 2020, arXiv
  e-prints, arXiv:2005.10324

\bibitem[{{Thompson} \& {Duncan}(1996)}]{thompson96ApJ:magnetar}
{Thompson}, C. \& {Duncan}, R.~C. 1996, \apj, 473, 322

\bibitem[{{Thompson} {et~al.}(2002){Thompson}, {Lyutikov}, \&
  {Kulkarni}}]{thompson02ApJ:magnetars}
{Thompson}, C., {Lyutikov}, M., \& {Kulkarni}, S.~R. 2002, \apj, 574, 332

\bibitem[{{Timokhin} \& {Arons}(2013)}]{2013MNRAS.429...20T}
{Timokhin}, A.~N. \& {Arons}, J. 2013, \mnras, 429, 20

\bibitem[Tong et al.(2013)]{tong2013ApJ} Tong, H., Xu, R.~X.,
  Song, L.~M., et al.\ 2013, \apj, 768,
  144. doi:10.1088/0004-637X/768/2/144

\bibitem[{{Turolla} {et~al.}(2015){Turolla}, {Zane}, \&
  {Watts}}]{turolla15:mag}
{Turolla}, R., {Zane}, S., \& {Watts}, A.~L. 2015, Reports on Progress in
  Physics, 78, 116901

\bibitem[{{van der Horst} {et~al.}(2012){van der Horst}, {Kouveliotou},
  {Gorgone}, {Kaneko}, {Baring}, {Guiriec}, {G{\"o}{\v g}{\"u}{\c s}},
  {Granot}, {Watts}, {Lin}, {Bhat}, {Bissaldi}, {Chaplin}, {Finger}, {Gehrels},
  {Gibby}, {Giles}, {Goldstein}, {Gruber}, {Harding}, {Kaper}, {von Kienlin},
  {van der Klis}, {McBreen}, {Mcenery}, {Meegan}, {Paciesas}, {Pe'er},
  {Preece}, {Ramirez-Ruiz}, {Rau}, {Wachter}, {Wilson-Hodge}, {Woods}, \&
  {Wijers}}]{vanderhorst12ApJ:1550}
{van der Horst}, A.~J., {Kouveliotou}, C., {Gorgone}, N.~M., {et~al.} 2012,
  \apj, 749, 122

\bibitem[{{Verner} {et~al.}(1996){Verner}, {Ferland}, {Korista}, \&
  {Yakovlev}}]{verner96ApJ:crossSect}
{Verner}, D.~A., {Ferland}, G.~J., {Korista}, K.~T., \& {Yakovlev}, D.~G. 1996,
  \apj, 465, 487

\bibitem[Wadiasingh \& Timokhin(2019)]{Wadiasingh2019ApJ}
  Wadiasingh, Z. \& Timokhin, A.\ 2019, \apj, 879,
  4. doi:10.3847/1538-4357/ab2240
  
\bibitem[{{Wilms} {et~al.}(2000){Wilms}, {Allen}, \& {McCray}}]{wilms00ApJ}
{Wilms}, J., {Allen}, A., \& {McCray}, R. 2000, \apj, 542, 914

\bibitem[{{Woods} {et~al.}(2004){Woods}, {Kaspi}, {Thompson}, {Gavriil},
  {Marshall}, {Chakrabarty}, {Flanagan}, {Heyl}, \&
  {Hernquist}}]{woods04ApJ:1E2259}
{Woods}, P.~M., {Kaspi}, V.~M., {Thompson}, C., {et~al.} 2004, \apj, 605, 378

\bibitem[{{Woods} {et~al.}(2007){Woods}, {Kouveliotou}, {Finger}, {G{\"o}{\v
  g}{\"u}{\c s}}, {Wilson}, {Patel}, {Hurley}, \& {Swank}}]{woods07ApJ:1806}
{Woods}, P.~M., {Kouveliotou}, C., {Finger}, M.~H., {et~al.} 2007, \apj, 654,
  470

\bibitem[{{Woods} {et~al.}(1999{\natexlab{a}}){Woods}, {Kouveliotou}, {van
  Paradijs}, {Finger}, {Thompson}, {Duncan}, {Hurley}, {Strohmayer}, {Swank},
  \& {Murakami}}]{woods99ApJ:1900}
{Woods}, P.~M., {Kouveliotou}, C., {van Paradijs}, J., {et~al.}
  1999{\natexlab{a}}, \apjl, 524, L55

\bibitem[{{Woods} {et~al.}(1999{\natexlab{b}}){Woods}, {Kouveliotou}, {van
  Paradijs}, {Hurley}, {Kippen}, {Finger}, {Briggs}, {Dieters}, \&
  {Fishman}}]{woods99ApJ:1627}
{Woods}, P.~M., {Kouveliotou}, C., {van Paradijs}, J., {et~al.}
  1999{\natexlab{b}}, \apjl, 519, L139

\bibitem[{{Younes} {et~al.}(2020{\natexlab{a}}){Younes}, {Baring},
  {Kouveliotou}, {Arzoumanian}, {Enoto}, {Doty}, {Gendreau},
  {G{\"o}{\u{g}}{\"u}{\textcommabelow s}}, {Guillot}, {G{\"u}ver}, {Harding},
  {Ho}, {van der Horst}, {Jaisawal}, {Kaneko}, {LaMarr}, {Lin}, {Majid},
  {Okajima}, {Pope}, {Ray}, {Roberts}, {Saylor}, {Steiner}, \&
  {Wadiasingh}}]{younes20arXiv200611358Y}
{Younes}, G., {Baring}, M.~G., {Kouveliotou}, C., {et~al.} 2020{\natexlab{a}},
  arXiv e-prints, arXiv:2006.11358

\bibitem[{{Younes} {et~al.}(2017{\natexlab{a}}){Younes}, {Baring},
  {Kouveliotou}, {Harding}, {Donovan}, {G{\"o}{\u{g}}{\"u}{\textcommabelow s}},
  {Kaspi}, \& {Granot}}]{younes17ApJ:1806}
{Younes}, G., {Baring}, M.~G., {Kouveliotou}, C., {et~al.} 2017{\natexlab{a}},
  \apj, 851, 17

\bibitem[{{Younes} {et~al.}(2020{\natexlab{b}}){Younes}, {Baring},
  {Kouveliotou}, {Wadiasingh}, {Huppenkothen}, \& {Harding}}]{younes20ApJ:1708}
{Younes}, G., {Baring}, M.~G., {Kouveliotou}, C., {et~al.} 2020{\natexlab{b}},
  \apjl, 889, L27

\bibitem[{{Younes} {et~al.}(2020{\natexlab{c}}){Younes}, {Guver}, {Enoto},
  {Arzoumanian}, {Gendreau}, {Hu}, {Ray}, {Kouveliotou}, {Guillot}, {Ho},
  {Ferrara}, \& {Malacaria}}]{younes20ATel13678}
{Younes}, G., {Guver}, T., {Enoto}, T., {et~al.} 2020{\natexlab{c}}, The
  Astronomer's Telegram, 13678, 1

\bibitem[{{Younes} {et~al.}(2017{\natexlab{b}}){Younes}, {Kouveliotou},
  {Jaodand}, {Baring}, {van der Horst}, {Harding}, {Hessels}, {Gehrels},
  {Gill}, {Huppenkothen}, {Granot}, {G{\"o}{\u g}{\"u}{\c s}}, \&
  {Lin}}]{younes17:1935}
{Younes}, G., {Kouveliotou}, C., {Jaodand}, A., {et~al.} 2017{\natexlab{b}},
  \apj, 847, 85

\bibitem[{{Younes} {et~al.}(2015){Younes}, {Kouveliotou}, \&
  {Kaspi}}]{younes15ApJ:1806}
{Younes}, G., {Kouveliotou}, C., \& {Kaspi}, V.~M. 2015, \apj, 809, 165

\bibitem[{{Younes} {et~al.}(2014){Younes}, {Kouveliotou}, {van der Horst},
  {Baring}, {Granot}, {Watts}, {Bhat}, {Collazzi}, {Gehrels}, {Gorgone},
  {G{\"o}{\u g}{\"u}{\c s}}, {Gruber}, {Grunblatt}, {Huppenkothen}, {Kaneko},
  {von Kienlin}, {van der Klis}, {Lin}, {Mcenery}, {van Putten}, \&
  {Wijers}}]{younes14ApJ:1550}
{Younes}, G., {Kouveliotou}, C., {van der Horst}, A.~J., {et~al.} 2014, \apj,
785, 52

\bibitem[{{Zhang} {et~al.}(2020){Zhang}, {Jiang}, {Men}, {Wang}, {Xu}, {Xu},
  {Niu}, {Zhou}, {Guan}, {Han}, {Jiang}, {Lee}, {Li}, {Lin}, {Niu}, {Wang},
  {Wang}, {Xu}, {Yu}, {Zhang}, \& {Zhu}}]{zhang20ATel13699}
{Zhang}, C.~F., {Jiang}, J.~C., {Men}, Y.~P., {et~al.} 2020, The Astronomer's
  Telegram, 13699, 1

\bibitem[Zhang et al.(2018)]{zhangYG18ApJ} Zhang, Y.~G.,
  Gajjar, V., Foster, G., et al.\ 2018, \apj, 866,
  149. doi:10.3847/1538-4357/aadf31
  
\bibitem[Zhong et al.(2020)]{zhong20ApJ} Zhong, S.-Q., Dai,
  Z.-G., Zhang, H.-M., et al.\ 2020, \apjl, 898,
  L5. doi:10.3847/2041-8213/aba262
  
\bibitem[{{Zhou} {et~al.}(2020){Zhou}, {Zhou}, {Chen}, {Wang}, {Vink}, \&
  {Wang}}]{zhou20arXiv200503517Z}
{Zhou}, P., {Zhou}, X., {Chen}, Y., {et~al.} 2020, arXiv e-prints,
  arXiv:2005.03517

\bibitem[Zhu et al.(2020)]{zhu20ATel14084} Zhu, W., Wang, B., Zhou,
  D., et al.\ 2020, The Astronomer's Telegram, 14084
  
\end{thebibliography}


\begin{center}
\begin{table*}
    \caption{Spectral properties of the persistent emission.}
    \label{tab:pers_sp_res}
    \begin{tabular}{cccccccc}
        \hline
        OBSID & MJD$^*$ & Exposure & Count Rate & kT &  Flux$^{**}$  & Radius \\
        & days & s & (cts/s) & (keV) &  & km \\ 
        \hline
        \hline
 	   3020560101 & 58967.165 & 233.70  &2.74  $\pm0.12$ & 0.79$_{-0.04}^{+0.04}$  & 2.2$_{-0.1}^{+0.1}$  & 2.1$_{-0.2}^{+0.2}$ \\
 		   3020560101 & 58967.229 & 231.10  &2.16  $\pm0.11$ & 0.73$_{-0.04}^{+0.04}$  & 1.7$_{-0.1}^{+0.1}$  & 2.2$_{-0.2}^{+0.2}$ \\
          	   3020560101 & 58967.293 & 727.80  &1.99  $\pm0.06$ & 0.78$_{-0.02}^{+0.03}$  & 1.54$_{-0.05}^{+0.05}$  & 1.8$_{-0.1}^{+0.1}$ \\
          	   3020560101 & 58967.681 & 805.90  &1.33  $\pm0.05$ & 0.65$_{-0.02}^{+0.02}$  & 0.98$_{-0.04}^{+0.04}$  & 2.1$_{-0.1}^{+0.2}$ \\
           	   3020560102 & 58968.583 & 934.09  &0.95  $\pm0.04$ & 0.62$_{-0.02}^{+0.03}$  & 0.68$_{-0.03}^{+0.03}$  & 1.9$_{-0.1}^{+0.2}$ \\
		   3020560103 & 58969.551 & 712.07  &0.68  $\pm0.04$ & 0.51$_{-0.02}^{+0.03}$  & 0.51$_{-0.03}^{+0.03}$  & 2.4$_{-0.3}^{+0.3}$ \\
                   3020560104 & 58980.615 & 944.04  &0.55  $\pm0.03$ & 0.47$_{-0.02}^{+0.02}$  & 0.44$_{-0.02}^{+0.02}$  & 2.7$_{-0.3}^{+0.3}$ \\
         	   3020560105 & 58988.922 & 915.07  &0.46  $\pm0.03$ & 0.49$_{-0.03}^{+0.03}$  & 0.37$_{-0.02}^{+0.02}$  & 2.2$_{-0.3}^{+0.3}$ \\		
          	   3020560106 & 58989.589 & 624.02  &0.54  $\pm0.03$ & 0.49$_{-0.03}^{+0.03}$  & 0.40$_{-0.03}^{+0.03}$  & 2.4$_{-0.3}^{+0.3}$ \\
          	   3020560107 & 58991.075 & 5176.37 &0.46  $\pm0.01$ & 0.49$_{-0.01}^{+0.01}$  & 0.34$_{-0.01}^{+0.01}$  & 2.2$_{-0.1}^{+0.1}$ \\
      	           3020560108 & 58992.032 & 3208.26 &0.57  $\pm0.02$ & 0.47$_{-0.02}^{+0.02}$  & 0.40$_{-0.01}^{+0.01}$  & 2.5$_{-0.2}^{+0.2}$ \\   
		   3020560109 & 58994.601 & 790.06  &0.49  $\pm0.03$ & 0.52$_{-0.03}^{+0.03}$  & 0.38$_{-0.02}^{+0.02}$  & 2.0$_{-0.3}^{+0.3}$ \\
		   3020560110 & 58997.896 & 1678.12 &0.31  $\pm0.02$ & 0.40$_{-0.02}^{+0.02}$  & 0.28$_{-0.02}^{+0.02}$  & 2.9$_{-0.3}^{+0.4}$ \\		
		   3020560111 & 58998.150 & 963.07  &0.34  $\pm0.03$ & 0.47$_{-0.03}^{+0.03}$  & 0.28$_{-0.02}^{+0.02}$  & 2.1$_{-0.3}^{+0.3}$ \\
		   3020560112 & 58999.249 & 1279.09 &0.41  $\pm0.03$ & 0.50$_{-0.03}^{+0.03}$  & 0.31$_{-0.02}^{+0.02}$  & 2.0$_{-0.3}^{+0.3}$ \\
		   3020560114 & 59001.130 & 2745.16 &0.46  $\pm0.02$ & 0.52$_{-0.02}^{+0.02}$  & 0.34$_{-0.01}^{+0.01}$  & 1.9$_{-0.2}^{+0.2}$ \\
		   3020560115 & 59002.992 & 972.06  &0.33  $\pm0.03$ & 0.54$_{-0.04}^{+0.04}$  & 0.26$_{-0.02}^{+0.02}$  & 1.5$_{-0.2}^{+0.3}$ \\
		   3020560116 & 59003.184 & 665.04  &0.33  $\pm0.03$ & 0.47$_{-0.04}^{+0.04}$  & 0.27$_{-0.03}^{+0.03}$  & 2.1$_{-0.4}^{+0.4}$ \\		
		   3020560117 & 59004.545 & 1368.08 &0.26  $\pm0.02$ & 0.48$_{-0.03}^{+0.03}$  & 0.21$_{-0.02}^{+0.02}$  & 1.8$_{-0.2}^{+0.3}$ \\
		   3020560118 & 59005.189 & 1201.07 &0.24  $\pm0.02$ & 0.44$_{-0.03}^{+0.03}$  & 0.20$_{-0.02}^{+0.02}$  & 2.1$_{-0.3}^{+0.3}$ \\
		   3020560119 & 59006.222 & 990.06  &0.19  $\pm0.02$ & 0.44$_{-0.04}^{+0.04}$  & 0.17$_{-0.02}^{+0.02}$  & 1.9$_{-0.3}^{+0.4}$ \\
		   3020560120 & 59020.891 & 883.04  &0.37  $\pm0.03$ & 0.44$_{-0.03}^{+0.03}$  & 0.31$_{-0.02}^{+0.02}$  & 2.5$_{-0.4}^{+0.4}$ \\
		   3020560121 & 59021.014 & 846.04  &0.27  $\pm0.03$ & 0.37$_{-0.03}^{+0.03}$  & 0.28$_{-0.03}^{+0.03}$  & 3.5$_{-0.6}^{+0.8}$ \\
		   3020560122 & 59022.445 & 1675.05 &0.36  $\pm0.02$ & 0.41$_{-0.02}^{+0.02}$  & 0.30$_{-0.02}^{+0.02}$  & 2.8$_{-0.3}^{+0.3}$ \\
		   3020560123 & 59023.101 & 3182.09 &0.21  $\pm0.01$ & 0.42$_{-0.02}^{+0.02}$  & 0.20$_{-0.01}^{+0.01}$  & 2.3$_{-0.2}^{+0.2}$ \\
		   3020560124 & 59024.961 & 1840.05 &0.27  $\pm0.02$ & 0.44$_{-0.02}^{+0.02}$  & 0.22$_{-0.01}^{+0.01}$  & 2.1$_{-0.3}^{+0.3}$ \\
		   3020560125 & 59025.931 & 1655.04 &0.26  $\pm0.02$ & 0.43$_{-0.02}^{+0.02}$  & 0.23$_{-0.02}^{+0.02}$  & 2.3$_{-0.3}^{+0.3}$ \\
		   3020560126 & 59027.614 & 847.01  &0.29  $\pm0.03$ & 0.42$_{-0.04}^{+0.04}$  & 0.24$_{-0.03}^{+0.03}$  & 2.6$_{-0.5}^{+0.7}$ \\
		   3020560127 & 59028.705 & 1703.04 &0.15  $\pm0.02$ & 0.42$_{-0.03}^{+0.03}$  & 0.15$_{-0.02}^{+0.02}$  & 2.0$_{-0.3}^{+0.3}$ \\
		   3020560128 & 59029.802 & 1692.03 &0.20  $\pm0.02$ & 0.42$_{-0.03}^{+0.03}$  & 0.18$_{-0.02}^{+0.02}$  & 2.2$_{-0.3}^{+0.4}$ \\
		   3020560129 & 59030.699 & 2150.65 &0.21  $\pm0.02$ & 0.43$_{-0.02}^{+0.02}$  & 0.19$_{-0.01}^{+0.01}$  & 2.1$_{-0.3}^{+0.3}$ \\
		   3020560130 & 59031.529 & 615.03  &0.33  $\pm0.04$ & 0.45$_{-0.04}^{+0.04}$  & 0.27$_{-0.03}^{+0.03}$  & 2.3$_{-0.5}^{+0.6}$ \\
		   3020560131 & 59032.500 & 1190.05 &0.16  $\pm0.02$ & 0.39$_{-0.03}^{+0.03}$  & 0.19$_{-0.02}^{+0.02}$  & 2.6$_{-0.4}^{+0.5}$ \\
		   3020560132 & 59033.598 & 1058.04 &0.27  $\pm0.02$ & 0.38$_{-0.02}^{+0.03}$  & 0.26$_{-0.02}^{+0.03}$  & 3.2$_{-0.5}^{+0.6}$ \\
		   3020560133 & 59034.824 & 819.03  &0.22  $\pm0.02$ & 0.40$_{-0.04}^{+0.04}$  & 0.21$_{-0.03}^{+0.03}$  & 2.6$_{-0.5}^{+0.7}$ \\
		   3020560134 & 59038.773 & 496.00  &0.13  $\pm0.04$ & 0.36$_{-0.06}^{+0.08}$  & 0.15$_{-0.04}^{+0.05}$  & 3$_{-1}^{+1}$ \\
		   3020560135 & 59040.246 & 361.01  &0.19  $\pm0.03$ & 0.34$_{-0.05}^{+0.05}$  & 0.28$_{-0.06}^{+0.10}$  & 4$_{-2}^{+3}$ \\
		   3020560136 & 59041.431 & 1267.99 &0.24  $\pm0.02$ & 0.41$_{-0.03}^{+0.03}$  & 0.22$_{-0.02}^{+0.02}$  & 2.4$_{-0.4}^{+0.4}$ \\
		   3020560137 & 59042.848 & 657.00  &0.14  $\pm0.03$ & 0.42$_{-0.05}^{+0.05}$  & 0.14$_{-0.02}^{+0.03}$  & 1.9$_{-0.5}^{+0.6}$ \\ 
		   3020560138 & 59045.692 & 3128.00 &0.12  $\pm0.02$ & 0.42$_{-0.02}^{+0.02}$  & 0.14$_{-0.01}^{+0.01}$  & 1.9$_{-0.2}^{+0.2}$ \\
		   3020560139 & 59046.269 & 1423.99 &0.21  $\pm0.02$ & 0.39$_{-0.03}^{+0.03}$  & 0.21$_{-0.02}^{+0.02}$  & 2.8$_{-0.4}^{+0.5}$ \\
		   3020560141 & 59049.029 & 776.01  &0.39  $\pm0.03$ & 0.39$_{-0.02}^{+0.02}$  & 0.36$_{-0.03}^{+0.03}$  & 3.6$_{-0.5}^{+0.6}$ \\
		   3020560142 & 59056.062 & 222.00  &0.42  $\pm0.07$ & 0.5$_{-0.1}^{+0.1}$  & 0.34$_{-0.06}^{+0.07}$  & 2.0$_{-0.8}^{+2}$ \\ 						 	   
         \hline
\tabularnewline
\end{tabular}
\\{\footnotesize{$^*$ Times are given as the middle of each observation. \\
$^{**}$ 0.3$-$10 keV flux values are unabsorbed and in units of $\times10^{-11}$~erg~s$^{-1}$~cm$^{-2}$. }}
\end{table*}
\end{center}

\pagebreak

\begin{center}
\begin{longtable}{l c c c c c c c}
\caption{Burst properties}\\
\label{bstTempProp}\\
\hline
\hline
Burst \# & $T_{\rm st,100}$ & $T_{\rm et,100}$ & $T_{\rm peak}$ & $T_{90}$ & $kT$ & $R^2$ & $\log_{10}F$ \\ 
      & UTC & UTC & UTC & s & keV & km$^2$ & erg~cm$^{-2}$~s$^{-1}$\\ 
\hline
1 & 00:41:21.230 & 00:41:21.479 & 00:41:21.260 & 0.194 & $3_{-1}^{+5}$ & $6_{-6}^{+10}$ & $-8.5_{-0.2}^{+0.2}$ \\ 
2 & 00:41:23.627 & 00:41:23.759 & 00:41:23.653 & 0.088 & \ldots & \ldots & \ldots \\ 
3 & 00:41:31.809 & 00:41:33.964 & 00:41:32.220 & 0.41 & $3.9_{-0.2}^{+0.2}$ & $600_{-40}^{+50}$ & $-6.26_{-0.01}^{+0.01}$ \\ 
4 & 00:41:49.245 & 00:41:49.748 & 00:41:49.321 & 0.216 & $1.9_{-0.4}^{+0.6}$ & $30_{-13}^{+21}$ & $-8.4_{-0.1}^{+0.1}$ \\ 
5 & 00:41:53.301 & 00:41:53.369 & 00:41:53.367 & 0.067 & \ldots & \ldots & \ldots \\ 
6 & 00:41:56.315 & 00:41:58.098 & 00:41:56.329 & 1.453 & $1.7_{-0.2}^{+0.2}$ & $15_{-4}^{+5}$ & $-8.81_{-0.05}^{+0.05}$ \\ 
7 & 00:42:00.202 & 00:42:00.482 & 00:42:00.253 & 0.225 & $2_{-1}^{+2}$ & $8_{-6}^{+14}$ & $-8.8_{-0.2}^{+0.2}$ \\ 
8 & 00:42:01.469 & 00:42:02.506 & 00:42:01.643 & 0.538 & $2.2_{-0.1}^{+0.1}$ & $120_{-16}^{+17}$ & $-7.61_{-0.02}^{+0.02}$ \\ 
9 & 00:42:14.191 & 00:42:14.793 & 00:42:14.484 & 0.462 & $1.4_{-0.2}^{+0.2}$ & $60_{-18}^{+25}$ & $-8.54_{-0.06}^{+0.06}$ \\ 
10 & 00:42:26.726 & 00:42:27.232 & 00:42:26.739 & 0.47 & $1.2_{-0.2}^{+0.3}$ & $32_{-14}^{+23}$ & $-9.0_{-0.1}^{+0.1}$ \\ 
11 & 00:42:43.612 & 00:42:43.976 & 00:42:43.711 & 0.188 & $2.1_{-0.4}^{+0.6}$ & $35_{-15}^{+22}$ & $-8.19_{-0.09}^{+0.09}$ \\ 
12 & 00:42:48.123 & 00:42:48.455 & 00:42:48.218 & 0.249 & $1.0_{-0.2}^{+0.2}$ & $94_{-48}^{+93}$ & $-9.0_{-0.1}^{+0.1}$ \\ 
13 & 00:42:51.848 & 00:42:53.764 & 00:42:52.058 & 0.985 & $1.96_{-0.08}^{+0.08}$ & $140_{-14}^{+15}$ & $-7.68_{-0.02}^{+0.02}$ \\ 
14 & 00:42:54.408 & 00:42:54.866 & 00:42:54.467 & 0.274 & $2.7_{-0.2}^{+0.2}$ & $165_{-23}^{+25}$ & $-7.2_{-0.02}^{+0.02}$ \\ 
15 & 00:43:01.767 & 00:43:03.795 & 00:43:01.939 & 1.521 & $2.3_{-0.4}^{+0.7}$ & $5_{-2}^{+3}$ & $-8.93_{-0.09}^{+0.09}$ \\ 
16 & 00:43:09.236 & 00:43:11.608 & 00:43:10.976 & 1.923 & $1.7_{-0.1}^{+0.2}$ & $20_{-4}^{+5}$ & $-8.72_{-0.04}^{+0.04}$ \\ 
17 & 00:43:16.781 & 00:43:17.206 & 00:43:16.911 & 0.331 & \ldots & \ldots & \ldots \\ 
18 & 00:43:22.462 & 00:43:24.047 & 00:43:22.492 & 1.401 & $1.7_{-0.1}^{+0.1}$ & $37_{-7}^{+8}$ & $-8.44_{-0.03}^{+0.04}$ \\ 
19 & 00:43:24.684 & 00:43:26.995 & 00:43:25.364 & 0.773 & $3.03_{-0.07}^{+0.07}$ & $470_{-22}^{+23}$ & $-6.62_{-0.01}^{+0.01}$ \\ 
20 & 00:43:33.269 & 00:43:34.182 & 00:43:33.334 & 0.809 & $1.2_{-0.2}^{+0.2}$ & $31_{-13}^{+21}$ & $-9.13_{-0.08}^{+0.09}$ \\ 
21 & 00:43:35.328 & 00:43:38.004 & 00:43:37.945 & 2.015 & $1.8_{-0.1}^{+0.2}$ & $19_{-4}^{+5}$ & $-8.69_{-0.04}^{+0.04}$ \\ 
22 & 00:43:40.402 & 00:43:40.590 & 00:43:40.490 & 0.113 & $1.4_{-0.3}^{+0.6}$ & $52_{-33}^{+68}$ & $-8.7_{-0.2}^{+0.2}$ \\ 
23 & 00:43:42.160 & 00:43:42.448 & 00:43:42.183 & 0.146 & \ldots & \ldots & \ldots \\ 
24 & 00:43:44.976 & 00:43:45.851 & 00:43:45.240 & 0.809 & $1.7_{-0.2}^{+0.3}$ & $20_{-7}^{+9}$ & $-8.68_{-0.07}^{+0.07}$ \\ 
25 & 00:44:00.064 & 00:44:01.836 & 00:44:00.180 & 1.366 & $1.7_{-0.2}^{+0.2}$ & $18_{-5}^{+6}$ & $-8.78_{-0.06}^{+0.06}$ \\ 
26 & 00:44:05.033 & 00:44:05.392 & 00:44:05.104 & 0.319 & $1.7_{-0.3}^{+0.5}$ & $20_{-10}^{+16}$ & $-8.7_{-0.1}^{+0.1}$ \\ 
27 & 00:44:08.027 & 00:44:10.291 & 00:44:08.368 & 1.207 & $3.40_{-0.07}^{+0.07}$ & $470_{-17}^{+17}$ & $-6.5_{-0.01}^{+0.01}$ \\ 
28 & 00:44:19.561 & 00:44:21.128 & 00:44:19.570 & 1.433 & $1.8_{-0.2}^{+0.2}$ & $18_{-5}^{+6}$ & $-8.71_{-0.05}^{+0.05}$ \\ 
29 & 00:44:25.351 & 00:44:27.499 & 00:44:26.236 & 0.955 & $1.6_{-0.2}^{+0.2}$ & $34_{-8}^{+10}$ & $-8.56_{-0.05}^{+0.05}$ \\ 
30 & 00:44:31.343 & 00:44:32.824 & 00:44:32.458 & 1.174 & $1.7_{-0.2}^{+0.2}$ & $27_{-7}^{+8}$ & $-8.63_{-0.05}^{+0.05}$ \\ 
31 & 00:44:39.437 & 00:44:40.107 & 00:44:40.056 & 0.417 & $1.47_{-0.3}^{+0.52}$ & $25_{-13}^{+21}$ & $-8.8_{-0.1}^{+0.1}$ \\ 
32 & 00:44:45.119 & 00:44:45.796 & 00:44:45.286 & 0.398 & $1.4_{-0.1}^{+0.2}$ & $77_{-22}^{+29}$ & $-8.47_{-0.06}^{+0.06}$ \\ 
33 & 00:44:48.153 & 00:44:49.369 & 00:44:48.974 & 1.041 & $1.7_{-0.1}^{+0.2}$ & $40_{-8}^{+10}$ & $-8.42_{-0.04}^{+0.04}$ \\ 
34 & 00:44:49.830 & 00:44:50.063 & 00:44:49.855 & 0.123 & $1.8_{-0.2}^{+0.3}$ & $120_{-39}^{+54}$ & $-7.85_{-0.07}^{+0.07}$ \\ 
35 & 00:44:51.116 & 00:44:52.185 & 00:44:51.209 & 0.954 & $1.6_{-0.2}^{+0.3}$ & $14_{-6}^{+9}$ & $-9.0_{-0.1}^{+0.1}$ \\ 
36 & 00:44:55.529 & 00:44:56.708 & 00:44:56.358 & 0.998 & $1.9_{-0.3}^{+0.4}$ & $10_{-4}^{+5}$ & $-8.88_{-0.08}^{+0.08}$ \\ 
37 & 00:44:59.515 & 00:45:00.664 & 00:44:59.898 & 0.3 & $2.1_{-0.1}^{+0.2}$ & $140_{-24}^{+27}$ & $-7.6_{-0.03}^{+0.03}$ \\ 
38 & 00:45:05.650 & 00:45:06.117 & 00:45:05.783 & 0.35 & $1.7_{-0.2}^{+0.3}$ & $44_{-15}^{+21}$ & $-8.4_{-0.07}^{+0.08}$ \\ 
39 & 00:45:09.942 & 00:45:11.478 & 00:45:11.175 & 0.854 & $2.0_{-0.1}^{+0.1}$ & $55_{-9}^{+10}$ & $-8.07_{-0.03}^{+0.03}$ \\ 
40 & 00:45:11.925 & 00:45:12.400 & 00:45:12.122 & 0.365 & $1.7_{-0.2}^{+0.3}$ & $42_{-14}^{+20}$ & $-8.42_{-0.07}^{+0.07}$ \\ 
41 & 00:45:20.704 & 00:45:20.911 & 00:45:20.845 & 0.145 & \ldots & \ldots & \ldots \\ 
42 & 00:45:21.060 & 00:45:22.732 & 00:45:21.543 & 1.259 & $1.5_{-0.1}^{+0.1}$ & $43_{-9}^{+11}$ & $-8.61_{-0.04}^{+0.04}$ \\ 
43 & 00:45:23.520 & 00:45:24.854 & 00:45:24.151 & 1.131 & $1.6_{-0.1}^{+0.2}$ & $38_{-8}^{+10}$ & $-8.51_{-0.04}^{+0.04}$ \\ 
44 & 00:45:28.567 & 00:45:29.696 & 00:45:28.849 & 0.915 & $1.5_{-0.2}^{+0.2}$ & $32_{-9}^{+12}$ & $-8.67_{-0.06}^{+0.06}$ \\ 
45 & 00:45:30.111 & 00:45:31.608 & 00:45:31.186 & 1.241 & $2.7_{-0.1}^{+0.1}$ & $79_{-7}^{+8}$ & $-7.53_{-0.02}^{+0.02}$ \\ 
46 & 00:45:33.014 & 00:45:34.145 & 00:45:33.729 & 0.921 & $2.0_{-0.5}^{+0.9}$ & $8_{-4}^{+7}$ & $-8.9_{-0.1}^{+0.1}$ \\ 
47 & 00:45:38.496 & 00:45:39.493 & 00:45:39.254 & 0.655 & $2.4_{-0.2}^{+0.2}$ & $59_{-9}^{+11}$ & $-7.81_{-0.03}^{+0.03}$ \\ 
48 & 00:45:41.129 & 00:45:42.806 & 00:45:42.233 & 1.484 & $1.1_{-0.1}^{+0.1}$ & $37_{-11}^{+15}$ & $-9.09_{-0.06}^{+0.06}$ \\ 
49 & 00:45:43.812 & 00:45:44.098 & 00:45:43.978 & 0.259 & $1.7_{-0.2}^{+0.3}$ & $62_{-19}^{+26}$ & $-8.24_{-0.06}^{+0.07}$ \\ 
50 & 00:45:44.961 & 00:45:47.429 & 00:45:46.842 & 2.015 & $1.8_{-0.1}^{+0.1}$ & $48_{-6}^{+7}$ & $-8.3_{-0.03}^{+0.03}$ \\ 
51 & 00:45:48.375 & 00:45:49.323 & 00:45:48.460 & 0.777 & $1.5_{-0.2}^{+0.3}$ & $23_{-8}^{+12}$ & $-8.9_{-0.08}^{+0.08}$ \\ 
52 & 00:45:49.616 & 00:45:50.493 & 00:45:49.738 & 0.513 & $1.5_{-0.1}^{+0.1}$ & $87_{-20}^{+25}$ & $-8.31_{-0.05}^{+0.05}$ \\ 
53 & 00:45:51.468 & 00:45:51.829 & 00:45:51.765 & 0.289 & $1.1_{-0.1}^{+0.2}$ & $124_{-53}^{+89}$ & $-8.69_{-0.07}^{+0.07}$ \\ 
54 & 00:45:55.816 & 00:45:57.065 & 00:45:56.514 & 0.962 & $2.2_{-0.3}^{+0.4}$ & $13_{-4}^{+5}$ & $-8.57_{-0.06}^{+0.06}$ \\ 
55 & 00:45:57.325 & 00:45:57.553 & 00:45:57.347 & 0.088 & $2.9_{-0.5}^{+0.8}$ & $70_{-25}^{+34}$ & $-7.49_{-0.07}^{+0.07}$ \\ 
56 & 00:45:58.054 & 00:45:59.256 & 00:45:58.064 & 0.544 & $1.4_{-0.2}^{+0.2}$ & $59_{-19}^{+26}$ & $-8.6_{-0.06}^{+0.07}$ \\ 
57 & 00:45:59.998 & 00:46:01.784 & 00:46:00.687 & 1.158 & $2.71_{-0.06}^{+0.06}$ & $400_{-17}^{+18}$ & $-6.81_{-0.01}^{+0.01}$ \\ 
58 & 00:46:03.706 & 00:46:07.398 & 00:46:06.445 & 2.511 & $2.05_{-0.06}^{+0.07}$ & $84_{-6}^{+7}$ & $-7.84_{-0.01}^{+0.01}$ \\ 
59 & 00:46:07.737 & 00:46:08.438 & 00:46:08.257 & 0.602 & $1.3_{-0.2}^{+0.3}$ & $37_{-15}^{+24}$ & $-8.83_{-0.08}^{+0.09}$ \\ 
60 & 00:46:11.078 & 00:46:12.430 & 00:46:12.379 & 1.277 & $1.1_{-0.1}^{+0.1}$ & $42_{-13}^{+18}$ & $-9.1_{-0.06}^{+0.06}$ \\ 
61 & 00:46:14.090 & 00:46:15.122 & 00:46:14.233 & 0.84 & $1.4_{-0.1}^{+0.1}$ & $53_{-13}^{+17}$ & $-8.62_{-0.05}^{+0.05}$ \\ 
62 & 00:46:15.566 & 00:46:19.285 & 00:46:18.015 & 2.852 & $1.97_{-0.05}^{+0.05}$ & $120_{-7}^{+8}$ & $-7.74_{-0.01}^{+0.01}$ \\ 
63 & 00:46:19.974 & 00:46:25.147 & 00:46:20.243 & 4.519 & $2.77_{-0.04}^{+0.04}$ & $270_{-7}^{+7}$ & $-6.95_{-0.01}^{+0.01}$ \\ 
64 & 00:46:26.731 & 00:46:28.005 & 00:46:27.415 & 1.129 & $1.3_{-0.1}^{+0.2}$ & $68_{-21}^{+31}$ & $-8.6_{-0.04}^{+0.04}$ \\ 
65 & 00:46:28.483 & 00:46:31.186 & 00:46:29.769 & 1.263 & $2.1_{-0.1}^{+0.1}$ & $80_{-9}^{+10}$ & $-7.86_{-0.02}^{+0.02}$ \\ 
66 & 00:46:33.057 & 00:46:38.340 & 00:46:33.658 & 3.7 & $1.97_{-0.05}^{+0.05}$ & $86_{-6}^{+6}$ & $-7.88_{-0.01}^{+0.01}$ \\ 
67 & 00:46:38.741 & 00:46:41.797 & 00:46:40.541 & 1.906 & $1.59_{-0.08}^{+0.09}$ & $55_{-8}^{+9}$ & $-8.38_{-0.03}^{+0.03}$ \\ 
68 & 00:46:42.755 & 00:46:44.073 & 00:46:43.237 & 0.61 & $2.98_{-0.08}^{+0.09}$ & $430_{-24}^{+25}$ & $-6.67_{-0.01}^{+0.01}$ \\ 
69 & 00:46:45.733 & 00:46:48.221 & 00:46:46.829 & 1.45 & $1.23_{-0.08}^{+0.09}$ & $70_{-14}^{+16}$ & $-8.69_{-0.04}^{+0.04}$ \\ 
70 & 00:46:48.223 & 00:46:49.132 & 00:46:48.887 & 0.697 & $2.0_{-0.3}^{+0.4}$ & $19_{-6}^{+9}$ & $-8.54_{-0.07}^{+0.07}$ \\ 
71 & 00:46:50.226 & 00:46:51.328 & 00:46:50.557 & 0.923 & $1.6_{-0.1}^{+0.1}$ & $54_{-10}^{+12}$ & $-8.35_{-0.04}^{+0.04}$ \\ 
72 & 00:46:54.959 & 00:46:57.714 & 00:46:56.705 & 2.204 & $1.76_{-0.06}^{+0.07}$ & $82_{-8}^{+8}$ & $-8.06_{-0.02}^{+0.02}$ \\ 
73 & 00:46:58.863 & 00:47:00.752 & 00:46:59.743 & 0.904 & $1.66_{-0.06}^{+0.07}$ & $180_{-18}^{+20}$ & $-7.81_{-0.02}^{+0.02}$ \\ 
74 & 00:47:01.372 & 00:47:02.979 & 00:47:02.017 & 1.086 & $1.9_{-0.1}^{+0.2}$ & $38_{-7}^{+8}$ & $-8.32_{-0.04}^{+0.04}$ \\ 
75 & 00:47:03.530 & 00:47:06.365 & 00:47:04.505 & 1.721 & $2.0_{-0.1}^{+0.1}$ & $66_{-7}^{+8}$ & $-7.98_{-0.02}^{+0.02}$ \\ 
76 & 00:47:08.250 & 00:47:11.064 & 00:47:09.756 & 2.044 & $1.3_{-0.1}^{+0.1}$ & $45_{-9}^{+11}$ & $-8.76_{-0.04}^{+0.04}$ \\ 
77 & 00:47:11.560 & 00:47:16.812 & 00:47:14.611 & 4.472 & $1.76_{-0.06}^{+0.06}$ & $50_{-4}^{+5}$ & $-8.28_{-0.02}^{+0.02}$ \\ 
78 & 00:47:18.051 & 00:47:19.906 & 00:47:18.948 & 0.91 & $2.01_{-0.07}^{+0.07}$ & $194_{-16}^{+17}$ & $-7.5_{-0.02}^{+0.02}$ \\ 
79 & 00:47:24.515 & 00:47:25.576 & 00:47:25.105 & 0.711 & $2.4_{-0.1}^{+0.1}$ & $210_{-17}^{+18}$ & $-7.27_{-0.01}^{+0.02}$ \\ 
80 & 00:47:26.949 & 00:47:27.578 & 00:47:27.151 & 0.409 & $1.4_{-0.2}^{+0.3}$ & $48_{-17}^{+25}$ & $-8.6_{-0.07}^{+0.08}$ \\ 
81 & 00:47:30.365 & 00:47:31.241 & 00:47:30.390 & 0.789 & $1.1_{-0.1}^{+0.2}$ & $56_{-20}^{+30}$ & $-8.98_{-0.07}^{+0.07}$ \\ 
82 & 00:47:31.729 & 00:47:32.925 & 00:47:32.145 & 0.579 & $2.0_{-0.1}^{+0.1}$ & $130_{-17}^{+19}$ & $-7.72_{-0.03}^{+0.03}$ \\ 
83 & 00:47:34.858 & 00:47:36.364 & 00:47:35.689 & 0.809 & $2.17_{-0.06}^{+0.07}$ & $300_{-20}^{+22}$ & $-7.21_{-0.01}^{+0.01}$ \\ 
84 & 00:47:38.281 & 00:47:39.484 & 00:47:39.312 & 0.861 & $1.3_{-0.1}^{+0.1}$ & $71_{-17}^{+21}$ & $-8.58_{-0.05}^{+0.05}$ \\ 
85 & 00:47:41.916 & 00:47:42.307 & 00:47:41.968 & 0.324 & $1.5_{-0.2}^{+0.3}$ & $80_{-27}^{+37}$ & $-8.35_{-0.07}^{+0.08}$ \\ 
86 & 00:47:43.310 & 00:47:43.516 & 00:47:43.459 & 0.157 & \ldots & \ldots & \ldots \\ 
87 & 00:47:44.125 & 00:47:46.512 & 00:47:44.331 & 1.941 & $1.7_{-0.1}^{+0.1}$ & $35_{-6}^{+7}$ & $-8.49_{-0.03}^{+0.03}$ \\ 
88 & 00:47:46.827 & 00:47:47.843 & 00:47:47.056 & 0.826 & $1.8_{-0.3}^{+0.4}$ & $15_{-6}^{+8}$ & $-8.81_{-0.08}^{+0.08}$ \\ 
89 & 00:47:51.486 & 00:47:52.608 & 00:47:52.328 & 0.719 & $2.02_{-0.08}^{+0.08}$ & $200_{-19}^{+20}$ & $-7.48_{-0.02}^{+0.02}$ \\ 
90 & 00:47:56.487 & 00:47:56.836 & 00:47:56.536 & 0.136 & $1.9_{-0.5}^{+1.0}$ & $25_{-15}^{+28}$ & $-8.5_{-0.1}^{+0.2}$ \\ 
91 & 00:47:57.478 & 00:47:58.779 & 00:47:57.573 & 0.748 & $2.45_{-0.07}^{+0.08}$ & $300_{-19}^{+21}$ & $-7.06_{-0.01}^{+0.01}$ \\ 
92 & 00:47:59.713 & 00:48:02.092 & 00:48:00.243 & 1.446 & $1.9_{-0.06}^{+0.06}$ & $170_{-13}^{+14}$ & $-7.63_{-0.01}^{+0.01}$ \\ 
93 & 00:48:07.963 & 00:48:08.779 & 00:48:08.251 & 0.453 & $2.1_{-0.2}^{+0.2}$ & $85_{-15}^{+18}$ & $-7.82_{-0.03}^{+0.04}$ \\ 
94 & 00:48:20.336 & 00:48:24.484 & 00:48:23.754 & 3.508 & $1.5_{-0.1}^{+0.1}$ & $26_{-4}^{+5}$ & $-8.76_{-0.03}^{+0.03}$ \\ 
95 & 00:48:26.817 & 00:48:28.266 & 00:48:27.020 & 1.129 & $1.79_{-0.07}^{+0.07}$ & $140_{-14}^{+15}$ & $-7.81_{-0.02}^{+0.02}$ \\ 
96 & 00:48:32.660 & 00:48:34.477 & 00:48:33.367 & 0.938 & $1.98_{-0.08}^{+0.08}$ & $150_{-14}^{+15}$ & $-7.63_{-0.02}^{+0.02}$ \\ 
97 & 00:48:43.468 & 00:48:46.029 & 00:48:44.858 & 0.644 & $2.55_{-0.07}^{+0.07}$ & $430_{-25}^{+26.0}$ & $-6.85_{-0.01}^{+0.01}$ \\ 
98 & 00:48:48.288 & 00:48:52.294 & 00:48:49.317 & 2.339 & $2.68_{-0.05}^{+0.06}$ & $210_{-9}^{+9}$ & $-7.1_{-0.01}^{+0.01}$ \\ 
99 & 00:48:52.729 & 00:48:57.134 & 00:48:56.199 & 3.788 & $1.89_{-0.05}^{+0.05}$ & $88_{-6}^{+6}$ & $-7.93_{-0.01}^{+0.01}$ \\ 
100 & 00:48:57.369 & 00:49:28.181 & 00:48:57.373 & 30.811 & $2.15_{-0.02}^{+0.02}$ & $180_{-5}^{+5}$ & $-7.43_{-0.01}^{+0.01}$ \\ 
101 & 00:49:32.031 & 00:49:33.119 & 00:49:32.702 & 0.821 & $1.27_{-0.08}^{+0.09}$ & $130_{-24}^{+28}$ & $-8.35_{-0.03}^{+0.04}$ \\ 
102 & 00:49:33.351 & 00:49:34.533 & 00:49:34.115 & 0.853 & $2.1_{-0.2}^{+0.2}$ & $42_{-7}^{+9}$ & $-8.1_{-0.03}^{+0.03}$ \\ 
103 & 00:49:35.236 & 00:49:36.404 & 00:49:35.948 & 0.799 & $1.99_{-0.07}^{+0.07}$ & $220_{-19}^{+20}$ & $-7.47_{-0.02}^{+0.02}$ \\ 
104 & 00:49:36.805 & 00:49:38.391 & 00:49:36.957 & 1.281 & $1.4_{-0.1}^{+0.1}$ & $35_{-9}^{+11}$ & $-8.77_{-0.05}^{+0.05}$ \\ 
105 & 00:49:38.688 & 00:49:41.148 & 00:49:40.293 & 2.106 & $1.6_{-0.1}^{+0.1}$ & $39_{-6}^{+7}$ & $-8.49_{-0.03}^{+0.03}$ \\ 
106 & 00:49:41.984 & 00:49:43.944 & 00:49:42.741 & 1.408 & $1.58_{-0.08}^{+0.09}$ & $72_{-10}^{+12}$ & $-8.28_{-0.03}^{+0.03}$ \\ 
107 & 00:49:45.692 & 00:49:49.384 & 00:49:46.772 & 1.321 & $2.72_{-0.07}^{+0.07}$ & $220_{-12}^{+13}$ & $-7.06_{-0.01}^{+0.01}$ \\ 
108 & 00:49:49.575 & 00:49:50.140 & 00:49:50.079 & 0.484 & $1.5_{-0.2}^{+0.2}$ & $50_{-15}^{+21}$ & $-8.57_{-0.06}^{+0.07}$ \\ 
109 & 00:49:50.658 & 00:49:53.730 & 00:49:52.271 & 2.119 & $1.92_{-0.08}^{+0.09}$ & $59_{-6}^{+7}$ & $-8.08_{-0.02}^{+0.02}$ \\ 
110 & 00:49:54.529 & 00:49:56.177 & 00:49:55.078 & 1.197 & $1.41_{-0.08}^{+0.08}$ & $89_{-14}^{+16}$ & $-8.36_{-0.03}^{+0.03}$ \\ 
111 & 00:49:56.997 & 00:50:02.615 & 00:50:01.371 & 3.538 & $2.49_{-0.05}^{+0.05}$ & $140_{-6}^{+6}$ & $-7.36_{-0.01}^{+0.01}$ \\ 
112 & 00:50:03.039 & 00:50:04.943 & 00:50:03.786 & 0.803 & $2.1_{-0.1}^{+0.1}$ & $87_{-11}^{+13}$ & $-7.83_{-0.03}^{+0.03}$ \\ 
113 & 00:50:05.322 & 00:50:05.908 & 00:50:05.664 & 0.44 & $1.28_{-0.08}^{+0.09}$ & $210_{-39}^{+47}$ & $-8.14_{-0.03}^{+0.04}$ \\ 
114 & 00:50:09.077 & 00:50:09.690 & 00:50:09.487 & 0.424 & $1.8_{-0.2}^{+0.3}$ & $35_{-11}^{+15}$ & $-8.37_{-0.07}^{+0.07}$ \\ 
115 & 00:50:11.166 & 00:50:11.870 & 00:50:11.182 & 0.624 & $1.2_{-0.1}^{+0.2}$ & $65_{-21}^{+30}$ & $-8.77_{-0.06}^{+0.07}$ \\ 
116 & 00:50:13.445 & 00:50:15.528 & 00:50:14.097 & 1.952 & $1.5_{-0.1}^{+0.1}$ & $43_{-7}^{+8}$ & $-8.54_{-0.03}^{+0.03}$ \\ 
117 & 00:50:16.502 & 00:50:17.943 & 00:50:17.529 & 1.172 & $1.56_{-0.08}^{+0.09}$ & $85_{-12}^{+14}$ & $-8.23_{-0.03}^{+0.03}$ \\ 
118 & 00:50:19.672 & 00:50:22.559 & 00:50:22.013 & 2.242 & $2.04_{-0.06}^{+0.06}$ & $120_{-8}^{+9}$ & $-7.71_{-0.01}^{+0.01}$ \\ 
119 & 00:50:30.649 & 00:50:30.973 & 00:50:30.726 & 0.237 & $1.2_{-0.3}^{+0.5}$ & $61_{-36}^{+71}$ & $-8.8_{-0.1}^{+0.2}$ \\ 
120 & 00:50:34.424 & 00:50:35.720 & 00:50:34.936 & 0.674 & $1.9_{-0.1}^{+0.1}$ & $150_{-18}^{+19}$ & $-7.71_{-0.02}^{+0.02}$ \\ 
121 & 00:50:39.025 & 00:50:41.308 & 00:50:39.645 & 1.926 & $1.9_{-0.1}^{+0.1}$ & $54_{-6}^{+7}$ & $-8.14_{-0.02}^{+0.02}$ \\ 
122 & 00:50:41.808 & 00:50:43.410 & 00:50:41.854 & 0.589 & $2.2_{-0.1}^{+0.1}$ & $180_{-18}^{+19}$ & $-7.41_{-0.02}^{+0.02}$ \\ 
123 & 00:50:44.090 & 00:50:45.374 & 00:50:45.171 & 1.159 & $1.4_{-0.1}^{+0.2}$ & $35_{-10}^{+13}$ & $-8.83_{-0.05}^{+0.06}$ \\ 
124 & 00:50:49.604 & 00:50:49.837 & 00:50:49.691 & 0.191 & $1.1_{-0.1}^{+0.2}$ & $180_{-67}^{+100}$ & $-8.54_{-0.07}^{+0.07}$ \\ 
125 & 00:50:50.202 & 00:50:51.121 & 00:50:50.751 & 0.748 & $1.4_{-0.1}^{+0.2}$ & $42_{-12}^{+16}$ & $-8.73_{-0.06}^{+0.06}$ \\ 
126 & 00:50:53.950 & 00:50:55.966 & 00:50:54.793 & 1.273 & $1.6_{-0.1}^{+0.2}$ & $6_{-7}^{+9}$ & $-8.55_{-0.04}^{+0.04}$ \\ 
127 & 00:50:58.262 & 00:50:58.915 & 00:50:58.869 & 0.605 & $1.3_{-0.2}^{+0.4}$ & $25_{-13}^{+23}$ & $-9.11_{-0.1}^{+0.1}$ \\ 
128 & 00:50:59.373 & 00:51:00.209 & 00:51:00.078 & 0.669 & $1.0_{-0.2}^{+0.3}$ & $46_{-26}^{+51}$ & $-9.3_{-0.1}^{+0.1}$ \\ 
129 & 00:51:00.731 & 00:51:01.874 & 00:51:01.562 & 0.975 & $1.4_{-0.2}^{+0.3}$ & $22_{-9}^{+13}$ & $-8.96_{-0.08}^{+0.09}$ \\ 
130 & 00:51:03.472 & 00:51:05.917 & 00:51:04.668 & 0.747 & $1.99_{-0.06}^{+0.06}$ & $300_{-22}^{+24}$ & $-7.32_{-0.01}^{+0.01}$ \\ 
131 & 00:51:07.203 & 00:51:07.937 & 00:51:07.723 & 0.674 & $1.1_{-0.1}^{+0.2}$ & $47_{-19}^{+31}$ & $-9.11_{-0.08}^{+0.08}$ \\ 
132 & 00:51:10.880 & 00:51:11.067 & 00:51:10.887 & 0.143 & $0.9_{-0.1}^{+0.2}$ & $270_{-123}^{+214}$ & $-8.64_{-0.08}^{+0.08}$ \\ 
133 & 00:51:12.408 & 00:51:13.300 & 00:51:12.758 & 0.421 & $0.8_{-0.1}^{+0.2}$ & $130_{-1}^{+110}$ & $-9.06_{-0.08}^{+0.08}$ \\ 
134 & 00:51:14.367 & 00:51:15.220 & 00:51:15.032 & 0.786 & $1.5_{-0.2}^{+0.2}$ & $25_{-8}^{+11}$ & $-8.81_{-0.07}^{+0.07}$ \\ 
135 & 00:51:15.791 & 00:51:16.934 & 00:51:16.746 & 0.942 & $1.6_{-0.2}^{+0.3}$ & $17_{-6}^{+8}$ & $-8.86_{-0.07}^{+0.07}$ \\ 
136 & 00:51:19.196 & 00:51:20.702 & 00:51:19.399 & 1.083 & $1.65_{-0.06}^{+0.06}$ & $170_{-17}^{+18}$ & $-7.83_{-0.02}^{+0.02}$ \\ 
137 & 00:51:22.604 & 00:51:24.764 & 00:51:23.832 & 1.493 & $2.0_{-0.2}^{+0.2}$ & $19_{-4}^{+5}$ & $-8.51_{-0.04}^{+0.04}$ \\ 
138 & 00:51:25.767 & 00:51:27.434 & 00:51:26.980 & 1.29 & $1.37_{-0.07}^{+0.08}$ & $98_{-14}^{+16}$ & $-8.37_{-0.03}^{+0.03}$ \\ 
139 & 00:51:30.529 & 00:51:33.113 & 00:51:32.226 & 1.532 & $1.49_{-0.08}^{+0.09}$ & $73_{-11}^{+12}$ & $-8.37_{-0.03}^{+0.03}$ \\ 
140 & 00:51:35.325 & 00:51:36.754 & 00:51:35.932 & 0.525 & $2.43_{-0.08}^{+0.09}$ & $340_{-25}^{+26}$ & $-7.02_{-0.01}^{+0.01}$ \\ 
141 & 00:51:54.916 & 00:51:56.684 & 00:51:55.467 & 0.757 & $2.41_{-0.07}^{+0.07}$ & $330_{-21}^{+22}$ & $-7.04_{-0.01}^{+0.01}$ \\ 
142 & 00:52:05.906 & 00:52:08.335 & 00:52:06.245 & 0.636 & $2.3_{-0.1}^{+0.1}$ & $130_{-14}^{+16}$ & $-7.47_{-0.02}^{+0.02}$ \\ 
143 & 00:52:09.678 & 00:52:10.788 & 00:52:10.355 & 0.875 & $1.2_{-0.1}^{+0.2}$ & $41_{-13}^{+18}$ & $-8.93_{-0.06}^{+0.06}$ \\ 
144 & 00:52:16.571 & 00:52:17.667 & 00:52:16.596 & 1.032 & $1.0_{-0.2}^{+0.2}$ & $30_{-14}^{+25}$ & $-9.4_{-0.1}^{+0.1}$ \\ 
145 & 00:52:18.925 & 00:52:20.712 & 00:52:19.884 & 1.352 & $1.67_{-0.08}^{+0.09}$ & $70_{-9}^{+11}$ & $-8.21_{-0.03}^{+0.03}$ \\ 
146 & 00:52:25.260 & 00:52:25.969 & 00:52:25.320 & 0.655 & $1.1_{-0.1}^{+0.2}$ & $57_{-22}^{+34}$ & $-9.02_{-0.08}^{+0.08}$ \\ 
147 & 00:52:28.128 & 00:52:28.881 & 00:52:28.554 & 0.565 & $1.5_{-0.3}^{+0.5}$ & $23_{-12}^{+21}$ & $-8.9_{-0.1}^{+0.1}$ \\ 
148 & 00:52:34.103 & 00:52:35.064 & 00:52:34.931 & 0.792 & $1.8_{-0.2}^{+0.3}$ & $25_{-7}^{+10}$ & $-8.58_{-0.06}^{+0.06}$ \\ 
149 & 00:52:36.287 & 00:52:37.256 & 00:52:36.998 & 0.885 & $2.2_{-0.3}^{+0.5}$ & $9_{-3}^{+5}$ & $-8.7_{-0.07}^{+0.07}$ \\ 
150 & 00:52:37.706 & 00:52:38.508 & 00:52:37.807 & 0.555 & $1.3_{-0.1}^{+0.1}$ & $95_{-25}^{+32}$ & $-8.5_{-0.05}^{+0.06}$ \\ 
151 & 00:52:39.462 & 00:52:40.935 & 00:52:40.158 & 0.811 & $1.0_{-0.1}^{+0.1}$ & $90_{-24}^{+32}$ & $-8.85_{-0.05}^{+0.05}$ \\ 
152 & 00:52:44.491 & 00:52:45.536 & 00:52:44.761 & 0.899 & $1.4_{-0.2}^{+0.2}$ & $32_{-9}^{+13}$ & $-8.81_{-0.06}^{+0.06}$ \\ 
153 & 00:52:48.433 & 00:52:49.067 & 00:52:48.753 & 0.513 & $1.6_{-0.2}^{+0.3}$ & $36_{-12}^{+16}$ & $-8.53_{-0.07}^{+0.07}$ \\ 
154 & 00:52:49.637 & 00:52:50.103 & 00:52:49.700 & 0.412 & $2.2_{-0.4}^{+0.8}$ & $14_{-7}^{+10}$ & $-8.5_{-0.1}^{+0.1}$ \\ 
155 & 00:52:53.507 & 00:52:54.099 & 00:52:53.591 & 0.549 & $1.3_{-0.2}^{+0.3}$ & $24_{-11}^{+19}$ & $-9.1_{-0.1}^{+0.1}$ \\ 
156 & 00:52:54.982 & 00:52:56.013 & 00:52:55.223 & 0.629 & $3.0_{-0.3}^{+0.4}$ & $28_{-6}^{+7}$ & $-7.87_{-0.04}^{+0.04}$ \\ 
157 & 00:53:09.712 & 00:53:10.469 & 00:53:09.842 & 0.505 & $1.4_{-0.1}^{+0.1}$ & $110_{-23}^{+28}$ & $-8.28_{-0.04}^{+0.04}$ \\ 
158 & 00:53:24.194 & 00:53:26.949 & 00:53:25.144 & 1.667 & $1.52_{-0.04}^{+0.04}$ & $220_{-17}^{+18}$ & $-7.85_{-0.01}^{+0.01}$ \\ 
159 & 00:53:28.982 & 00:53:29.949 & 00:53:29.106 & 0.896 & $1.5_{-0.2}^{+0.3}$ & $17_{-6}^{+9}$ & $-8.93_{-0.08}^{+0.08}$ \\ 
160 & 00:53:30.310 & 00:53:30.390 & 00:53:30.342 & 0.064 & \ldots & \ldots & \ldots \\ 
161 & 00:53:37.836 & 00:53:40.169 & 00:53:38.865 & 1.438 & $2.03_{-0.06}^{+0.06}$ & $180_{-13}^{+13}$ & $-7.51_{-0.01}^{+0.01}$ \\ 
162 & 00:53:41.691 & 00:53:42.150 & 00:53:41.866 & 0.23 & $2.0_{-0.4}^{+0.6}$ & $28_{-12}^{+18}$ & $-8.4_{-0.1}^{+0.1}$ \\ 
163 & 00:53:45.002 & 00:53:45.823 & 00:53:45.267 & 0.582 & $1.4_{-0.2}^{+0.2}$ & $45_{-14}^{+19}$ & $-8.63_{-0.06}^{+0.07}$ \\ 
164 & 00:53:48.619 & 00:53:49.493 & 00:53:48.630 & 0.655 & $1.0_{-0.1}^{+0.2}$ & $52_{-22}^{+35}$ & $-9.11_{-0.08}^{+0.08}$ \\ 
165 & 00:53:51.979 & 00:53:54.304 & 00:53:52.330 & 0.518 & $2.1_{-0.1}^{+0.1}$ & $210_{-21}^{+23}$ & $-7.42_{-0.02}^{+0.02}$ \\ 
166 & 00:53:58.858 & 00:53:59.338 & 00:53:59.009 & 0.258 & $1.0_{-0.1}^{+0.2}$ & $130_{-49}^{+76}$ & $-8.71_{-0.08}^{+0.08}$ \\ 
167 & 00:54:03.982 & 00:54:04.455 & 00:54:04.045 & 0.309 & $1.5_{-0.3}^{+0.4}$ & $37_{-17}^{+27}$ & $-8.7_{-0.1}^{+0.1}$ \\ 
168 & 00:54:06.986 & 00:54:08.459 & 00:54:07.803 & 1.124 & $1.3_{-0.2}^{+0.3}$ & $14_{-7}^{+11}$ & $-9.3_{-0.1}^{+0.1}$ \\ 
169 & 00:54:10.816 & 00:54:11.717 & 00:54:10.899 & 0.524 & $2.1_{-0.2}^{+0.3}$ & $43_{-10}^{+12}$ & $-8.10_{-0.04}^{+0.05}$ \\ 
170 & 00:54:12.364 & 00:54:14.865 & 00:54:13.657 & 2.053 & $1.4_{-0.1}^{+0.1}$ & $22_{-5}^{+7}$ & $-9.00_{-0.05}^{+0.05}$ \\ 
171 & 00:54:27.094 & 00:54:28.184 & 00:54:27.206 & 0.291 & $2.4_{-0.2}^{+0.3}$ & $76_{-15}^{+18}$ & $-7.70_{-0.04}^{+0.04}$ \\ 
172 & 00:54:28.759 & 00:54:29.723 & 00:54:29.083 & 0.596 & $1.4_{-0.1}^{+0.1}$ & $75_{-18}^{+22}$ & $-8.46_{-0.05}^{+0.05}$ \\ 
173 & 00:54:31.305 & 00:54:31.503 & 00:54:31.380 & 0.132 & \ldots & \ldots & \ldots \\ 
174 & 00:54:45.094 & 00:54:45.445 & 00:54:45.166 & 0.239 & $1.4_{-0.3}^{+0.5}$ & $52_{-28}^{+52}$ & $-8.6_{-0.1}^{+0.1}$ \\ 
175 & 00:54:45.893 & 00:54:47.504 & 00:54:46.135 & 1.076 & $1.9_{-0.1}^{+0.2}$ & $34_{-7}^{+8}$ & $-8.36_{-0.04}^{+0.04}$ \\ 
176 & 00:54:48.694 & 00:54:50.134 & 00:54:49.606 & 1.212 & $1.9_{-0.2}^{+0.3}$ & $13_{-4}^{+5}$ & $-8.72_{-0.06}^{+0.06}$ \\ 
177 & 00:54:52.418 & 00:54:54.214 & 00:54:53.356 & 1.549 & $2.4_{-0.7}^{+2}$ & $3_{-2}^{+3}$ & $-9.1_{-0.1}^{+0.2}$ \\ 
178 & 00:54:56.294 & 00:55:00.410 & 00:54:57.506 & 1.75 & $2.9_{-0.1}^{+0.1}$ & $160_{-8}^{+9}$ & $-7.15_{-0.01}^{+0.01}$ \\ 
179 & 00:55:04.966 & 00:55:05.956 & 00:55:05.902 & 0.947 & $2.6_{-0.7}^{+2}$ & $3_{-2}^{+3}$ & $-9.0_{-0.1}^{+0.1}$ \\ 
180 & 00:55:07.227 & 00:55:08.359 & 00:55:07.428 & 0.995 & $1.5_{-0.2}^{+0.2}$ & $30_{-9}^{+12}$ & $-8.71_{-0.06}^{+0.07}$ \\ 
181 & 00:55:15.923 & 00:55:16.818 & 00:55:16.313 & 0.696 & $1.1_{-0.2}^{+0.2}$ & $43_{-18}^{+30}$ & $-9.11_{-0.08}^{+0.09}$ \\ 
182 & 00:55:21.964 & 00:55:23.251 & 00:55:22.728 & 0.703 & $1.7_{-0.1}^{+0.2}$ & $65_{-13}^{+15}$ & $-8.21_{-0.04}^{+0.04}$ \\ 
183 & 00:55:32.652 & 00:55:33.519 & 00:55:33.087 & 0.656 & $1.5_{-0.2}^{+0.3}$ & $20_{-8}^{+11}$ & $-8.89_{-0.08}^{+0.09}$ \\ 
184 & 00:55:36.705 & 00:55:37.475 & 00:55:37.363 & 0.546 & \ldots & \ldots & \ldots \\ 
185 & 00:55:39.902 & 00:55:40.236 & 00:55:39.985 & 0.222 & $1.9_{-0.2}^{+0.3}$ & $64_{-19}^{+26}$ & $-8.04_{-0.06}^{+0.06}$ \\ 
186 & 00:55:42.231 & 00:55:43.249 & 00:55:42.913 & 0.91 & $1.4_{-0.2}^{+0.2}$ & $31_{-9}^{+13}$ & $-8.8_{-0.06}^{+0.06}$ \\ 
187 & 00:55:54.569 & 00:55:58.916 & 00:55:56.351 & 2.408 & $1.9_{-0.1}^{+0.1}$ & $51_{-5}^{+6}$ & $-8.16_{-0.02}^{+0.02}$ \\ 
188 & 00:56:03.231 & 00:56:05.963 & 00:56:04.480 & 2.209 & $1.0_{-0.1}^{+0.1}$ & $33_{-10}^{+15}$ & $-9.36_{-0.06}^{+0.06}$ \\ 
189 & 00:56:14.911 & 00:56:17.318 & 00:56:16.715 & 1.829 & $2.4_{-0.3}^{+0.4}$ & $8_{-2}^{+3}$ & $-8.67_{-0.05}^{+0.05}$ \\ 
190 & 00:56:18.852 & 00:56:22.785 & 00:56:20.144 & 2.813 & $1.7_{-0.1}^{+0.1}$ & $30_{-4}^{+5}$ & $-8.59_{-0.03}^{+0.03}$ \\ 
191 & 00:56:24.810 & 00:56:25.919 & 00:56:24.827 & 0.808 & $1.4_{-0.1}^{+0.1}$ & $95_{-17}^{+21}$ & $-8.37_{-0.04}^{+0.04}$ \\ 
192 & 00:56:27.624 & 00:56:27.778 & 00:56:27.635 & 0.139 & \ldots & \ldots & \ldots \\ 
193 & 00:56:36.565 & 00:56:37.742 & 00:56:36.597 & 1.082 & $1.6_{-0.3}^{+0.5}$ & $10_{-5}^{+7}$ & $-9.1_{-0.1}^{+0.1}$ \\ 
194 & 00:56:42.355 & 00:56:44.586 & 00:56:42.937 & 1.133 & $2.3_{-0.6}^{+1}$ & $4_{-2}^{+3}$ & $-9.1_{-0.1}^{+0.1}$ \\ 
195 & 00:56:45.939 & 00:56:46.110 & 00:56:46.009 & 0.085 & $1.2_{-0.3}^{+0.5}$ & $130_{-74}^{+150}$ & $-8.5_{-0.1}^{+0.2}$ \\ 
196 & 00:56:48.694 & 00:56:51.315 & 00:56:49.842 & 0.712 & $2.27_{-0.06}^{+0.07}$ & $380_{-24}^{+25}$ & $-7.05_{-0.01}^{+0.01}$ \\ 
197 & 00:57:00.538 & 00:57:01.365 & 00:57:00.921 & 0.604 & $2.0_{-0.4}^{+0.6}$ & $14_{-6}^{+8}$ & $-8.7_{-0.1}^{+0.1}$ \\ 
198 & 00:57:05.060 & 00:57:05.354 & 00:57:05.089 & 0.084 & \ldots & \ldots & \ldots \\ 
199 & 00:57:20.313 & 00:57:20.770 & 00:57:20.368 & 0.293 & $2.2_{-0.3}^{+0.4}$ & $33_{-11}^{+14}$ & $-8.17_{-0.06}^{+0.07}$ \\ 
200 & 00:57:22.727 & 00:57:23.894 & 00:57:23.647 & 0.823 & $1.8_{-0.2}^{+0.3}$ & $15_{-5}^{+7}$ & $-8.74_{-0.07}^{+0.07}$ \\ 
201 & 00:57:28.352 & 00:57:29.815 & 00:57:28.429 & 1.228 & $1.3_{-0.3}^{+0.6}$ & $8.6_{-5}^{+11}$ & $-9.6_{-0.2}^{+0.2}$ \\ 
202 & 00:57:35.366 & 00:57:36.718 & 00:57:35.854 & 1.059 & $1.4_{-0.1}^{+0.1}$ & $53_{-11}^{+13}$ & $-8.56_{-0.04}^{+0.04}$ \\ 
203 & 00:57:45.709 & 00:57:46.407 & 00:57:46.063 & 0.54 & $1.6_{-0.3}^{+0.4}$ & $24_{-10}^{+15}$ & $-8.7_{-0.1}^{+0.1}$ \\ 
204 & 00:57:48.155 & 00:57:48.759 & 00:57:48.299 & 0.091 & $2.9_{-0.4}^{+0.6}$ & $100_{-29}^{+36}$ & $-7.32_{-0.05}^{+0.05}$ \\ 
205 & 00:57:53.779 & 00:57:54.492 & 00:57:54.070 & 0.578 & $1.5_{-0.3}^{+0.3}$ & $29_{-10}^{+15}$ & $-8.81_{-0.08}^{+0.08}$ \\ 
206 & 00:57:56.452 & 00:57:58.414 & 00:57:57.069 & 0.878 & $1.7_{-0.1}^{+0.1}$ & $91_{-13}^{+15}$ & $-8.06_{-0.03}^{+0.03}$ \\ 
207 & 00:58:02.249 & 00:58:03.098 & 00:58:03.009 & 0.499 & \ldots & \ldots & \ldots \\ 
208 & 00:58:05.090 & 00:58:05.783 & 00:58:05.625 & 0.508 & $1.4_{-0.3}^{+0.6}$ & $11.15_{-7}^{+14}$ & $-9.3_{-0.1}^{+0.2}$ \\ 
209 & 00:58:12.219 & 00:58:12.485 & 00:58:12.225 & 0.228 & $1.9_{-0.5}^{+1}$ & $12.42_{-8}^{+17}$ & $-8.8_{-0.2}^{+0.2}$ \\ 
210 & 00:58:18.286 & 00:58:18.534 & 00:58:18.326 & 0.211 & $1.9_{-0.3}^{+0.5}$ & $38_{-15}^{+22}$ & $-8.27_{-0.08}^{+0.09}$ \\ 
211 & 00:58:30.109 & 00:58:31.671 & 00:58:31.214 & 1.429 & $0.9_{-0.1}^{+0.2}$ & $36_{-16}^{+28}$ & $-9.51_{-0.08}^{+0.08}$ \\ 
212 & 00:58:42.989 & 00:58:46.172 & 00:58:45.280 & 2.232 & $1.3_{-0.04}^{+0.04}$ & $200_{-18}^{+19}$ & $-8.20_{-0.02}^{+0.02}$ \\ 
213 & 00:58:47.121 & 00:58:47.949 & 00:58:47.429 & 0.591 & $1.4_{-0.4}^{+0.8}$ & $9_{-6}^{+13}$ & $-9.3_{-0.2}^{+0.2}$ \\ 
214 & 00:58:50.421 & 00:58:51.319 & 00:58:50.622 & 0.748 & $1.0_{-0.2}^{+0.2}$ & $37_{-17}^{+30}$ & $-9.3_{-0.1}^{+0.1}$ \\ 
215 & 00:58:52.656 & 00:58:52.921 & 00:58:52.678 & 0.133 & $1.6_{-0.2}^{+0.2}$ & $220_{-54}^{+69}$ & $-7.78_{-0.05}^{+0.05}$ \\ 
216 & 00:58:53.605 & 00:58:54.700 & 00:58:53.978 & 0.547 & $1.3_{-0.2}^{+0.2}$ & $55_{-18}^{+25}$ & $-8.66_{-0.07}^{+0.07}$ \\ 
217 & 00:58:56.906 & 00:58:58.271 & 00:58:57.484 & 0.825 & $1.9_{-0.1}^{+0.1}$ & $160_{-16}^{+18}$ & $-7.68_{-0.02}^{+0.02}$ \\ 
  \hline
218 & 03:55:37.447 & 03:55:39.450 & 03:55:37.683 & 1.532 & $0.75_{-0.08}^{+0.09}$ & $83_{-29}^{+44}$ & $-9.45_{-0.06}^{+0.06}$ \\ 
219 & 03:56:21.396 & 03:56:21.828 & 03:56:21.540 & 0.383 & $1.2_{-0.2}^{+0.3}$ & $32_{-17}^{+30}$ & $-9.1_{-0.1}^{+0.1}$ \\ 
220 & 05:26:40.266 & 05:26:41.375 & 05:26:40.345 & 1.029 & $2.3_{-0.6}^{+2}$ & $2_{-1}^{+3}$ & $-9.4_{-0.2}^{+0.2}$ \\ 
221 & 05:27:55.174 & 05:27:55.914 & 05:27:55.325 & 0.527 & \ldots & \ldots & \ldots \\ 
222 & 06:51:54.178 & 06:51:54.370 & 06:51:54.208 & 0.165 & $0.5_{-0.1}^{+0.1}$ & $970_{-560}^{+1300}$ & $-9.0_{-0.1}^{+0.1}$ \\ 
223 & 07:00:16.717 & 07:00:16.941 & 07:00:16.773 & 0.168 & $1.7_{-0.2}^{+0.3}$ & $53_{-19}^{+26}$ & $-8.4_{-0.1}^{+0.1}$ \\ 
\hline
\end{longtable}
\begin{list}{}{}
\item{{\bf Notes.} Times of bursts are from 2020-04-28. Fluxes are in the
  0.5-10~keV range.}
\end{list}
\end{center}

\end{document}

We report on NICER observations of the Magnetar SGR~1935+2154, covering its 2020 burst storm and long-term persistent emission evolution up to $\sim90$ days post outburst. During the first 1120~seconds taken on April 28 00:40:58 UTC we detect over 217 bursts, corresponding to a burst rate of $>0.2$~bursts~s$^{-1}$. Three hours later the rate is at 0.008~bursts~s$^{-1}$, remaining at a comparatively low level thereafter. The $T_{90}$ burst duration distribution peaks at 840~ms; the distribution of waiting times to the next burst is fit with a log-normal with an average of 2.1~s. The 1-10 keV burst spectra are well fit by a blackbody, with an average temperature and area of $kT=1.7$~keV and $R^2=53$~km$^2$. The differential burst fluence distribution over $\sim3$ orders of magnitude is well modeled with a power-law form $dN/dF\propto F^{-1.5\pm0.1}$. The source persistent emission pulse profile is double-peaked hours after the burst storm. We find that the bursts peak arrival times follow a uniform distribution in pulse phase, though the fast radio burst associated with the source aligns in phase with the brighter peak. We measure the source spin-down from heavy-cadence observations covering days 21 to 39 post-outburst, $\dot\nu=-3.72(3)\times10^{-12}$~Hz~s$^{-1}$; a factor 2.7 larger than the value measured after the 2014 outburst. Finally, the persistent emission flux and blackbody temperature decrease rapidly in the early stages of the outburst, reaching quiescence 40 days later, while the size of the emitting area remains unchanged.